\documentclass[a4paper,11pt]{article}
\usepackage{pos}
\usepackage{amsmath}
\usepackage{enumerate}
\usepackage[font=small]{caption,subfig}
\usepackage{tikz}
\usetikzlibrary{decorations.markings}

\font\FermiSmallfont=cmssq8 scaled 1200

\def\UCIppthead#1{
\null 
\begin{center}\vskip -0.7truein{\hbox to 5.9truein {
\hfill
\vbox to 1in {\vfill \FermiSmallfont
              \hbox{#1}
              \vfill}
}}\vskip-0.0truein\end{center}}

\title{Neutrinos in Astrophysics and Cosmology}

\author*{Kevork N.\ Abazajian}

\affiliation{Center for Cosmology, University of California, Irvine, Department of Physics \& Astronomy \\
  4129 F.\ Reines Hall, Irvine, CA 92697-4575, USA}

\emailAdd{kevork@uci.edu}

\abstract{I introduce the consequences of neutrino mass and mixing in the dense environments of the early Universe and in astrophysical environments. Thermal and matter effects are reviewed in the context of a two-neutrino formalism, with methods of extension to multiple neutrinos. The observed large neutrino mixing angles place the strongest constraint on cosmological lepton (or neutrino) asymmetries, while new sterile neutrinos provide a wealth of possible new physics, including lepton asymmetry generation as well as candidates for dark matter. I also review cosmic microwave background and large-scale structure constraints on neutrino mass and energy density. Lastly, I review how X-ray astronomy has become a branch of neutrino physics in searches for keV-scale sterile neutrino dark matter radiative decay. }

\FullConference{%
  TASI 2020 -- The Obscure Universe: Neutrinos and Other Dark Matters\\
  1-26 June, 2020\\
  Boulder, Colorado, USA}

\begin{document}

\UCIppthead{UCI-HEP-TR-2021-09}

\maketitle

\section{Introduction}
The discovery that neutrinos have mass and ``mix'' is one of the
biggest discoveries in particle physics in the past 25
years.\footnote{In fact, I was a graduate student attending TASI in
  1998 (June 1-26) at the same time Super-Kamiokande announced their
  discovery of atmospheric neutrino oscillations at the {\it XVIIIth
    International Conference on Neutrino Physics and Astrophysics} in
  Takayama, Japan, on June 7, 1998 \cite{Kajita1998}. We TASI students got to hear
  from lecturers who attended the conference about the details of the
  announcement and returned to participate in that TASI. Combined with
  the discovery of dark energy and the accelerating expansion of the
  Universe, it was a good year to be starting a research path toward
  neutrinos and cosmology. } There are numerous consequences of
massive, mixed neutrinos, which, despite their weak coupling and small
cross-section, have tremendous impact in the dense environments of the
early Universe and supernovae. Because of their weak couplings, they
are produced in thermal abundance only in high-lepton and baryon-number density
environments, and provide a probe of these novel cosmological and
astrophysical conditions. In this chapter, I will follow the
exposition in the summer lectures as I gave them.

As I delve into the topic of neutrinos in astrophysics and cosmology,
their impact in high-density environments will beg the use of methods
from condensed matter physics, and the quantum physics of their
oscillations in these environments borrows from
atomic-molecular-optical physics. Then, the effects of massive
neutrinos in cosmic microwave background (CMB) and large-scale structure (LSS)
requires the tools of Boltzmann evolution in a General Relativistic
background for the understanding of linear structure growth, though
detailed calculations of the Boltzmann equations need numerical
methods. Massive neutrino effects on nonlinear cosmological structure
necessarily require numerical methods. Lastly, the possibility that a
new, sterile, neutrino state exists with properties that make it a
partial or full component of dark matter combines much of these tools
to understand: high-density Boltzmann calculation of production
out-of-equilibrium, linear structure formation calculation effects,
and nonlinear simulations to tie to many observations of structures in
our nonlinear Universe. As an added bonus, I will discuss what turns
out to be one of the most powerful tools to probe the weaker-than-weak
sterile neutrinos that may be or contribute to the dark matter:
X-ray astronomy.

\section{Oscillations with Matter and Thermal Effects}
To start with a topic that should be familiar to anyone that has
worked even a little with neutrino oscillations, let us recall that
the flavor eigenstates of neutrinos are not identical with that of
their mass eigenstates,
\begin{align}
|\nu_\alpha \rangle =& \phantom{ - }\, \cos \theta |\nu_1\rangle + \sin \theta | \nu_2
 \rangle  \cr
|\nu_\beta \rangle =& - \sin \theta {|\nu_1\rangle} + \cos \theta {| \nu_2
 \rangle}\, .
\end{align}
An immediate consequence is that a neutrino produced in a flavor state
will evolve with time, $t$,
\begin{equation}
  e^{i(\vec{k} \cdot \vec{x} - \omega t)} =
  e^{i(\vec{k} \cdot \vec{x} - \sqrt{m_i^2 + k^2}t)} .
\end{equation}
The neutrino mass is small relative to its energy, $\sqrt{m_i^2+k^2}
\sim k\left(1 + {m_i^2/2k^2}\right) $, so the states evolve with a
relative phase dependent on $\delta m^2 = m_2^2 - m_1^2$,
\begin{align}
  |\nu(t) \rangle =& e^{i(\vec{k} \cdot \vec{x} - kt -(m_1^2+m_2^2)t/4k)} \cr
 & \times \left[\cos \theta |\nu_1 \rangle e^{i \delta m^2 t/4k}
+ \sin \theta |\nu_2 \rangle e^{-i \delta m^2 t/4k} \right]\, .
\end{align}
The overall phase, in front of the brackets above, does not matter, so
it is then simple to then calculate the probability of finding the
neutrino in a different flavor state in the future:
\begin{align}
  P_{\nu_\alpha} (t) &= | {\langle \nu_\alpha} {| \nu(t) \rangle} |^2
\cr
&= 1 - {\sin^2 2 \theta} \sin^2 \left({\delta m^2 t \over 4 
  k}\right)\cr
& =1 -{\sin^2 2 \theta }\sin^2 \left({\delta m^2c^4 L\over 4 
\hbar c E} \right) \, , \label{oscillation}
\end{align}
where the $\sin^22\theta$ is the amplitude of the oscillation and
$\sin^2(\delta m^2t/4k)$ is the phase of the observable
oscillation. In the last line above, I converted to include units and the
observable length $L$ and energy $E$ of the neutrinos. For a general
$n$ flavor state system, the oscillation probability is the
generalized form of
\begin{align}
  P({\nu_\alpha\rightarrow\nu_\beta}) =\ &\delta_{\alpha\beta} - 4
  \sum_{i>j}{\rm Re}\left(U_{\alpha i}^\ast U_{\beta i} U_{\alpha j}
  U_{\beta j}^\ast\right)\ \sin^2 \left(\delta
  m^2_{ij}\frac{L}{4E}\right)\cr &+ 2 \sum_{i>j}{\rm
  Im}\left(U_{\alpha i}^\ast U_{\beta i} U_{\alpha j} U_{\beta
  j}^\ast\right)\ \sin \left(\delta m^2_{ij}\frac{L}{2E}\right)\, ,
\end{align}
where $U_{\alpha i}$ are the elements of the unitary matrix defining
the mixing between the flavor and mass eigenstates.

\subsection{The Density Matrix}
Now, let us shift to discuss the way in which a two-neutrino system
evolves in a high-density environment. The necessary formalism to
incorporate matter-affected oscillations, collisions and
active-active forward scattering is the density-matrix or
matrix-of-densities
\begin{equation}
  \rho({\mathbf p}, t) \equiv \langle\psi_\alpha|\hat{\rho}({\mathbf p},
t)|\psi_\beta \rangle \doteq  \label{densitymatrix}
\begin{pmatrix}
\rho_{\alpha\alpha}({\mathbf p}, t) & \rho_{\alpha\beta}({\mathbf p},
t)\cr \rho_{\beta\alpha}({\mathbf p}, t) & \rho_{\beta\beta}({\mathbf
  p}, t)\end{pmatrix} \, .
\end{equation}
This is given as abstract density operator, expanded in a given basis,
usually flavor, as
\begin{equation}
\hat \rho({\mathbf p}, t) \equiv \sum_j p_j
|\psi_j\rangle\langle\psi_j|\, .
\end{equation}
Evolution of the density matrix gives the quantities of interest,
namely the number densities of a given flavor.

\begin{figure}[htbp]
\begin{center}
\begin{tikzpicture}[x=0.5cm,y=0.5cm,z=0.3cm,>=stealth]
\draw[->] (xyz cs:x=-7.5) -- (xyz cs:x=7.5) node[above] {$y$};
\draw[->] (xyz cs:y=-2.5) -- (xyz cs:y=7.5) node[right] {$z$};
\draw[->] (xyz cs:z=7.5) -- (xyz cs:z=-7.5) node[left] {$x$};

\draw[->,line width=3.2pt] (xyz cs:y=0) -- (xyz cs:y=5.5) node[right] {$\mathbf P$};

\node[align=center] at (-3,3) (ori) {\text{pure $\vert \nu_\alpha\rangle$}};

\end{tikzpicture}
\caption{The $\mathbf P$ vector formulation of neutrino oscillations, with flavor asymmetry purely in the “upper” state. }
\label{pvector}
\end{center}
\end{figure}
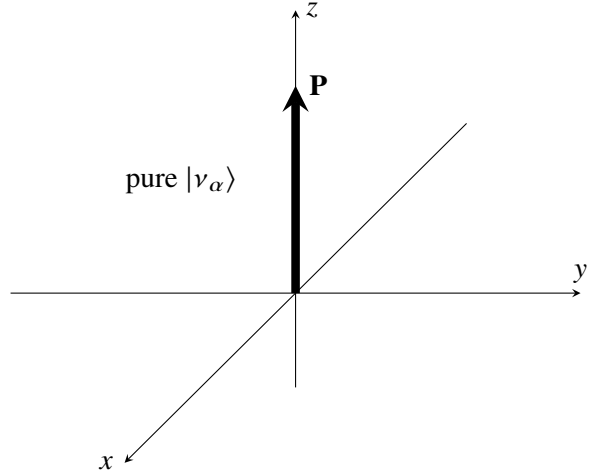

\begin{figure}[htbp]
\begin{center}
\begin{tikzpicture}[x=0.5cm,y=0.5cm,z=0.3cm,>=stealth]
\draw[->] (xyz cs:x=-7.5) -- (xyz cs:x=7.5) node[above] {$y$};
\draw[->] (xyz cs:y=-7.5) -- (xyz cs:y=7.5) node[right] {$z$};
\draw[->] (xyz cs:z=7.5) -- (xyz cs:z=-7.5) node[left] {$x$};

\draw[->,line width=3.2pt] (xyz cs:y=0) -- (xyz cs:y=-5.5) node[right] {$\mathbf P$};

\node[align=center] at (-3,3) (ori) {\text{pure $\vert \nu_\beta\rangle$}};

\end{tikzpicture}
\caption{The $\mathbf P$ vector formulation of neutrino oscillations, with flavor asymmetry purely in the “lower” state. }
\label{pvectoralt}
\end{center}
\end{figure}
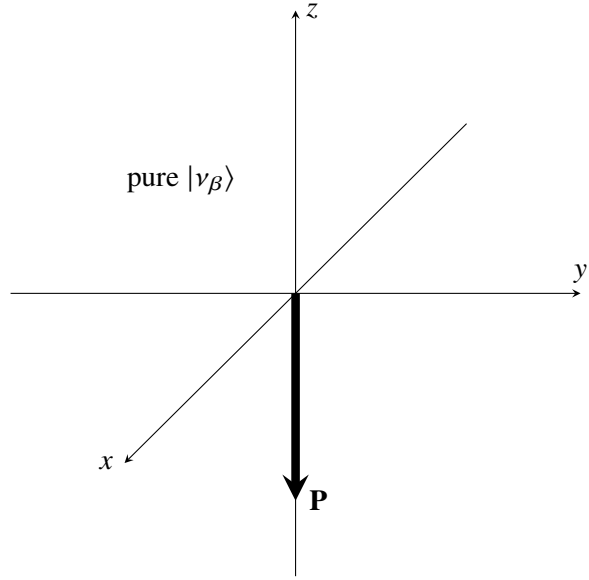

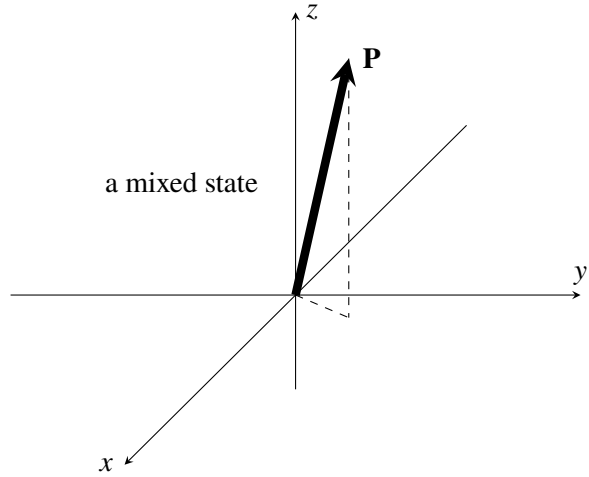
\begin{figure}[htbp]
\begin{center}
\begin{tikzpicture}[x=0.5cm,y=0.5cm,z=0.3cm,>=stealth]
\draw[->] (xyz cs:x=-7.5) -- (xyz cs:x=7.5) node[above] {$y$};
\draw[->] (xyz cs:y=-2.5) -- (xyz cs:y=7.5) node[right] {$z$};
\draw[->] (xyz cs:z=7.5) -- (xyz cs:z=-7.5) node[left] {$x$};

\draw[->,line width=3.2pt] (0,0,0) -- (2.,6.87,-1 ) node[right] {$\mathbf P$};
\draw[dashed] (0,0,0) -- (2,0,-1);
\draw[dashed] (2,6.87,-1) -- (2,0,-1);
\node[align=center] at (-3,3) (ori) {\text{a mixed state}};

\end{tikzpicture}
\caption{The $\mathbf P$ vector formulation of neutrino oscillations, in a mixed state. }
\label{pvectormix}
\end{center}
\end{figure}

Why is the density
matrix necessary? Collisions and forward scattering by neutrinos
themselves require it. But, before I get into the forward
self-potential and scattering, let us look at the evolution of neutrinos
in this formulation. A two-state density matrix system has been of
interest in quantum mechanics for some time for studies in optical and
solid state systems. The density matrix must preserve probability (or
number density). That allows for a geometric interpretation of mixed
flavor states. Let us define a general mixed state as
\begin{align}
  |\psi\rangle =& \cos(\theta/2)|\nu_\alpha\rangle + e^{i\phi}
  \sin(\theta/2)|\nu_\beta\rangle\cr
   =& \cos(\theta/2)|\nu_\alpha\rangle + (\cos\phi + i\sin\phi)
   \sin(\theta/2)|\nu_\beta\rangle \, .
\end{align}
The parameters $\theta$ (different from the mixing angle!) and $\phi$, can be
reinterpreted in spherical coordinates as respect to longitude from
the $z$-axis and the longitude with respect to the $x$-axis They specify a
point in spherical coordinates on the unit sphere, at the end of the
vector $\mathbf P$, called the ``Bloch vector,''
\begin{equation}
  \mathbf{P} = (\sin\theta\cos\phi,\sin\theta\sin\phi,\cos\theta)\, .
\end{equation}
In this formulation, one can represent a pure $\nu_\alpha$ flavor
state as an upward ($z$) pointing $\mathbf P$ vector, as in Fig.~\ref{pvector}.
And, the alternate state is simply pointed oppositely, as in Fig.~\ref{pvectoralt}.
So then a mixed state would be one pointed in a direction not
perfectly aligned with the $z$-axis, with arbitrary orientation, as in Fig.~\ref{pvectormix}.

Going back to the density matrix, Eq.\ \eqref{densitymatrix}, I can
write it in terms of the $\mathbf P$ Bloch vector, and I have
\begin{align}
\rho(p) = \frac{1}{2}\left[P_0(p)+{\mathbf P}(p)\cdot{\boldsymbol{\sigma}}\right] = 
\frac{1}{2}
\begin{pmatrix}
P_0(p)+P_z(p) & P_x(p) - i P_y(p)\cr
P_x(p)+i P_y(p)  & P_0(p)-P_z(p)\end{pmatrix}\, ,
                   \label{eq:pvector}
\end{align}
where ${\boldsymbol{\sigma}} \equiv (\sigma_1,\sigma_2, \sigma_3)$ is
a vector composed of the Pauli spin matrices. The time evolution of
the density matrix is such that
\begin{equation}
  i \hbar \frac{\partial \rho}{\partial t} = [H,\rho]\, ,\label{densityHeisenberg}
\end{equation}
or
\begin{equation}
  \rho(t) = e^{-i H t/\hbar}\rho(0) e^{i H t/\hbar}\,. 
\end{equation}
In this formulation, I continue to recover vacuum oscillations
with probability densities evolving as
Eq. \eqref{oscillation}. The symmetries of the time evolution equation
Eq.\ \eqref{densityHeisenberg} turn out to be commensurate with the
evolution equation of the spin precession of a ``dipole'' 
($\mathbf P$) around a ``magnetic field'' ($\mathbf V$),
\begin{equation}
\partial_t{\mathbf P}(p) = {\mathbf{V}(p)}\times {\mathbf P}(p).
\end{equation}
For vacuum evolution, the ``potential'' or ``magnetic field'' maps on
to provide the $\delta m^2$-dependent time evolution term, 
\begin{equation}
{\mathbf V}(p) =
{\mathbf \Delta}(p) \, ,
\end{equation}
where
\begin{equation}
{\mathbf \Delta}(p) \equiv \frac{\delta m^2}{2p}(\sin
2\theta_0,0,-\cos 2\theta_0)\, .\label{eq:vacpot}
\end{equation}
In this relation, $\theta_0$ is the vacuum mixing angle. The ``magnetic field'' orientation that the flavor state evolves
around is determined by the size of the vacuum mixing angle $\theta_0$. For maximal mixing, $\theta_0=45^\circ$, ${\mathbf V}(p)$ is oriented with the
$x$-axis, as in Fig.~\ref{pvectormax}:
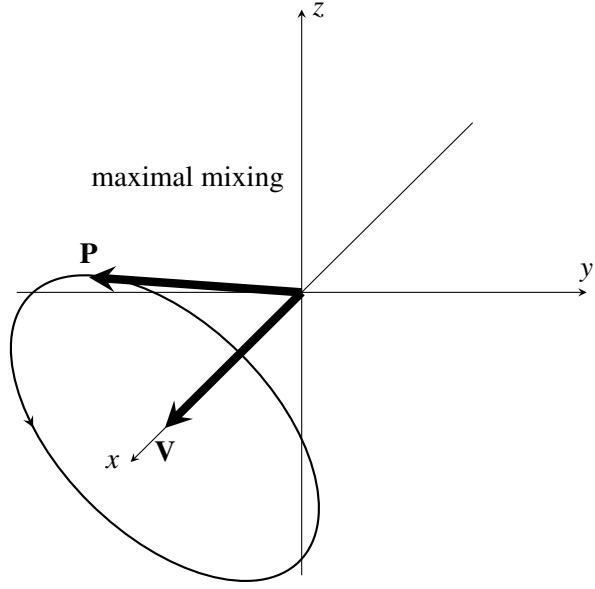
\begin{figure}[htbp]
\begin{center}
\begin{tikzpicture}[x=0.5cm,y=0.5cm,z=0.3cm,>=stealth]

\begin{scope}[decoration={
	markings,
	mark=at position 0.4 with {\arrow{>}} }
      ]

\draw[->] (xyz cs:x=-7.5) -- (xyz cs:x=7.5) node[above] {$y$};
\draw[->] (xyz cs:y=-7.5) -- (xyz cs:y=7.5) node[right] {$z$};
\draw[->] (xyz cs:z=7.5) -- (xyz cs:z=-7.5) node[left] {$x$};

\draw[->,line width=3.2pt] (0,0,0) -- (0,0,-6 ) node[below] {$\mathbf V$};
\draw[->,line width=3.2pt] (0,0,0) -- (-2.,4,-6 ) node[above] {$\mathbf P$};
\draw[thick, rotate around={45:(0,0,-6)}, postaction={decorate}] (0,0,-6) ellipse (2.8  and 5);

\node[align=center] at (-3,3) (ori) {\text{maximal mixing}};
\end{scope}
\end{tikzpicture}
\caption{The $\mathbf P$ flavor vector evolution and $\mathbf V$ “potential” vector in the case of maximal mixing. }
\label{pvectormax}
\end{center}
\end{figure}
Here, the neutrino is in an initially mixed state. If it was in a flavor state initially, as in typical neutrino experiments, the $\mathbf P$ vector would be oriented toward the $\pm z$-axis initially, and would precess in the $y$-$z$ plane. I show the initial mixed case example as it is clearer to visualize.

\subsection{Matter and Thermal Effects}
Neutrinos propagating through space are “dressed” by loop diagrams that become important at finite temperature and density. At high matter density and high temperatures, they have significant contributions \cite{Venumadhav:2015pla}. This is often called forward scattering of neutrinos; the commonality with photon forward scattering is just that $p =p’$ after the scattering.
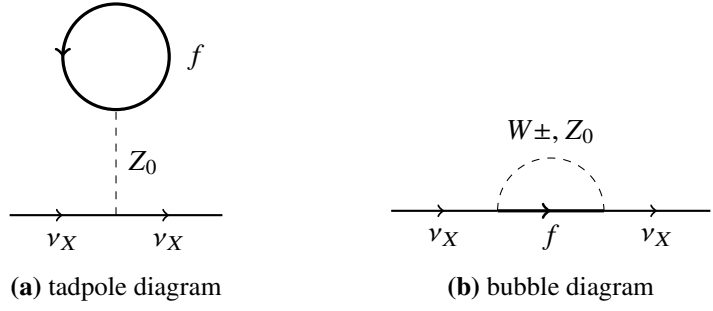
\begin{figure}[t]
\begin{center}  
\subfloat[\label{fig:tadpole}tadpole diagram]{
    \begin{tikzpicture}[scale=0.7]
    \begin{scope}[decoration={
	markings,
	mark=at position 0.5 with {\arrow{>}} }
      ]
      \draw[black, thick, postaction={decorate}] (-2, 0) -- (0, 0);
      \node at (-1,-0.5) {${\nu_X}$};
      \draw[black, thick, postaction={decorate}] (0, 0) -- (2, 0);
      \node at (1,-0.5) {${\nu_X}$};
      \draw[black, dashed] (0, 0) -- (0, 2);
      \node at (0.5, 1) {$Z_0$};
      \draw[black, very thick, postaction={decorate}] (0, 3) circle (1cm);
      \node at (1.5, 3) {$f$};
    \end{scope}
    \end{tikzpicture}
    }
  \hspace*{2cm}%
  \subfloat[\label{fig:bubble}bubble diagram]{
    \begin{tikzpicture}[scale=0.7]
      \begin{scope}[decoration={
	markings,
	mark=at position 0.5 with {\arrow{>}} }
      ]
      \draw[black, thick, postaction={decorate}] (-3, 0) -- (-1, 0);
      \node at (-2,-0.5) {${\nu_X}$};
      \draw[black, very thick, postaction={decorate}] (-1, 0) -- (1, 0);
      \draw[black, thick, postaction={decorate}] (1, 0) -- (3, 0);
      \node at (2,-0.5) {${\nu_X}$};
      \draw[black, dashed] (-1, 0) arc (180:0:1cm);
      \node at (0, 1.5) {$W\pm,Z_0$};
      \node at (0, -0.5) {$f$};
      \end{scope}
    \end{tikzpicture}
    }    \end{center}

  \caption{\label{fig:oneloop}Shown here are the lowest order contributions to a propagating active neutrino's self energy in the presence of finite matter density and temperature. In (a), $f$ is any species with weak charge, with both neutral and charged leptons, quarks, mesons, or nucleons. In (b), $f = \nu_X, X^-$. (This figure is adapted from Ref.~\cite{Venumadhav:2015pla}.) } 
\end{figure}
Both diagrams in Fig.~\ref{fig:oneloop} contribute in finite density environments, while only bubble diagrams contribute to the effects of finite temperature (plasma) backgrounds. Both effects contribute flavor-dependent background potentials to the neutrino evolution (therefore oriented in the $z$-axis in our $\mathbf P$-vector system). The finite density background arises due to asymmetries in weak-charged particles,
\begin{equation}
V^B(p) = 
\begin{cases}
\sqrt{2}G_F\left[(n_{e^-}-n_{e^+}) - n_{n}/2\right] & \text{for\ } \nu_\alpha\rightleftharpoons\nu_s,
\\
\sqrt{2}G_F\left(n_{e^-}-n_{e^+}\right) & \text{for\ }
\nu_e\rightleftharpoons\nu_{\mu,\tau},
\\
0 & \text{for\ }
\nu_\mu\rightleftharpoons\nu_\tau.
\end{cases}\label{eq:vd}
\end{equation}
Here, $n_{e^\pm}$ are the densities of positrons/electrons and $n_n$ is the density of neutrons. This potential is the one responsible for the Mikheev-Smirnov-Wolfenstein (MSW) effect in the Sun \cite{Wolfenstein:1977ue,Mikheev:1986gs}.

The thermal potential is \cite{notzold:1987ik}
\begin{align}
 V^T (p) = &  - \frac{8\sqrt{2} G_{\rm F} p_\nu}{3 m_{\rm Z}^2}
\left(\langle E_{\nu_\alpha} \rangle n_{\nu_\alpha} + \langle
E_{\bar\nu_\alpha} \rangle n_{\bar\nu_\alpha}\right) \cr 
 &  
-\frac{8\sqrt{2} G_{\rm F} p_\nu}{3 m_{\rm W}^2} \left(\langle E_\alpha
\rangle n_\alpha + \langle E_{\bar\alpha} \rangle
n_{\bar\alpha}\right)\, ,
\end{align}
where $p=p_\nu$ is the momentum of the neutrino, $\langle E_{\nu_\alpha}\rangle$ $(\langle E_{\alpha}\rangle)$ is the average energy of the background neutrinos (charged weak particles), as well as the respective antiparticles. 
These background and thermal potentials are flavor-dependent, but identical for all active neutrinos with symmetric backgrounds. In the early Universe, the $e^\pm$ background persists longer than the other charged leptons, therefore the thermal potential from $e^\pm$ on $\nu_e/\bar{\nu}_e$ is important in determining the flavor evolution in any neutrino-flavor asymmetric early Universe history \cite{abazajian:2002qx,wong:2002fa}. Including the full thermal effects for active neutrinos’ mixing with other active neutrinos requires handling the neutrino self-potential, and that will be discussed more below. For active-sterile neutrino mixing, the evolution simplifies, as the thermal potential arises only for the active neutrinos, and a thermal potential contributes to the $z$-axis component of ${\mathbf V}(p)$,
\begin{equation}
{\mathbf V}(p)  =
  {\mathbf \Delta}(p) +
\left[ V^B(p)+ V^T(p)\right]\hat{\mathbf z}\, .
\end{equation}

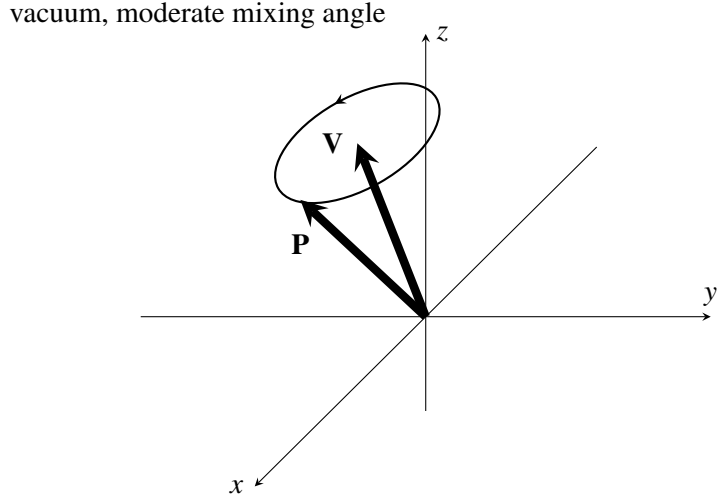
\begin{figure}[htbp]
\begin{center}

\begin{tikzpicture}[x=0.5cm,y=0.5cm,z=0.3cm,>=stealth]
      \begin{scope}[decoration={
	markings,
	mark=at position 0.25 with {\arrow{>}} }
      ]
\draw[->] (xyz cs:x=-7.5) -- (xyz cs:x=7.5) node[above] {$y$};
\draw[->] (xyz cs:y=-2.5) -- (xyz cs:y=7.5) node[right] {$z$};
\draw[->] (xyz cs:z=7.5) -- (xyz cs:z=-7.5) node[left] {$x$};

\draw[->,line width=3.2pt] (0,0,0) -- (0,6.4,-3 ) node[left] {$\mathbf V$};
\draw[->,line width=3.2pt] (0,0,0) -- (0,6.4,-5.5 ) node[below=8pt] {$\mathbf P$};
\draw[thick,rotate around={30:(0,6.4,-3)},postaction={decorate}] (0,6.4,-3) ellipse (2.4  and 1.2);

\node[align=center] at (-6,8) (ori) {\text{vacuum, moderate mixing angle}};
\end{scope}
\end{tikzpicture}
\caption{The $\mathbf P$ flavor vector evolution and $\mathbf V$ “potential” vector in the case of moderate mixing. }
\label{pvectormod}
\end{center}
\end{figure}

\textbf{Non-maximal vacuum oscillations:} The case of non-maximal vacuum oscillations can be contrasted to high-asymmetry (which is equivalent to small vacuum mixing). That is, the matter-affected mixing is small. The potential is defined entirely by the vacuum term, Eq.~\eqref{eq:vacpot}, and shown in Fig.~\ref{pvectormod}.
The high-density (strong matter asymmetry) case evolves with the $\mathbf V$ vector then orienting to a position away from the $z$-axis, determined by the size of the vacuum mixing angle.

This can also be formulated as an \textit{matter-affected} mixing angle
\begin{equation}
 {\mathbf \Delta_m}(p) \equiv \frac{\delta m^2}{2p}(\sin
2\theta_m,0,-\cos 2\theta_m)\, ,
\end{equation}
which is equivalent to the total potential $\mathbf V(p)$. Here, the matter-affected mixing angle is then
\begin{equation}
\sin^2 2\theta_m = \frac{\Delta^2 (p) \sin^2 2\theta}{{\Delta^2 (p)}
\sin^2 2\theta + \left[{\Delta (p)} \cos 2\theta - {V^B} -
{V^T(p)}\right]^2}\, .
\label{eq:matterangle}
\end{equation}

\begin{figure}[htbp]
\begin{center}
\begin{tikzpicture}[x=0.5cm,y=0.5cm,z=0.3cm,>=stealth]
\begin{scope}[decoration={
	markings,
	mark=at position 0.2 with {\arrow{>}} }
      ]

\draw[->] (xyz cs:x=-7.5) -- (xyz cs:x=7.5) node[above] {$y$};
\draw[->] (xyz cs:y=-2.5) -- (xyz cs:y=7.5) node[above] {$z$};
\draw[->] (xyz cs:z=7.5) -- (xyz cs:z=-7.5) node[left] {$x$};

\draw[->,line width=3.2pt] (0,0,0) -- (0,6.4,0 ) node[left] {$\mathbf V$};
\draw[->,line width=3.2pt] (0,0,0) -- (2.,6.,-1 ) node[right] {$\mathbf P$};
\draw[thick,postaction={decorate}] (0,6.4,0) ellipse (2.4  and 1.2);

\node[align=center] at (-4,4) (ori) {\text{strong matter asymmetry}};
\end{scope}
\end{tikzpicture}
\caption{The $\mathbf P$ flavor vector evolution with a matter asymmetry dominating the  $\mathbf V$ “potential” vector. }
\label{pvectorsol}
\end{center}
\end{figure}
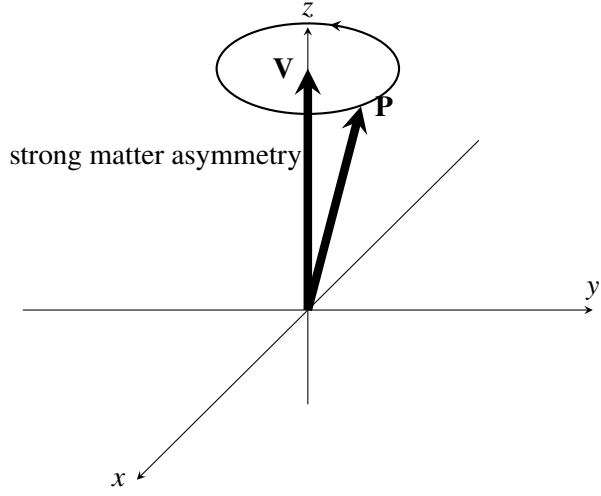

\textbf{Solar Neutrinos}: 
If you have a very high asymmetry background, without a high neutrino density, \textit{such as the interior of the Sun}, the potential $\mathbf V(p)$ is dominated by the finite-density portion $\mathbf V^B$, and it pins precession to the $z$-axis. In this case, the projection of $\mathbf P$ on the $z$-axis is fixed, and flavor does not evolve, as in Fig.~\ref{pvectorsol}.    
So, for the solar neutrino solution, the evolution of the background potential goes from a the high-density background to that of a non-maximal vacuum state. 

As the matter density (asymmetry) is reduced for a neutrino exiting the Sun, its evolution crosses the $x$-$y$-plane (at a critical $\theta_c\approx \pi/4$), to settle on a vacuum oscillation on the other side of the plane. This is flavor-conversion level crossing that is present in the MSW mechanism for matter-affected oscillation solutions to the solar neutrino problem, and connects this vector formalism to the matter-affected mixing and level-crossing structure of solar neutrino physics. For an excellent review of solar neutrino oscillations, see Ref.~\cite{Robertson:2012ib}.

\section{Neutrinos in the Early Universe}

\label{nuearly}
 For high temperature backgrounds, the thermal potential is identical for all active flavor states, and the self-potential must be included to understand its evolution \cite{dolgov:2002ab,abazajian:2002qx,wong:2002fa}.  For active-sterile mixing, only the active neutrinos are “dressed” with the thermal effects of Fig.~\ref{fig:oneloop}, and a large thermal asymmetry is present that pins the vector $\mathbf V$ to the $z$-axis. As the thermal potential decreases from early to later times, the $\mathbf P$ can oscillate to non-zero probabilities of “interacting” (or failing to) as a sterile neutrino in the interaction-rich plasma, and therefore entering a purely sterile neutrino state. That sterile neutrino state could propagate and achieve an amplitude to convert back to an active state, but that reverse rate is minimal when the usual case of $n_{\nu_\alpha} \gg n_{\nu_s}$ holds. 
 
In the early Universe, a new background asymmetry can exist from lepton asymmetries housed in the active neutrinos. In the case of active-active mixing, this leads to synchronized oscillations, discussed below. In the case of active-sterile oscillations, the physics is simpler, with a finite density potential in the early Universe potentially dominated by asymmetries in the neutrino-based lepton number. That asymmetry is present in a ``lepton potential'' $V^L$, which takes the form
\begin{equation}
V^L = \frac{2 \sqrt{2} \zeta (3)}{\pi^2}\,G_{\rm F} T^3 \left({\cal
L}^\alpha \pm \frac{\eta}{4}\right), \label{vl}
\end{equation}
where I take \lq\lq $+$\rq\rq\ for $\alpha=e$ and \lq\lq$-$\rq\rq\ for
$\alpha=\mu,\tau$.
Here, the lepton number ${\cal{L}^\alpha}$ is in terms
of the lepton numbers in each active neutrino species,
\begin{equation}
\label{L}
{\cal{L}^\alpha} \equiv 2 L_{\nu_\alpha} +
\sum_{\beta\neq\alpha}{L_{\nu_\beta}},
\end{equation}
with the final sum over active neutrino flavors other than
$\nu_\alpha$.  The parameter $\eta\equiv n_{\rm b}/n_\gamma$ is the baryon to
photon ratio.
An $\alpha$-type neutrino asymmetry is defined as 
\begin{equation}
L_{\nu_{\alpha}}\equiv\frac{n_{\nu_{\alpha}}-n_{\bar{\nu}_{\alpha}}}{n_\gamma},\label{leptonnumber}
\end{equation}
where the photon number density $n_{\gamma}=2\zeta(3)T^{3}/\pi^{2}$. Here, the asymmetry in the charged leptons and baryons is also incorporated in the last term of Eq.~\eqref{vl}, and I will use $V^L$ instead of $V^B$ for the early Universe active-sterile mixing case. The matter-affected mixing angle in the early-Universe for active-sterile mixing changes from Eq.~\eqref{eq:matterangle} to
\begin{equation}
\sin^2 2\theta_m = \frac{\Delta^2 (p) \sin^2 2\theta}{{\Delta^2 (p)}
\sin^2 2\theta + \left[{\Delta (p)} \cos 2\theta - {V^L} -
{V^T(p)}\right]^2}\, .
\label{eq:matterangleEU}
\end{equation}

While in the scattering-dominated regime, $T\gg 5 \rm\ MeV$, the conversion rate of active neutrinos to sterile neutrinos is the product of half of the {\it total} interaction rate, $\Gamma_\alpha$, of the neutrinos
with the plasma and the probability that an active neutrino has transformed to a sterile, which is also suppressed by a damping term arising from the quantum-Zeno effect, $D(p) = \Gamma(p)/2$, discussed below,
\begin{equation}
\Gamma(\nu_\alpha \rightarrow \nu_s) = \frac{(\Gamma_\alpha(p)/2){ \Delta^2 (p)} \sin^2 2\theta}{{ \Delta^2 (p)}
\sin^2 2\theta + D^2(p) + \left[{ \Delta (p)} \cos 2\theta - { V^L} -
{ V^T(p)}\right]^2}\, .
\label{conversrate}
\end{equation}
In this relation’s denominator rest both the thermal potential and damping term, which  simultaneously suppress production with increasing temperature $\Gamma\propto T^{-7}$. At late times, the thermal and damping term are subdominant to the vacuum term in Eq.~\eqref{conversrate}, and production decreases with increasing temperature as $\Gamma\propto T^{5}$.   Since production is a decreasing function with increasing temperature at high temperatures, and decreasing function with decreasing temperature at lower temperatures, the behavior is
\begin{equation}
\frac{\Gamma}{H}\propto 
\begin{cases}
T^{-9}\quad\text{High $T$}\cr
T^{3}\quad\text{Low $T$}\, ,
\end{cases}
\label{productionpower}
\end{equation}
with respect to the Hubble expansion rate in the radiation dominated era, $H^2 = (8\pi/3)G\rho\propto T^4$.  Using this formulation of the mixing and the Hubble expansion above the weak decoupling era, one can determine a mixing angle and mass-squared difference boundary where this conversion rate will “thermalize” the sterile neutrino when the rate becomes greater than the Hubble expansion rate $\Gamma \ge H$. 

This allows for placing a constraint on the presence of sterile neutrinos with large mixings and large mass differences with active neutrinos if one wants to avoid their thermalization to become a “fourth” neutrino state entering into the era of big bang nucleosynthesis \cite{Steigman:1977kc,Langacker:1989sv,Enqvist:1991qj}, as it would alter the primordial helium abundance outside of its observed abundance. The boundary in mass and mixing angle space is approximately \cite{abazajian:2002bj}
\begin{equation}
\delta m_{\alpha s}^2 \sin^4 2\theta_{\rm BBN} \lesssim
\begin{cases}
5\times 10^{-6},\qquad\text{for $\alpha=e$} 
\\
3\times 10^{-6},\qquad\text{for $\alpha=\mu,\tau$}
\end{cases}
\,,
\label{oldconstraints}
\end{equation}
where $\sin^4 2\theta_{\rm BBN}$ is the effective vacuum mixing angle of the active-sterile system of interest. For practical purposes, this rules out the regions of parameter space of the large mixing angle solution of the solar neutrino problem as being active-sterile mixing, as well as active-sterile-active oscillation solutions to the Liquid Scintillator Neutrino Detector \cite{dibari:2001ua,abazajian:2002bj}, since thermalization of the sterile neutrino would alter the primordial helium abundance to deviate significantly from its inferred value from observations. 

The helium abundance is altered by an extra neutrino in the thermal background because the Hubble expansion rate in the radiation-dominated era is dependent on the total density of the background including photons, $\rho_\gamma$, electrons and positrons, $\rho_{e^\pm}$, active neutrinos, $\rho_{\nu_\alpha}$, and any extra sterile neutrinos, $\rho_{\nu_s}$: $H\propto(\rho_\gamma+ \rho_{e^\pm}+\rho_{\nu_\alpha}+\rho_{\nu_s}+...)^{1/2}$. Since almost all neutrons in the early Universe are incorporated to helium nuclei, the neutron-to-proton ratio set when weak rates freeze out is key to determining the primordial helium abundance. The weak rates are 
\begin{align}
\lambda_{n\rightarrow p} &\begin{cases} n+e^+ \rightarrow p+\bar{\nu}_e\cr
n+\nu_e\rightarrow p+e^-\cr
n\rightarrow p+e^-+\bar{\nu}_e
\end{cases}\, ,\cr
\lambda_{p\rightarrow n} &\begin{cases} p+\bar{\nu}_e \rightarrow n+e^+\cr
p+e^- \rightarrow n+\nu_e\cr
p+e^- +\bar{\nu}_e\rightarrow n
\label{weakrates}
\end{cases}\, .
\end{align}
Here, neutron decay is not important until after freeze-out, and, of course, three-body fusion in the last line is very slow. 

\begin{figure}
\begin{center}
\includegraphics[width=4.5in]{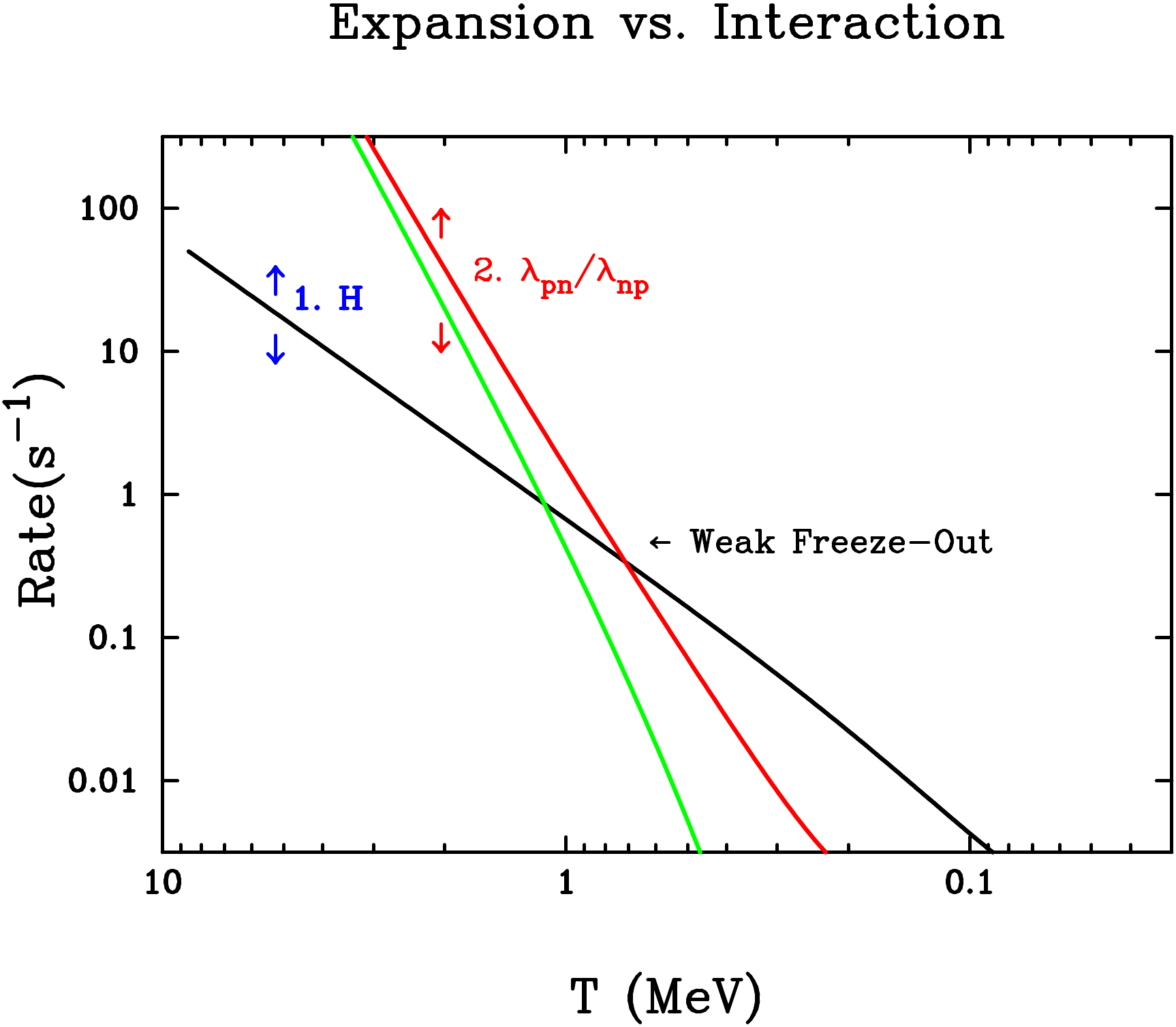}
\caption{\label{rateplot} Primordial nucleosynthesis’ primary response to the presence of new particles like light sterile neutrinos in this radiation-dominated era is the freezing-out of the weak rates earlier when those rates $\lambda_{n\rightarrow p}$ and $\lambda_{p\rightarrow n}$, Eq.~\eqref{weakrates}, fall below the Hubble rate. With the significant presence of sterile neutrinos, the Hubble rate is augmented, forcing freeze-out of the weak rates with a higher neutron-to-proton ratio. Recall, that via the Friedmann Equation, $H\propto(\rho_\gamma +\rho_{e^\pm}+\rho_{\nu_\alpha}+\rho_{\nu_s}+...)^{1/2}$. Since almost all neutrons are incorporated in helium, the helium abundance is increased. }
\end{center}
\end{figure}

If the weak rates freeze-out earlier, they are closer to each other in rate, and therefore more neutrons are present at weak freeze-out.  In equilibrium, given the mass difference of the neutron-proton, $\Delta m_{np} \approx 1.3\ \mathrm{MeV}$, the ratio of neutrons to protons is given by the relevant Boltzmann factor, $n/p \approx e^{-\Delta m_{np}/T}$. With the weak rates falling below the sterile-neutrino augmented Hubble expansion factor earlier,  at higher temperatures, the neutron-to-proton ratio $n/p$ is larger and allows for a greater production of helium. See Fig.~\ref{rateplot}. Using detailed calculations and recent inferred primordial helium abundances, Ref.~\cite{Cyburt:2015mya} finds 
\begin{equation}
N_\mathrm{eff} < 3.2 \quad (2\sigma)\, ,
\label{neffbbn}
\end{equation}
where 
\begin{equation}
N_{\rm eff} \equiv \frac{\rho_{\rm rel}-\rho_\gamma}{\rho_\nu}\, ,
\label{neff}
\end{equation}
is the abundance of all non-photon relativistic species relative to a single thermal active neutrino density. In this case, it is the density at weak freeze-out. Sometimes, $N_{\rm eff}$ is called “neutrino number,” and denoted $N_\nu$, but is truly much more general than energy density in neutrinos. Note that this bound, Eq.~\eqref{neffbbn}, is model dependent. Several ways of evading it are listed in Ref.~\cite{abazajian:2002bj}, including a pre-existing small lepton number \cite{foot:1995bm}, which suppresses the matter-affected mixing angle, a dynamically-generated lepton number from a predominantly sterile fifth mass eigenstate (with requisite mass less than the predominantly active mass $m_s<m_\alpha$) \cite{dibari:2001ua}, a new Majoron field \cite{Bento:2001xi}, a low reheating temperature Universe \cite{Giudice:2000ex,Kawasaki:2000en,Hasegawa:2020ctq},  baryon-antibaryon inhomogeneities \cite{Giovannini:2002qw}, extended quintessence \cite{Chen:2000xxa}, or even CPT violating neutrinos \cite{Murayama:2000hm,Barenboim:2001ac}.  Note that partial thermalization of sterile neutrinos may even be a good thing, and help solve tensions with the amplitude of fluctuations on small scales, dubbed $\sigma_8$, \cite{Jacques:2013xr}. The $\sigma_8$ tension is discussed more below.

\subsection{The Neutrino Self-Potential}
\begin{figure}
\begin{center}
 \begin{tikzpicture}[scale=0.7]
      \begin{scope}[decoration={
	markings,
	mark=at position 0.5 with {\arrow{>}} }
      ]
      \draw[black, thick, postaction={decorate}]  (0, 0) -- (-3, 2) ;
      \node at (-2,2) {${\nu_\alpha(p)}$};
      \draw[black, thick, postaction={decorate}] (0, 0) -- (3, 2);
            \node at (2,2) {${\nu_\beta(q)}$};

      \draw[black, thick, postaction={decorate}] (3, -2) -- (0,0);
      \node at (2,-2) {${\nu_\beta(p)}$};

      \draw[black, thick, postaction={decorate}] (-3, -2) -- (0,0);
      \node at (-2,-2) {${\nu_\alpha(q)}$};
      
      \fill (0,0) circle[radius=3pt];
      \end{scope}
    \end{tikzpicture}

  \caption{\label{fig:selfpotential} Shown here is the exchange diagram that leads to forward scattering of active neutrinos off their background, which mixes different momenta modes for all active neutrino with each other. This has implications for asymmetric oscillations in the early Universe \cite{dolgov:2002ab,abazajian:2002qx,wong:2002fa} and in the high-density neutrino background in supernovae \cite{Qian:1994wh}. } 
     \end{center}

\end{figure}
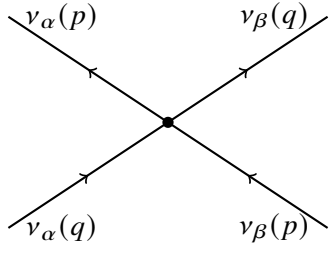

Active neutrinos can exchange their momentum through weak interactions of the form shown in Fig.~\ref{fig:selfpotential}, so that same, or different, flavors can exchange momenta $\nu_\alpha \rightleftharpoons \nu_\beta$. The self-potential enters the spatially homogeneous and isotropic early Universe as the potential ${\mathbf V}^S(p)$,
\begin{equation}
{\mathbf V}(p) =  {\mathbf \Delta}(p) + \left[{V^B(p)}+{V^T(p)}\right]\hat{\mathbf z}+ { {{\mathbf V}^S(p)}}\, ,
\end{equation}
where
\begin{equation}
 {\mathbf V}^S(p) =2\sqrt{2}G_F n_{\nu_\alpha}({\mathbf J} - \bar{\mathbf J})\, .
\end{equation}
And the $\mathbf J$ vectors are the sum or integral of the asymmetric active neutrino background,
\begin{equation}
{\mathbf J}\equiv \sum_p {\mathbf P}_p, \qquad \bar{\mathbf J}\equiv \sum_p \bar{\mathbf P}_p\, .
\end{equation}
This self-potential is crucial in any asymmetries that exist in neutrino flavors. Due to the asymmetries in neutrino flavor in supernovae, and their very high densities, the self-potential is a dominant driver of the evolution. Matter asymmetries provide a $V^B$ term, but neutrino flavor evolution is also strongly governed by the self-potential. Unlike the early Universe, supernovae are far from homogeneous and isotropic, so the directional-dependence of the weak interactions in such a system complicates flavor evolution immensely. The several lines of study of neutrino flavor evolution in supernovae would make for a large set of lectures unto themselves \cite{Fuller:1993ry,Qian:1994wh,Pastor:2002we,Hannestad:2006nj,Dasgupta:2007ws,Cherry:2011fn,Johns:2020qsk}. For a review of collective neutrino oscillations and their effects in core-collapse supernovae, see Ref.~\cite{Duan:2010bg}.

{\bf Degenerate nucleosynthesis and syncrhonized oscillations:} A remarkable piece of physics occurs in the early Universe if any asymmetries exist between neutrino flavors. Thermal processes pairwise produce active neutrinos and antineutrinos, but novel processes could set up an asymmetry in neutrinos and antineutrinos via leptogenesis. Such models are consistent with constraints from primordial nucleosynthesis because a small asymmetry in the electron neutrino sector can offset the neutron-to-proton ratio increase from earlier weak freeze-out caused by the energy density in a much larger asymmetry in the muon or tau neutrino flavors (see initial work by Ref.~\cite{Kohri:1996ke}, and related work in Refs.~\cite{Orito:2000zb,Orito:2002hf}).  The equilibrium distribution functions of the neutrinos are generally
\begin{align}
		f_\nu(p)&=\frac{1}{\exp \left(p/T_\nu -
		\xi_\nu\right)+1}~, \\
		f_{\bar{\nu}}(p)&=\frac{1}{\exp
		\left(p/T_\nu + \xi_\nu\right)+1}~,
\end{align}
where the degeneracy parameters are the chemical potentials of the neutrinos scaled by the temperature $\xi_\nu\equiv \mu_\nu/T_\nu$, as then they are invariants with expansion. 
The effect from electron asymmetry affects the weak rates so that $n/p = \exp(-\Delta m_{np}/T - \xi_{\nu_e})$, while the $\nu_\mu$ and $\nu_\tau$ deneracies affect only the energy density in the neutrinos, forcing an earlier weak freeze-out,
\begin{equation}
\rho_{\nu}  + \rho_{\bar{\nu}} = {7 \over 8} {\pi^2 \over 15} \sum_i T_{\nu_i}^4
 \left[1 + {15 \over 7} \left(
{\xi_{\nu_i} \over \pi}\right)^4 + {30 \over 7} \left(
{\xi_{\nu_i} \over \pi}\right)^2\right]~~,
\label{density}
\end{equation}
where of course $ \xi_{\nu_e}$ also contributes. 

It was determined that the establishment of the large mixing angle solution to the solar neutrino problem prohibits the preservation of the order-of-magnitude difference between the electron and muon/tau neutrinos’ asymmetries \cite{dolgov:2002ab}. This is certainly not due to vacuum oscillations, as the self-potential dominates the other potentials $V^{T,B}$, and vacuum term $\mathbf \Delta$, by several orders of magnitude when the requisite degenerate asymmetries could initially be present. An originally larger flavor asymmetry in the $\nu_\mu$ and/or $\nu_\tau$ is actually preserved due to the presence of the thermal potential of electrons and positrons in the background contributing to  $V^{B}$ asymmetrically between the flavors. The disappearance of the $e^\pm$  background relaxes the potential sum ${\mathbf \Delta}+{V}^B\hat {\mathbf z}$ to its vacuum orientation of  $\mathbf \Delta$, which, for the large mixing angle solution is large and oriented closely along the $x$-axis. 

However, the neutrino self-potential $\mathbf V^S=\alpha ({\mathbf J} - \bar{\mathbf J})$ is orders of magnitude larger,
$$\partial_t{\mathbf P}_p = + {\mathbf A}_p\times {\mathbf P}_p +
  \alpha ({\mathbf J} - \bar{\mathbf J}) \times {\mathbf P}_p$$
$$\partial_t{\bar{\mathbf P}}_p = -{\mathbf A}_p\times \bar{\mathbf
P}_p + \alpha ({\mathbf J} -  \bar{\mathbf J})  \times \bar{\mathbf P}_p\, .$$
If one were to ignore for a moment the much larger self potential, each neutrino momentum mode would follow the vacuum plus thermal potentials, $\mathbf A_p \equiv \mathbf \Delta+  V^{B}\hat{\mathbf z}$,
$$\partial_t{\mathbf P}_p = + {\mathbf A}_p\times {\mathbf P}_p $$
$$\partial_t{\bar{\mathbf P}}_p = -{\mathbf A}_p\times \bar{\mathbf
P}_p $$
\begin{equation}\nonumber
{\mathbf A}_p = {{ \mathbf \Delta}(p)}
 +{ V^T(p)} \hat{\mathbf z}\, ,
\end{equation}
and, therefore, the flavors would all transform as $(-){\mathbf A}_p$ goes from oriented along the $z$-axis to the vacuum orientation ${\mathbf \Delta}(p)$, Eq.~\eqref{eq:vacpot}.

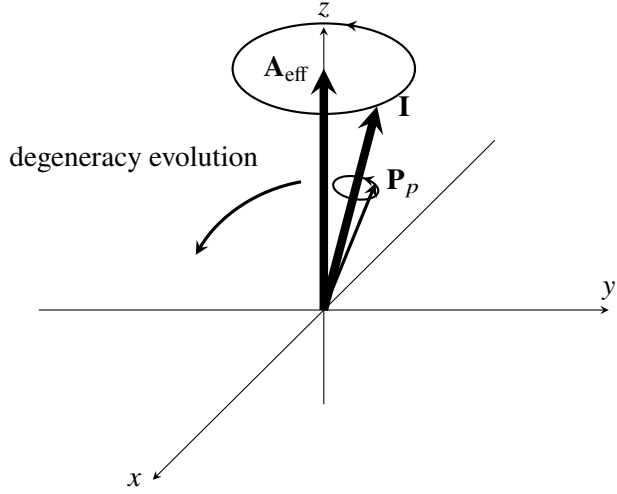
\begin{figure}[htbp]
\begin{center}
\begin{tikzpicture}[x=0.5cm,y=0.5cm,z=0.3cm,>=stealth]
\begin{scope}[decoration={
	markings,
	mark=at position 0.2 with {\arrow{>}} }
      ]

\draw[->] (xyz cs:x=-7.5) -- (xyz cs:x=7.5) node[above] {$y$};
\draw[->] (xyz cs:y=-2.5) -- (xyz cs:y=7.5) node[above] {$z$};
\draw[->] (xyz cs:z=7.5) -- (xyz cs:z=-7.5) node[left] {$x$};

\draw[->,line width=3.2pt] (0,0,0) -- (0,6.4,0 ) node[left] {$\mathbf A_\mathrm {eff}$};
\draw[->,line width=3.2pt] (0,0,0) -- (2.,6.,-1 ) node[right] {$\ \mathbf I$};
\draw[->,line width=1.2pt] (0,0,0) -- (1.9,3.9,-0.9 ) node[right] {$\mathbf P_p$};
\draw[thick,postaction={decorate}] (0,6.4,0) ellipse (2.4  and 1.2);
\draw[thick,rotate around={-10:(1.2,3.6,-0.6 )},postaction={decorate}] (1.2,3.6,-0.6 ) ellipse (0.6  and 0.3);
\draw[->,line width=1.2pt] (0,4,-1) arc (100:150:4);

\node[align=center] at (-5,4) (ori) {\text{degeneracy evolution}};
\end{scope}
\end{tikzpicture}

  \caption{\label{aeff} This is the flavor evolution of the collective neutrino oscillations with an effective potential $\mathbf A_\mathrm {eff}$ and collective flavor vector $\mathbf I$. } 
     \end{center}

\end{figure}

It was shown in Refs.~\cite{abazajian:2002qx,wong:2002fa} that the system’s collective momenta modes, aggregated into the vector ${\mathbf I}\equiv {\mathbf J} - \bar{\mathbf J}$ actually follows an effective ${\mathbf A}_{\rm eff}$,
\begin{equation}  
{\partial_t{\mathbf I} = {\mathbf A}_{\rm eff}\times {\mathbf I}}\, ,
\end{equation}
where
\begin{equation}
\nonumber
{{\mathbf A}_{\rm eff} \simeq \frac{1}{{\mathbf I}^2} \int {\mathbf
A}_p ({\mathbf P}_p + {\mathbf{\overline P}}_p) \cdot{\mathbf I}}\, .
\end{equation}
Since the system starts with a thermal asymmetry, $\mathbf I||(\mathbf P_p+\bar{\mathbf P}_p)$, the system’s $\mathbf P_p$ remarkably follow the “synchronization potential” ${\mathbf A}_{\rm eff}$, which in turn follows the momentum mode of $p_\mathrm{sync}$ with orientation $\theta_\mathrm{sync}$,
\begin{equation}
\nonumber
{\mathbf A}_{\mathrm{eff}} \equiv
{ \Delta_{\mathrm{sync}} }\left( \sin 2 \theta_{\mathrm{sync}}  {\mathbf{\hat x}}
- \cos 2 \theta_{\mathrm{sync}} {\mathbf{\hat z}} \right)\, .
\end{equation}
Here,
\begin{equation}
{\frac{p_{\rm sync}}{T} = \pi \sqrt{1+\xi^2/2\pi^2}\simeq \pi} \, ,
\nonumber
\end{equation}
and the collective flavor evolution is shown in Fig.~\ref{aeff}.

Recall that the average momentum for a Ferm-Dirac distribution is
\begin{equation}
{\langle p/T \rangle = \frac{7 \pi^4}{180 \zeta(3)} \simeq 3.15}\, . \label{averagep}
\end{equation}
In summary, the synchronized system follows $p_\mathrm{sync}\approx 3.14$ as it follows ${\mathbf A}_{\rm eff}$. Momentarily ignoring the synchronization self-potential, the system’s momenta modes $\mathbf P_p$ evolve independently, and on average would have the behavior of the average momentum, Eq.~\eqref{averagep}. Though the self-potential dominates by orders of magnitude, the system effectively behaves as if it is not there since the momentum that the synchronization follows is practically identical to the average Fermi-Dirac momentum when the self-potential is not present \cite{abazajian:2002qx,wong:2002fa}.

In the end, the synchronized evolution orients the final asymmetry to be slightly non-symmetric at an orientation given by the large but not maximal mixing angle of the solar neutrino solution (so that the final ${\mathbf A}_{\rm eff}$ orientation is not directly in the $x$-axis direction). Therefore, the precise determination of the solar mixing angle by laboratory experiments will provide a constraint on the lepton asymmetry in the Universe. Even without a precise determination of the solar mixing angle, however, the large difference in asymmetry between the electron neutrinos and mu/tau neutrinos is largely eliminated by the synchronized evolution, and degenerate primordial nucleosynthesis is no longer viable. The constraints on $\nu_e$ asymmetry are directly correlated with the uncertainty on the primordial helium abundance mass fraction, $Y_p$:
$\Delta \xi_e \approx {\Delta Y_p}/{Y_p}. $
Conservatively, errors ${\Delta Y_p}/{Y_p}\approx 0.1 \Rightarrow \Delta \xi_e \lesssim 0.1$ or, more accurate accounting gives $0.03\lesssim \xi_e \lesssim 0.07$ \cite{abazajian:2002qx}. The final asymmetry that can exist from any early era is given by the final asymmetry vector orientation, which is determined by the solar neutrino angle. Given the restriction on the final degeneracy parameter and related lepton number $\left[L_\alpha \equiv \frac{\nu_\alpha-\nu_{\bar\alpha}}{n_\gamma}\right]$,$|L_e| \lesssim 0.1$.
The final asymmetry can be
\begin{eqnarray}
\nonumber 
\xi_e^f &=& 
\left(\frac{1-\cos 2\theta_0}{2}\right)\xi_{\mu^\ast}^i \,, \\
\xi_{\mu^\ast}^f &=& 
\left(\frac{1+\cos 2\theta_0}{2}\right)\xi_{\mu^\ast}^i \,,
\nonumber
\end{eqnarray}
where $\mu^\ast$ is the effective 2-neutrino solution of the $\nu_\mu/\nu_\tau$ neutrino flavor system with $\nu_e$. Therefore, the possible final lepton number in $\nu_\mu/\nu_\tau$ is relatively very small $ |L_\mu+L_\tau| \lesssim 0.5$.

\subsection{Collisions}
The evolution of neutrinos in the early Universe is of course not only determined by thermal and matter effects. They are also undergoing rapid collisions, $\nu_\alpha + X \leftrightharpoons \nu_\alpha +X $, where $X$ is any weak-charge carrying particle, including charged leptons, hadrons, other neutrinos, and above the QCD-transition, quarks themselves. The collision term, ${\mathcal C}[{\mathbf P}(p)]$, in the precession equation formulation is included as
\begin{equation}
\partial_t{\mathbf P}(p) = {\mathbf V}(p)\times {\mathbf P}(p) 
+ {\mathcal C}[{\mathbf P}(p)]\, .
\end{equation}
In full, the collision term is 
\begin{align}
{\mathcal C}[{\mathbf P}(p)] &\approx - D(p) {\mathbf P}_T(p) 
 +\int{dp^\prime
    d(p,p^\prime){\mathbf P}_T(p^\prime)} \\
& \qquad - {\mathbf C}(p)P_0(p)+
    \int{dp^\prime {\mathbf c}(p,p^\prime) P_0 (p^\prime)}\, ,
\end{align} 
where ${\mathbf P}_T \equiv P_x(p)\hat{\mathbf x}+ P_y(p)\hat{\mathbf y}$.
Here, the length of the polarization vector $P_0(p)$ could change with collisions, but is constant with detailed balance. The second and fourth terms of momentum exchange are also symmetric under detailed balance, and therefore typically ignored. The primary term is the “damping” term, $- D(p) {\mathbf P}_T(p)$, that acts against any evolution of ${\mathbf P}(p)$ away from the $z$-axis. That is, collisions reset the flavor evolution of neutrinos so that they are no longer in a mixed state, as expected. This is sometimes referred to as the “quantum-Zeno” effect (which is also known as the Turing paradox). An illustration of this is shown in Fig.~\ref{zeno}.
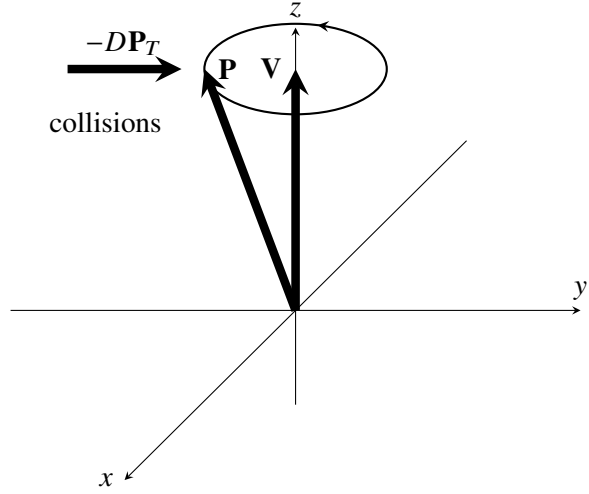
\begin{figure}[htbp]
\begin{center}
\begin{tikzpicture}[x=0.5cm,y=0.5cm,z=0.3cm,>=stealth]
\begin{scope}[decoration={
	markings,
	mark=at position 0.2 with {\arrow{>}} }
      ]

\draw[->] (xyz cs:x=-7.5) -- (xyz cs:x=7.5) node[above] {$y$};
\draw[->] (xyz cs:y=-2.5) -- (xyz cs:y=7.5) node[above] {$z$};
\draw[->] (xyz cs:z=7.5) -- (xyz cs:z=-7.5) node[left] {$x$};

\draw[->,line width=3.2pt] (0,0,0) -- (0,6.4,0 ) node[left] {$\mathbf V$};
\draw[->,line width=3.2pt] (0,0,0) -- (-2.4,6.4,0 ) node[right] {$\mathbf P$};
\draw[thick,postaction={decorate}] (0,6.4,0) ellipse (2.4  and 1.2);

\draw[->,line width=3.2pt] (-6,6.4,0) -- (-3,6.4,0 )  node[pos=0.5, above]{$-D\mathbf P_T$};

\node[align=center] at (-5,5) (ori) {\text{collisions}};
\end{scope}
\end{tikzpicture}

  \caption{\label{damping} Collisions damp flavor oscillations, with the $-D\mathbf P_T$ term, pinning the flavor asymmetry vector $\mathbf P$ to the pure flavor state oriented with the $z$-axis, which it will do even if the potential $\mathbf V$ does not align with the $z$-axis. } 
     \end{center}

\end{figure}

\begin{figure}[htbp]
\begin{center}
\includegraphics[width=4.5in]{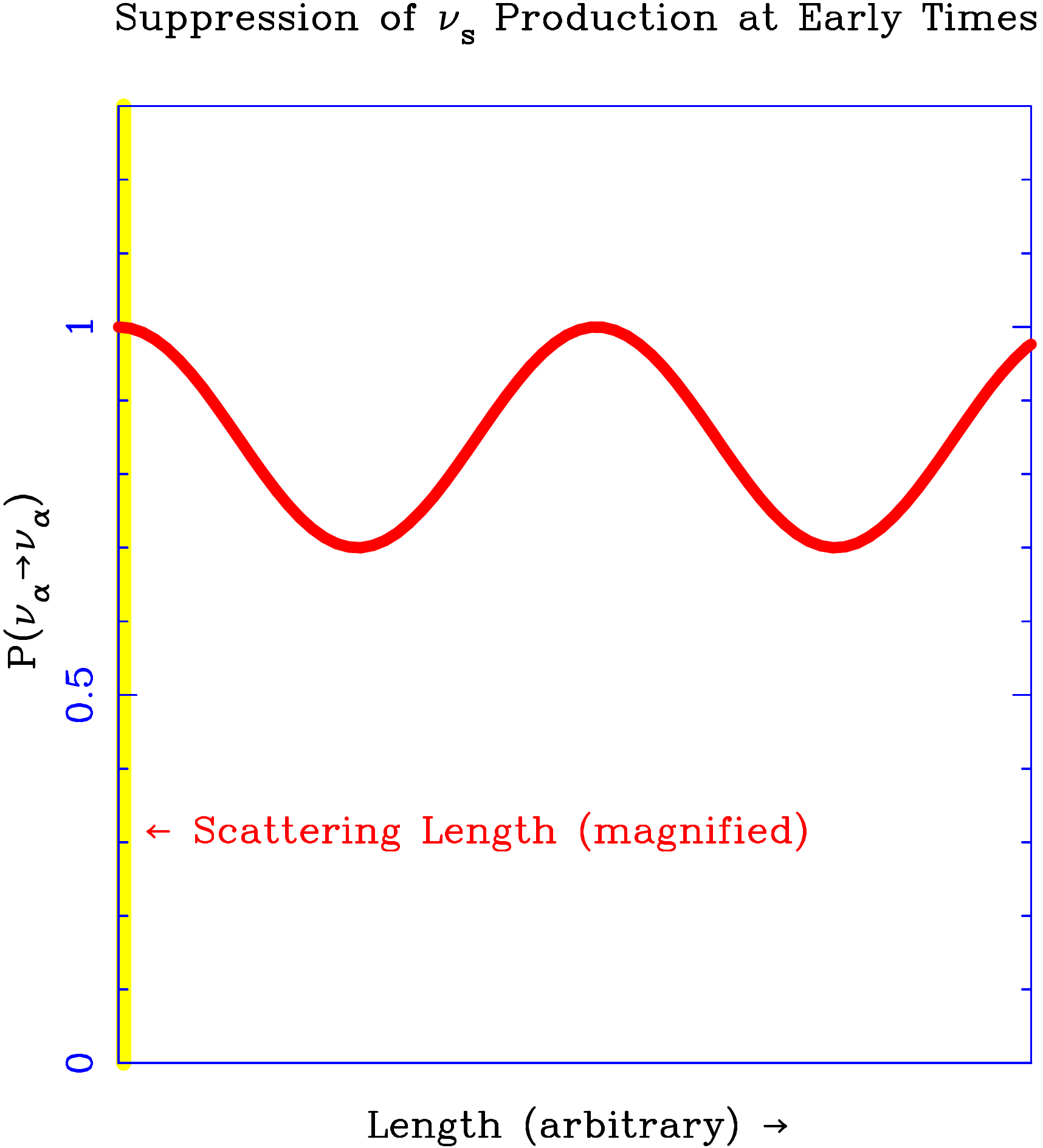}
\caption{Shown is an exaggerated oscillation conversion probability for a monochromatic neutrino in the early Universe for active-sterile mixing. When the scattering rate is rapid enough, it “damps” flavor conversion in what is sometimes referred to as the quantum “Zeno” effect.}
\label{zeno}
\end{center}
\end{figure}

In addition to reseting the flavor state evolution of mixed neutrinos, the collisions also “test” if the neutrino is in a flavor state that can interact or is in a weak-interaction singlet or sterile state. If it can interact, it is an active neutrino at the collision. If the mixed, evolved neutrino has a nonzero probability to be a sterile state, it could “non-scatter” to enter into the sterile neutrino state. This is how sterile neutrinos are populated in the early Universe. The overall sterile neutrino conversion rate can be derived from the density-matrix formalism to that of the quasi-classical form in Eq.~\eqref{conversrate} \cite{Bell:1998ds}.

\subsection{Lepton Number Generation and Destruction in the Early Universe}

In the case of a mostly sterile mass eigenstate with mass less than one of the mass eigenstates more closely associated with an active neutrino, the phenomenon of lepton number generation can occur. On the other hand, if there is any previously-generated lepton number, either through the aforementioned mixing, or via another mechanism, lepton number can also be destroyed by resonant flavor evolution in the early Universe. The evolution of a neutrino state in the early Universe is in many ways similar to that through the solar envelope with decreasing density with time on its radial trajectory. However, in the early Universe the background potential can include not only the charged lepton and baryon asymmetry in Eq.~\eqref{eq:vd}, but also contributions from any asymmetry in the neutrino background. That is, for active-sterile mixing, one must use the lepton potential Eq.~\eqref{vl}.
Because of the very small baryon number relative to photon number, $n_b/n_\gamma \sim 10^{-10}$, the effects of the baryon asymmetry are usually negligible. I have written the asymmetries as lepton number asymmetries, and they are positive (negative) for neutrinos (antineutrinos) in Eq.~\eqref{leptonnumber}. An excellent presentation of the physics and dynamics of relic neutrino evolution in the presence of sterile neutrinos is in Ref.~\cite{Bell:1998ds}.

Resonance occurs when the term in brackets in the denominator of Eq.~\eqref{eq:matterangleEU} is nil.  The resonance is therefore momentum dependent.  The resonance for neutrinos and antineutrinos is the same if there is no asymmetry, $L_\nu$, but will be different for finite $L_\nu$.  The positions of the resonances can cause either a {\it stable evolution} or an {\it unstable evolution}.  

As there is a small matter asymmetry from the baryon asymmetry, with a baryon to photon ratio of order $10^{-10}$, the positions of the resonance (maximal mixing) in the distribution of the neutrinos will be different for neutrinos and antineutrinos due to the sign difference of $V^B$.

The evolution of the presence of the resonance between active and sterile neutrinos proceeds as follows \cite{Foot:1995qk,Shi:1996ic,Foot:1996qc}:
\begin{itemize}
\item{\bf For Neutrinos:} the momentum position ($p/T$) of resonance increases if
$L_\nu$ increases. 
\item{\bf For Antineutrinos:} the momentum position ($p/T$) of the resonance decreases
as $L_\nu$ increases.
\end{itemize}
This dependance causes a stability if the resonances are below the distribution peak ($p/t \sim 2.2$), and an instability if the resonances are above the distribution peak.

Below the distribution peak, the neutrino resonance samples a greater number of neutrinos than the antineutrino resonance (due to the greater abundance at higher $p/T$).  This has the property of lowering $L_\nu$, driving the resonances of $\nu/\bar\nu$ closer together, and stabilizing the evolution.

Above the distribution peak, the neutrino resonance samples a lower number of neutrinos than the antineutrino  resonance (due to the lower abundance at higher $p/T$) for conversion into sterile neutrinos.  This has the property of increasing $L_\nu$, driving the resonances further apart, and amplifying the difference between the $\nu/\bar\nu$ resonances.  This nonlinearly
amplifies the magnitude of $L_\nu$.  
The same amplification occurs above the thermal peak for $L_\nu <0$,
as the positions of the $\nu/\bar\nu$ resonances are switched. The maximal production of an asymmetry is of order 

\begin{figure}[htbp]
\begin{center}
\includegraphics[width=5.5in]{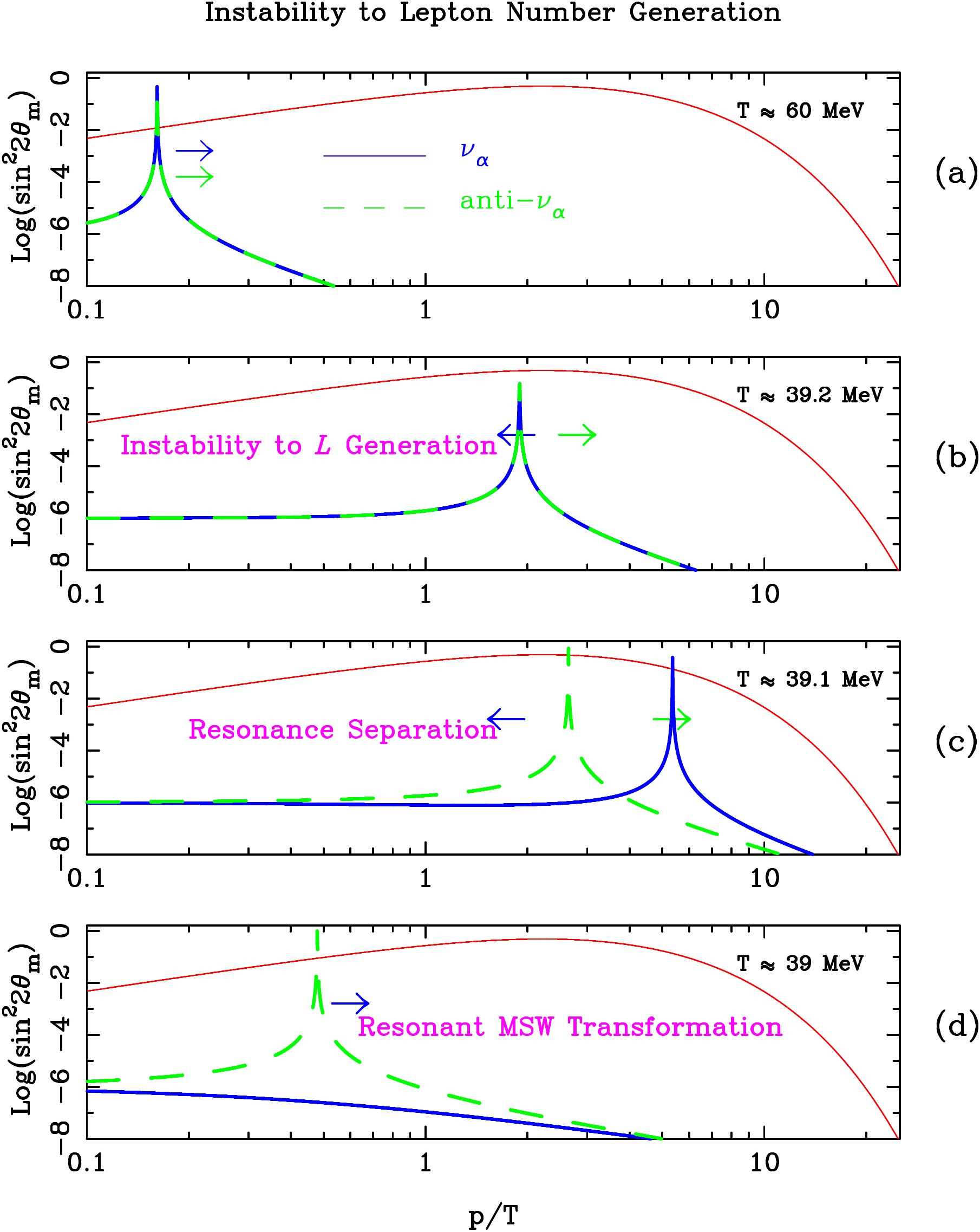}
\caption{Shown are the steps for resonance evolution in the case of spontaneous lepton number generation in the early Universe due to active-sterile neutrino mixing. See text for details.}
\label{resonanceevols}
\end{center}
\end{figure}

For the case where the mostly-sterile mass eigenstate is smaller in mass than the active state, $\delta m^2_{s\alpha} < 0$, the evolution of the resonant conversion proceeds like the following \cite{Foot:1995qk,Shi:1996ic,Foot:1996qc}.  A schematic picture of the evolution is in Fig.~\ref{resonanceevols}, with steps (a)-(d) being:
\begin{enumerate}[(a)]
\item Initially, one can assume that $L_{\nu_\alpha}=0$, so the
effective potential $V$ is dominated by the thermal term $V^T$.
Therefore, the $\nu_\alpha$ and $\bar\nu_\alpha$ have the same
resonance. The population of the sterile neutrinos proceeds symmetrically.
\item At this point, just to the high energy side of the thermal
distribution, the lower temperatures bring the $V^B$ term closer to
$V^T$. An instability sets in, and drives the production of lepton
number since the resonances are sampling different portions of the
distribution and have different efficiencies.
\item The instability drives a separation of the neutrino and
anti-neutrino resonances.  The asymmetry $L$ starts being produced nonlinearly.
\item Only the neutrino resonance remains within the populated portion
of the energy spectrum. A resonant MSW conversion of the active
neutrinos to sterile neutrinos proceeds. In this case, $L_\nu < 0$ is
generated. It can be equally likely that the anti-neutrino resonance
is left and $L >0$ is generated.
\end{enumerate}
This process is clearly nonlinear, and it has been argued to be physically chaotic \cite{Shi:1996ic}. It can lead to fractal structure of the sign of the final lepton asymmetry produced, so that if sterile neutrinos are discovered, some regions of parameter space are so that it even arbitrary precision of the active-sterile neutrino system will not provide a prediction of the sign of the lepton number that is produced \cite{Abazajian:2008dz}.

{\bf Lepton number destruction:} In addition to this possibility of lepton number generation, the resonant evolution in the presence of sterile neutrinos described above can destroy, partially or completely, any previously existing neutrino flavor asymmetry. The asymmetry could have been produced by active-sterile neutrino mixing or an earlier phenomenon. If the asymmetry is produced by active-sterile neutrino mixing in a four-neutrino framework, the consequences are complicated by many contingencies of the neutrino mass levels and ordering, and can either adversely affect primordial nucleosynthesis by increasing the relic neutrino density and altering the weak rates with asymmetries in $\nu_e/\bar\nu_e$. Multi-flavor mixing can offset deleterious effects of an increased radiative energy density in sterile neutrinos with an asymmetry in $\nu_e/\bar\nu_e$ affecting the weak rates, similar to degenerate primordial nucleosynthesis described previously. The neutrino momentum distributions can be nonthermal due to incomplete conversion and the non-thermal nature of the MSW mechanism \cite{Bell:1998sr,Shi:1999kg}. The conversion of a previously generated neutrino asymmetry can be resonantly converted to sterile neutrinos with $\sim$keV-scale masses so that MSW-resonance generated sterile neutrinos could be all or a portion of cosmological dark matter \cite{shi:1998km}. In this case for keV scale neutrino dark matter, the resonant conversion occurs at $\gtrsim$100 MeV, and lepton number cannot be generated from active-sterile oscillations prior to that temperature scale due to limits on the magnitude of $\delta m^2_{\alpha s}$. Recall the sterile neutrino is less massive than the active neutrino for resonant lepton number production, and the active neutrino mass is constrained by the CMB and LSS, discussed next.

\section{Neutrinos from the CMB and Large-Scale Structure}

The standard cosmological model that emerged over 20 years ago successfully describes observations from Galactic scales to the cosmic horizon using a handful of parameters: the densities of dark matter, dark energy, baryonic matter, the Hubble expansion rate, the spectrum of primordial perturbations arising from inflation, as well as the amplitude of those perturbations. It has become a standard part of the Particle Data Group’s Review of Particle Physics \cite{10.1093/ptep/ptaa104}. There are excellent pedagogical review papers on the effects of massive neutrinos and any possible extra neutrinos on cosmological observables, particularly the CMB and LSS \cite{lesgourgues:2006nd,Abazajian:2016hbv}. Therefore, I will provide only a brief summary here of the effects of neutrino mass and number. 

 The energy density of neutrinos in thermal equilibrium in the early Universe is 
\begin{equation}
\rho_\nu = \frac{7}{8}\ \frac{\pi^2}{30} g T_\nu^4,
\label{neutrinodens}
\end{equation}
where $g$ are the number spin states of the neutrinos.  If the neutrinos are relativistic---which is only true at very early times for massive neutrinos---one can define the  ``effective'' number of neutrinos relative
to a relativistic single  thermal neutrino density, Eq.~\eqref{neutrinodens}, as in Eq.~\eqref{neff}.
$N_\mathrm{eff}$ is an effective “specific density” for relativistic cosmological constituent description of 
the total radiation density $\rho_{\rm rel}$, explicitly not including cosmic photons.

The neutrinos are kept in thermal equilibrium by weak interactions such as $\nu_\alpha+e^\pm \leftrightarrow \nu_\alpha+e^\pm$. As these rates drop below the expansion rate, a “weak decoupling” occurs and the neutrinos are decoupled from the $e^\pm$ and photon background. Since electrons and positrons annihilate after this decoupling, the neutrinos have a different, cooler, temperature than the photons, which carry forward the energy density and entropy of the $e^\pm$ background. The temperature ratio of photons to neutrinos becomes approximately 
\begin{equation}
\frac{T_\gamma}{T_\nu} = \left(\frac{11}{4}\right)^{1/3} \simeq 1.40.
\end{equation}

However, the neutrinos are thermal and the highest energies in their distribution are still partially coupled to the $e^\pm$ background when it annihilates away. A small amount of heating of the active neutrinos then occurs. Detailed calculations of this heating find that the standard active neutrino density in the relativistic era is such that $N_{\rm eff} = 3.046\ \text{to}\ 3.052$ \cite{lopez:1998vk,mangano:2005cc,Grohs:2015tfy}. Note, fully describing this out-of-equilibrium energy transfer to the neutrinos requires their full momentum space distribution function, and, to include mixing, would require the momentum dependent $3\times 3$ full density matrix for the active neutrinos alone. 

Eventually, the neutrinos’ mass becomes small relative to their temperature, and they act as non-relativistic matter. Therefore, they are a known component of dark matter, though certainly not a large fraction of it. The density of massive neutrinos relative to the cosmological critical density is 
\begin{equation}
\Omega_\nu = \frac{\Sigma m_\nu n_\nu}{\rho_{\rm crit}} \approx
\frac{\Sigma m_\nu}{94.10{\rm eV}\ h^2},\label{eq:omeganu}
\end{equation}
where $\Omega_\nu=\rho_\nu/\rho_{\rm crit}$  and $\Sigma m_\nu $ is the sum of the three standard neutrino mass
eigenstates and the Hubble constant is parameterized as $h\equiv
H_0/(100\rm\ km\ s^{-1}\ Mpc^{-1})$. The numerical value on the r.h.s. above 
is set  for the case $N_{\rm eff}
=N_\nu = 3$. Including non-thermal effects for the slight heating from $e^\pm$ modifies the relation by $\sim$1\%. 

\subsection{Neutrino Mass from the CMB and LSS}

\subsubsection{$\Sigma m_\nu$ Effects on LSS}

\begin{figure}[htbp]
\begin{center}
\includegraphics[width=5.5in]{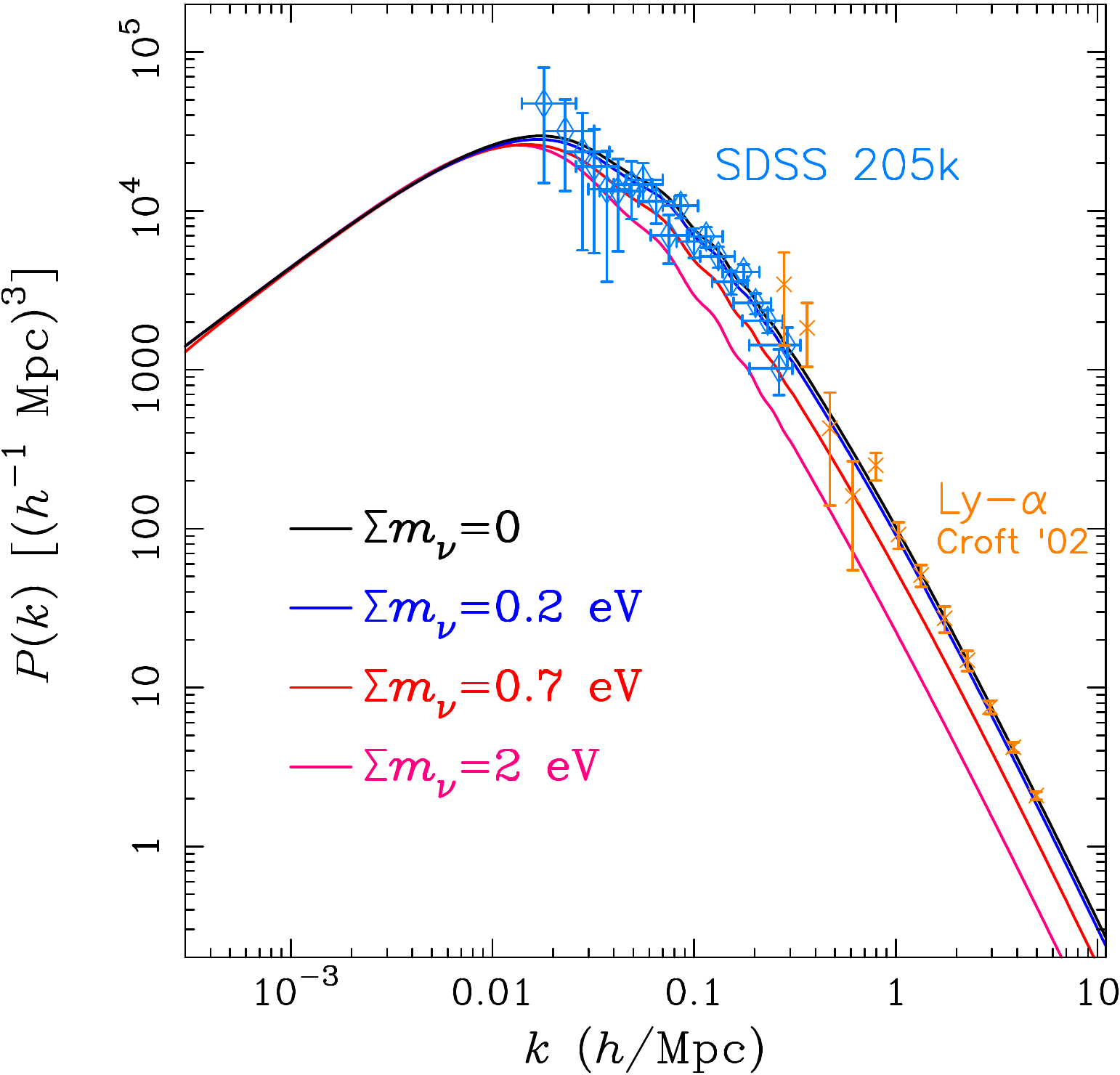}
\caption{The matter power spectrum for several cases of the sum of neutrino masses is plotted, along with some of the earliest cosmological LSS data that placed stringent constraints, from the 205k main galaxy sample of the SDSS \cite{tegmark:2003uf} as well as Keck HIRES and LRIS data from Croft et al. \cite{Croft:2000hs}. It is readily apparent that cosmological measurements can compete with terrestrial laboratory sensitivity absolute to neutrino mass.}
\label{pkmnu}
\end{center}
\end{figure}

The active neutrinos cannot contribute a large fraction to the matter density in the Universe because they are light and free-stream due to their thermal velocities, and they suppress the growth of structure along the way. The suppression of growth occurs due to two physical effects: 
\begin{enumerate}
\item the free streaming of neutrinos in the radiation dominated era to late times so that they do not cluster as cold dark matter does, which is about one-fourth of the suppression, and 
\item the lack of participation of neutrinos in the linear growth of LSS, while they do contribute to the total matter density which affects the “friction” Hubble term in the growth equation of LSS. This is about three-fourths of the suppression effect \cite{Eisenstein:1997jh}. 
\end{enumerate}
On the largest scales, the neutrinos cluster as normal matter. The total effect of suppression of clustering on scales below the free-streaming scale is given by 
\begin{equation}
\frac{\Delta P_\delta}{P_\delta} \approx  -8
\frac{\Omega_\nu}{\Omega_m}\approx -8\frac{\Sigma m_\nu}{94 \ \Omega_m h^2\ {\rm eV}}\,.
\end{equation}
This was first described in Hu, Eisenstein, and Tegmark \cite{Hu:1997mj} as a consequence of inferring total matter clustering as traced by Luminous Red Galaxies in the original Sloan Digital Sky Survey, in a “cosmic” complementarity with expected cosmological parameter limits from planned CMB experiments \cite{Eisenstein:1998hr}. In Fig.~\ref{pkmnu}, I plot the matter power spectrum $P(k)$ for fixed large amplitudes and increasing the neutrino mass, along with some of the first data that constrained neutrino mass from the Sloan Digital Sky Survey (SDSS) as well as the inferred clustering of matter from early Lyman-$\alpha$ forest data \cite{tegmark:2003uf,Croft:2000hs,croft:1999mm}.  The forecast sensitivities of CMB via the Wilkinson Microwave Anisotropy Probe and Planck Observatory with SDSS Galaxy Redshift Survey data were remarkably accurate when compared to the eventual results, including neutrino mass sensitivities \cite{Eisenstein:1998hr,Tegmark:1999ke}. 
The sensitivity to neutrino mass from among the latest LSS galaxy surveys, combined with the latest Planck Observatory CMB observations place the limits on neutrino mass at \cite{Cuesta:2015iho,Canac:2016smv}
$$
 \Sigma m_\nu < 0.13\rm\ eV\ (95\%\ CL)\, .
$$
I discuss future sensitivities in the next section.

\subsubsection{$\Sigma m_\nu$ Effects on the CMB}

Massive neutrinos also have direct effects on the CMB anisotropies. The physics of the effects of massive neutrinos on the CMB was first explored in Ref.~\cite{Dodelson:1995es}. Though neutrinos with mass $\ll 1\rm\ eV$ are relativistic at CMB decoupling, their transition to matter-like affects the angular diameter distances to the last scattering surface as well as the scale of the integrated Sachs-Wolfe effect. There are degeneracies with the Hubble expansion rate and cosmological constant density, though the primary CMB can still place direct constraints on neutrino mass, which are at the level of $\Sigma m_\nu < 0.26\rm\ eV$ (95\% CL) when using both temperature and polarization anisotropy data from the Planck Observatory data of the primary CMB \cite{Aghanim:2018eyx}.
The primary CMB is also affected by intervening LSS, called lensing the CMB. The modification to LSS can then be imprinted on the CMB itself \cite{Kaplinghat:2003bh}. 

The degeneracy of effects of massive neutrinos on expansion history at later times, via angular diameter distance measures, are strongly constrained by the baryon acoustic oscillation (BAO)  scale measurements in galaxy redshift surveys. Even without LSS data, CMB plus BAO give the limit of \cite{Aghanim:2018eyx}
$$
 \Sigma m_\nu < 0.12\rm\ eV\ \ (95\%\ CL)\, .
$$
This puts CMB plus geometric measure constraints on neutrino mass at the same level of sensitivity as CMB plus LSS, described above. 

With the limits now at the $\approx$120 meV scale, they are within a factor of two of the minimal level of neutrino mass given by oscillation experiments, $\Sigma m_\nu \approx 58\rm\ meV$. In combination, future CMB and LSS measures promise to get close to detection threshold of this minimal atmospheric plus solar total neutrino mass. Such future observations include galaxy surveys, the lensing of the CMB, galaxy-galaxy weak lensing, as well as mapping of hydrogen by the 21 cm background (for a review, see Ref.~\cite{Abazajian:2016hbv}).  

\subsection{Neutrino Number from the CMB and LSS}

\subsubsection{$N_\mathrm{eff}$ Effects on  LSS}

Radiation energy density that is parametrized in Eq.~\eqref{neff} effectively does one thing: alter the presence of radiation energy density relative to other cosmologically-relevant mass-energy components. The primary outcome is the alteration of the moment of matter-radiation equality as radiation energy density scales as the scale factor, $a$, to the inverse fourth power, $a^{-4}$, and matter density as the inverse third power, $a^{-3}$. The effect of an increase or decrease in $N_\mathrm{eff}$ on the scale of matter-radiation equality can be kept constant with a commensurate alteration in the matter density, along with a requisite change in $\Lambda$ density to keep matter-$\Lambda$ equality constant. All of those geometric scales set by the equality benchmarks are strongly constrained by cosmological data, and are how $N_\mathrm{eff}$ can be masked by the degeneracies with the other density parameters. However, the alteration in matter density has to be within the dark matter component, $\Omega_\mathrm{DM}$ as the baryon density $\Omega_b$, via $\Omega_b h^2$, is independently strongly constrained by the CMB’s acoustic peaks. Therefore, secondary effects are what break the degeneracies between these densities. Namely, any increase in  $N_\mathrm{eff}$ will decrease the ratio of baryon to dark matter $\Omega_b/\Omega_\mathrm{DM}$, which is itself well constrained by LSS. The radiation dominated epoch is when the $k$-modes below the peak of $P(k)$ in Fig.~\ref{pkmnu} enter the horizon and grow. During this epoch, baryons remain coupled to the photon background through Couloumb interactions, and do not cluster. The enhancement of $\Omega_\mathrm{DM}$, in order to increase with $N_\mathrm{eff}$, in the ratio $\Omega_b/\Omega_\mathrm{DM}$ enhances clustering of that dark matter component, and therefore augments the small-scale power, below the peak of $P(k)$ in Fig.~\ref{pkmnu}, when keeping the matter-radiation and matter-$\Lambda$ scales equal. Decreasing the ratio $\Omega_b/\Omega_\mathrm{DM}$ also damps the BAO feature since there is less baryon density. Overall, this allows both the BAO feature and shape of $P(k)$ to constraint $N_\mathrm{eff}$. The constraint from Planck 2018 with BAO data is \cite{Aghanim:2018eyx}
$$
N_\mathrm{eff} = 2.99^{+0.34}_{-0.33}\ \ (95\%\ \mathrm{CL})\, .
$$
However, the effects on the shape of the matter power spectrum remain of great sensitivity when including luminous red galaxy (LRG) power spectrum data to constrain the inferred matter power spectrum \cite{Reid:2009xm}, placing constraints from matter clustering from CMB+LSS(LRG) that are comparable to that from CMB+BAO \cite{Canac:2016smv}.

\subsubsection{$N_\mathrm{eff}$ Effects on the CMB}

Similar to keeping matter-radiation and matter-$\Lambda$ scales fixed with alterations in those densities commensurate with a modification in $N_\mathrm{eff}$ to avoid LSS constraints, this set of degeneracies exist in the CMB. However, since $\Omega_b h^2$ is very well constrained by the CMB, a modification of the Hubble parameter $h$ is done to keep $\Omega_b h^2$ fixed. Any change in $N_\mathrm{eff}$ then alters the anisotropic stress tensor of the neutrino background, changing the growth of perturbations and therefore altering the position and amplitude of the acoustic peaks of the CMB. In addition, the diffusion damping scale of photons is altered with $N_\mathrm{eff}$. The acoustic peaks’ change is directly proportional to changes in $h$ while the diffusion scale goes as mean-free-path time scales, as $\sqrt{h}$. The differential scaling of these two major features of the CMB---their acoustic peaks and their damping scale---provides the sensitivity of the CMB to $N_\mathrm{eff}$. The constraint from Planck 2018 alone is nearly as good as that including the BAO data \cite{Aghanim:2018eyx},
$$
N_\mathrm{eff} = 2.92^{+0.36}_{-0.37}\ \ (95\%\ \mathrm{CL})\, .
$$

\subsection{Tension Data Sets: Evidence for Neutrino Mass and Extra Neutrino Number?}
Two significant anomalies have persisted in cosmology the past few years. First, the most consistent anomaly is the measurement of a local Hubble expansion rate, $H_0$, that is statistically significantly higher than that inferred from measures of the CMB and BAO. This tension arose after Planck 2015 data’s inferred $H_0$ \cite{Ade:2015xua} was inconsistent with local measures from 2011 at 3$\sigma$ \cite{Addison:2015wyg}. The inconsistency increased to 3.4$\sigma$ with an increased precision of the Hubble Space Telescope (HST) local $H_0$ measurement \cite{Riess:2016jrr}, and to 3.7$\sigma$ in 2018 \cite{Riess:2018uxu}. In late 2020, the inconsistency has increased to 4.2$\sigma$ \cite{Riess:2020fzl}. One hypothesis for the origin of the tension is an increase $N_\mathrm{eff}$ to ameliorate the early and late Universe expansion  discrepancy. From the previous section, recall that the desire is to decrease early-Universe $h$ to keep $\Omega_b h^2$ fixed while increasing $N_\mathrm{eff}$ and other densities. However, an increase in $N_\mathrm{eff}$ affects the other CMB and LSS observables described above, and relieving the “tension” with local $H_0$ measures increases internal tension in CMB and LSS observables. Therefore, the statistical evidence for $N_\mathrm{eff}$ as being a solution to the early-to-late Universe $H_0$ tension is not present \cite{Canac:2016smv}. In addition, an order-one increase in $N_\mathrm{eff}$ will come into conflict with primordial element abundance constraints, described above. Evidence actually suggests that the $H_0$ tension is more likely in effects arising from non-standard dark energy \cite{DiValentino:2016hlg,Poulin:2018cxd,Keeley:2019esp}. 

The second tension data set, less consistent across all the data, is the inferred amplitude of perturbations at 8 Mpc~$h^{-1}$ scales, dubbed $\sigma_8$. This scale, which is quite large of course on human scales, is a measure of cosmologically relatively small scale perturbations, and is the scale at approximately where the gravitational growth of perturbations is entering the nonlinear regime. Some specific sets of cosmological data infer an amplitude of clustering at small scales that is smaller than that inferred with zero neutrino mass and the inferred cosmological parameters from the CMB. One way to reduce this amplitude is to increase neutrino mass, as shown in Fig.~\ref{pkmnu}. Several cluster and lensing data sets indicate a lower $\sigma_8$ than that inferred from CMB data, which could be indicative of significant active neutrino masses, extra massive sterile neutrinos, or both \cite{Battye:2013xqa,Wyman:2013lza,Dvorkin:2014lea,Beutler:2014yhv}. The lower $\sigma_8$ could also indicate partially thermalized sterile neutrinos from the eV \cite{Jacques:2013xr} to tens of eV scales \cite{Abazajian:2019ejt}. If this persists with new datasets, it would be a significant cosmological discovery. However, these results are inconsistent with the standard BAO and LSS results described above, and, for the case of thermalized sterile neutrinos, would be inconsistent with $N_\mathrm{eff}$ constraints from primordial nucleosynthesis, Eq.~\eqref{neffbbn}. 

\section{Sterile Neutrino Dark Matter}

\subsection{Oscillation-based Production Mechanisms}
The production of sterile neutrino dark matter via matter-affected oscillations largely follows the physics described in Section \ref{nuearly}. What is different for the dark matter scale is that, in order for it to be consistent with structure formation, the sterile neutrino particle mass must be $\gtrsim$keV, and that pushes up the oscillation-based production, whether through a non-resonant or resonant production to be at early Universe temperatures of $T\gtrsim 100\rm\,MeV$ (while $\sim$eV-scale sterile neutrinos thermalize at  $T\sim 10\rm\,MeV$). This is all because of how eV to keV scales populate as in Eq.~\eqref{conversrate}, which in the non-resonant production model is the Dodelson-Widrow mechanism \cite{dodelson:1993je}. In Eq.~\eqref{conversrate}, the competing scalings of Eq.~\eqref{productionpower} have the production peak at a specific temperature in the non-resonant case \cite{dodelson:1993je,Barbieri:1989ti}:
\begin{equation}
T_\mathrm{max} \approx 133\,\mathrm{MeV}\left(\frac{m_s}{1\,\mathrm{keV}}\right)^{1/3}\, .
\end{equation}
This ties in the production of keV-scale sterile neutrino dark matter to the time-temperature scale of the quark-hadron transition, at $T\sim 170\,\rm MeV$, and allows for inferences of the nature of the quark-hadron transition if sterile neutrino dark matter is detected \cite{abazajian:2002yz}. The mixing angle and masses needed for dark matter production were also of interest for producing high-velocity pulsar kicks in supernova explosions from asymmetric active-sterile neutrino mixing in core collapse supernovae \cite{kusenko:1998bk,fuller:2003gy}. 

For the case of resonant production, sterile neutrinos are produced from the destruction of a preexisting lepton number via the Shi-Fuller mechanism \cite{shi:1998km}.  The resonance is where the matter-affected mixing angle, Eq.~\eqref{eq:matterangleEU} is maximized. The position of the
resonance in momentum space, $\epsilon_{\rm res} \equiv p/T|_{\rm res}$, evolves as  
\begin{align}
\label{eres}
\epsilon_{\rm res} &\approx {\frac{\delta m^2}{
{\left( 8\sqrt{2}\zeta(3)/\pi^2\right)} G_{\rm F} T^4 {{L}} }}\\
&\approx 3.65 {\left({\frac{\delta m^2}{(7\,{\rm
keV})^2}}\right)} {\left({\frac{{10}^{-3}}{{{L}}}}\right)}
{\left({\frac{170\,{\rm MeV}}{T}}\right)}^4 \nonumber\ ,
\end{align}
depending on the requisite parameters. Note, the resonance is near the peak of the Fermi-Dirac distribution at temperatures at and just above the quark-hadron transition. The resonance evolves from low to high momenta as the Universe cools, and as lepton number is destroyed. Therefore, low-momentum sterile neutrinos are enhanced in production, and “cooler” sterile neutrino dark matter is produced in the Shi-Fuller mechanism than in the Dodelson-Widrow mechanism. Therefore, the free-streaming scale, which is relevant for structure formation signals and constraints, is different for sterile neutrino dark matter produced from the two mechanisms. Note that both Dodelson-Widrow and Shi-Fuller mechanisms could produce sterile neutrino dark matter as a fraction of the total dark matter density. Therefore, the sterile neutrino dark matter’s effects on structure formation are dependent on the nature of the other dark matter component, and reduced as the fraction of dark matter as sterile neutrinos are reduced. 

There is a large set of work on sterile neutrino dark matter production in low-reheating-temperature Universe scenarios \cite{Gelmini:2004ah,Gelmini:2019clw}, where the lower reheating temperature and plasma interaction rate allows for larger mixing angles. Such scenarios also allow for production of sterile neutrinos as a fraction of the total dark matter, and can be detected in X-ray observations, discussed below. These production mechanisms occur at large mixing angles that can be probed in laboratory experiments \cite{Smith:2016vku,Benso:2019jog}.

\subsection{Other Production Mechanisms}
Sterile neutrinos could certainly have other couplings than the mass terms generating neutrino mass. For example, a generic scalar $S$-particle can exist with an interaction Lagrangian
\begin{equation}
\mathcal{L}_\mathrm{int} =
\frac{y}{2}\overline{\left(\nu_R\right)^c}\nu_R S+ h.c. \, .
\end{equation}
$S$-particles can be created in the early Universe and could decay into sterile neutrinos at some later point. Therefore, the production is decoupled from the neutrino oscillation parameters, for cases where the mixing is much less than that needed for Dodelson-Widrow or Shi-Fuller production. 
In general, a scalar-decay sterile neutrino dark matter could have smaller or larger free-streaming scales for the same particle mass in oscillation production, and a good review of the mechanisms is in Ref.~\cite{Adhikari:2016bei}. 

Regardless of the mechanism, if the active-sterile mixing is large
enough for Dodelson-Widrow or Shi-Fuller production to be significant,
the production from Dodelson-Widrow or Shi-Fuller processes have to be included. In the various particle-decay mechanisms, the
``parent'' $S$-particle can itself be in or out of thermal equilibrium
at the time the sterile neutrinos are produced. The decay into the
sterile neutrino dark matter has been considered via several scalar
particles as a single particle decay channel or via one part of
multiple possible decay channels
\cite{shaposhnikov:2006xi,Kusenko:2006rh,Petraki:2007gq,Bezrukov:2009yw,Kusenko:2012ch,Merle:2015oja}.
In scalar decay cases, the production of sterile neutrinos is governed
by a Boltzmann kinetic equation describing their distribution
\cite{shaposhnikov:2006xi}
\begin{equation}
  \frac{\partial f}{\partial t} - H p \frac{\partial f}{\partial p} =
  \frac{2m_S \Gamma}{p^2}\int_{p+m^2_s/4p}^\infty{f(E)\ dE}\, ,
\end{equation}
where the sterile neutrino distribution is $f(p,t)$ as a function of momentum $p$ and time $t$, given a scalar particle of mass $m_S$. The rate $\Gamma$ is the partial width for the
scalar decay into sterile neutrinos. For production when the degrees of freedom in the background plasma is constant,
this kinetic equation is analytically integrable. And, one can get the late
time number density of sterile neutrinos,
\begin{equation}
n_0 \approx \frac{3\Gamma M_\mathrm{Pl} \zeta(5)}{3.32\pi m_S^2
  \sqrt{g_\ast}}\ T^3, 
\end{equation}
where $g_\ast$ is the number of statistical degrees of freedom during
the production epoch, $M_\mathrm{Pl}$ is the Planck mass, and
$\zeta(5)$ is the zeta function. One can also calculate the average momentum of the sterile
neutrinos, which is $\langle p\rangle = \pi^6/(378\zeta(5))T \approx 2.45 T$. In this model, the sterile neutrino dark matter  is colder than thermal $\langle p\rangle \approx 3.15 T$, but not
overwhelmingly colder \cite{Petraki:2007gq,Petraki:2008ef}. 
With very high-scale production mechanisms, like that in Ref.~\cite{Kusenko:2010ik}, the sterile neutrino dark matter can be very cold even for $\sim$keV-scale masses, and escape many structure formation constraints. The variety of sterile neutrino production models’ relation with structure formation constraints is discussed in Ref.~\cite{Abazajian:2019ejt}.

There have been explorations where the parent particle is a vector \cite{Shuve:2014doa} or fermion \cite{Abada:2014zra}. In all of these models, as well as in
oscillation-production cases, any further decays of particles, including related more massive sterile neutrinos, can further ``cool'' the sterile neutrino dark matter relative to the thermal bath
\cite{Asaka:2006ek,Petraki:2008ef,Boyanovsky:2008nc,Patwardhan:2015kga}. The sterile neutrinos usually cannot be  arbitrarily cooled, and can still be constrained by measures of its effects on structure formation
\cite{abazajian:2006yn,Menci:2017nsr}.

\subsection{Structure Formation Implications}

Since sterile neutrino dark matter is produced via thermal or quasi-thermal processes, and is relatively light, they behave on the spectrum of “warm” dark matter. The original Dodelson-Widrow case is ruled out as 100\% of the dark matter from Local Group galaxy counts plus X-ray constraints \cite{horiuchi:2013noa}, which I discuss below. It is important to keep in mind, however, that structure formation constraints are only beginning to constrain models with mixed cold plus warm dark matter (CWDM). Sterile neutrinos  could be a significant fraction of dark matter ($\sim$20\%) are only partially constrained \cite{Abazajian:2017tcc}. 

\begin{figure}[htbp]
\begin{center}
\includegraphics[width=5.8in]{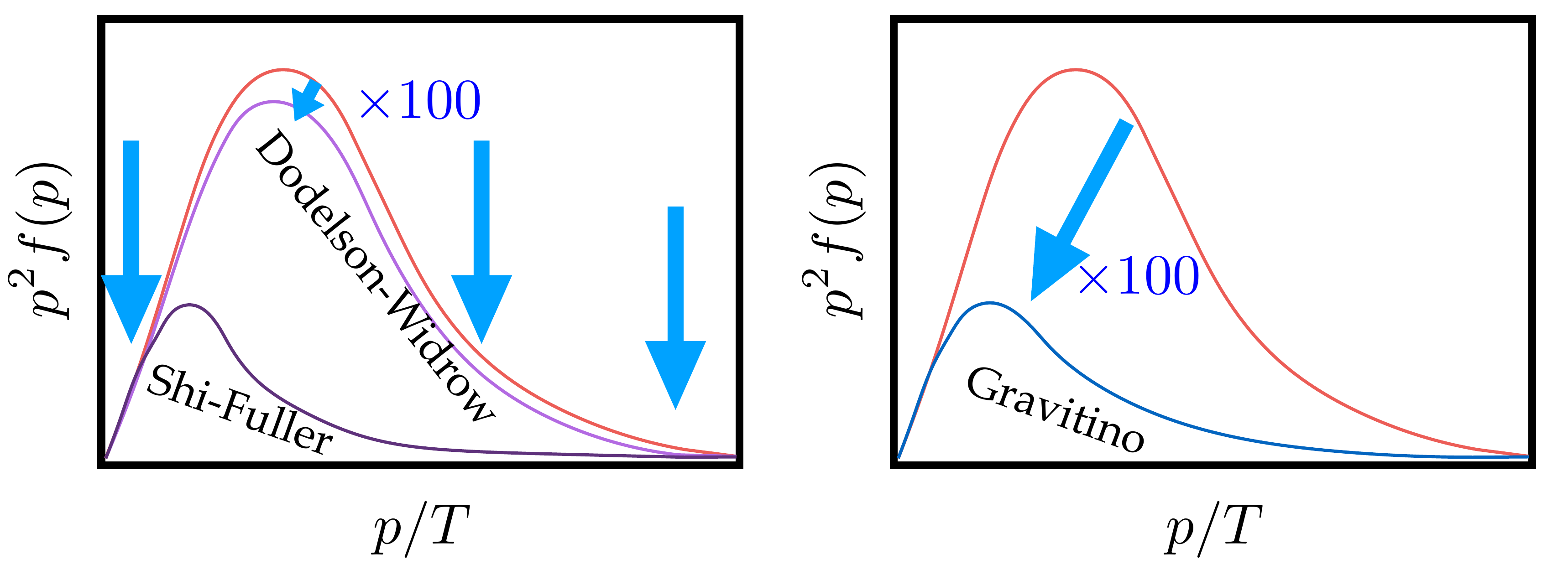}
\caption{Shown on the left are sketched distributions of Dodelson-Widrow (light purple) versus Shi-Fuller (purple) momentum distributions versus a thermal distribution (red), on arbitrary scales. On the right, I show a sketch of a “thermal” warm dark matter candidate’s distribution, like that of a gravitino, relative to thermal (red). The Dodelson-Widrow distribution reflects the thermal neutrino background, with a small distortion due to the intervening quark-hadron transition during or after the sterile neutrino’s production \cite{abazajian:2005xn}. The Shi-Fuller mechanism has resonant that moves from low to high momenta, which enhances production of low momenta modes relative to Dodelson-Widrow. The Gravitino is produces at very high-temperature scales ($T\gg 100\,\mathrm{GeV}$), and is suppressed and cooled to an abundance consistent with the observed dark matter abundance by the disappearance of degrees of freedom in the plasma from Standard Model and beyond Standard Model particles. The neutrino and gravitino amplitudes are multiplied by approximately a factor of $\sim$100, relative to the thermal distributions, to facilitate their display.  }
\label{distsketch}
\end{center}
\end{figure}

For the case of all sterile neutrinos being the dark matter, the upper scale of suppression of structure is the physical distance that the sterile neutrinos, or other warm dark matter, travels until matter-radiation equality, 
\begin{equation}
\lambda_\mathrm{FS} = \int_0^{t_\mathrm{EQ}} \frac{v(t)dt}{a(t)} \approx 1.2\,\mathrm{Mpc}\, \left(\frac{1\,\mathrm{keV}}{m_s}\right)\, \left(\frac{\langle p/T\rangle}{3.15}\right)\, .
\end{equation}
So, clearly the physical impact is dependent not only on particle mass, but its momentum distribution, reflected in $\langle p/T \rangle$. The momentum distribution is determined by how the dark matter particle is produced from the thermal bath. Recall that axions are very light, $\ll 1\, \mathrm{eV}$, but created in a zero-momentum mode, and therefore act as cold dark matter (for a brief yet up-to-date axion review, see section 91 of the Review of Particle Physics~\cite{10.1093/ptep/ptaa104}). For warm-to-cold-produced dark matter, the connection between particle mass depends on the production mechanism and what happens to the background plasma between the production and matter-radiation equality \cite{colombi:1995ze}. In Fig.~\ref{distsketch}, I show a sketch and description of the relative momentum distributions from the standard Dodelson-Widrow, Shi-Fuller, and a “thermal” warm dark matter gravitino momentum distributions. Sterile neutrinos can be much colder than that shown in Fig.~\ref{distsketch} if they are produced at high temperature, similar to gravitinos, or if there is a large entropy transfer by new or standard particles into the plasma after the sterile neutrino dark matter is produced \cite{Abazajian:2019ejt}.

Overall, the mapping between different particle masses, given the different particle production mechanism’s momenta distributions, are as follows. The thermal warm dark matter particle maps to a Dodelson-Widrow sterile neutrino as \cite{abazajian:2005xn}
\begin{equation}
m_s|_\text{Dodelson-Widrow} \approx 4.46\,\mathrm{keV}\, \left(\frac{m_\text{thermal}}{1\,\mathrm{keV}}\right)^{4/3}\,.
\label{thermaldw}
\end{equation}
The Shi-Fuller mechanism generally has a particle mass that is smaller than Dodelson-Widrow for a given fixed  free-streaming scale due to the Shi-Fuller mechanism’s colder momentum distribution, i.e., $m_s|_\text{Shi-Fuller} < m_s|_\text{Dodelson-Widrow}$, for the same free-streaming scale. There is not a single-function mapping because the lepton-number (or mixing angle) has to be specified for the Shi-Fuller mechanism as well as the particle mass. One interesting case is that the particle masses 
\begin{equation}
\boxed{m_\mathrm{thermal}\approx 2\,\mathrm{keV}}\, \Leftrightarrow\,\boxed{ m_s|_\text{Dodelson-Widrow} \approx 11\,\mathrm{keV}}\, \Leftrightarrow\,\boxed{ m_s|_\text{Shi-Fuller} \approx 7\,\mathrm{keV}}
\end{equation}
have approximately the same free-streaming scale when the Shi-Fuller neutrino has a mixing angle matching that of the putative 3.55 keV X-ray signal, described below \cite{abazajian:2014gza}. Note that the $m_\mathrm{thermal}\approx 2\,\mathrm{keV}$ scale was of interest for solving problems in galaxy formation \cite{boylankolchin:2011de,boylankolchin:2011dk,Bullock:2017xww} at dwarf galaxy scales \cite{lovell:2011rd,anderhalden:2012jc,Horiuchi:2015qri,Bozek:2015bdo}. For a review, see Ref.~\cite{Abazajian:2017tcc}. 

There are many ways to constrain the amount of small-scale structure present. Among the first were Local Group galaxy counts \cite{polisensky:2010rw,horiuchi:2013noa} and clustering in the the Lyman-$\alpha$ forest \cite{abazajian:2005xn,seljak:2006qw}.  Challenges in modeling the Lyman-$\alpha$ forest have led to arguments to use Local Group galaxy counts as the most robust constraints \cite{Cherry:2017dwu} and the latest Local Group constraints from the Dark Energy Survey find $m_\mathrm{thermal}> 6.5\,\mathrm{keV}$ (95\% CL) \cite{Nadler:2020prv}. The most stringent claimed constraints currently come from substructure inferred from imaging of strong lensing systems \cite{Gilman:2019nap,Nadler:2021dft}. The latest lensing limits are quite stringent at $m_\mathrm{thermal}> 9.7\,\mathrm{keV}$ (95\% CL), which corresponds to $m_s|_\text{Dodelson-Widrow} \gtrsim 92\,\mathrm{keV}$. The approximation in the Dodelson-Widrow sterile neutrino mass is due to the fact that the mapping relation, Eq.~\eqref{thermaldw}, has not been calibrated at these high particle masses. Note, however, that these lensing systems may be complicated by intervening structure along the line-of-sight that is not always modeled \cite{D_Aloisio_2010,Richardson:2021onm}. On another front, future constraints are thought to be robustly possible from analyzing the disruption stellar streams by dark matter subhalo structure \cite{Hermans:2020skz}. 

The thermal dark matter particle mass scales probed now are very significantly above those that could be responsible for alleviating problems in dwarf galaxy formation, $m_\mathrm{thermal}\approx 2\,\mathrm{keV}$. Studies of suppression of substructure are now of more interest in focussing on how much fractional suppression is allowed, since, as mentioned previously, constraints on mixed cold plus warm dark matter scenarios are only starting to be explored, and can also can alleviate problems in dwarf galaxy formation. Current constraints find that up to 20\% of the dark matter can be a warm dark matter particle with the free streaming scale of a 1 keV thermal warm dark matter particle \cite{anderhalden:2012jc}, and that fraction becomes commensurately higher as the particle mass increases, the particle distribution $\langle p/T\rangle$ decreases, or its free streaming scale gets smaller. In extended models, more significant new physics can occur. For example, Ref.~\cite{deGouvea:2019phk} explores a sterile neutrino dark matter model with self-interactions that could be explored in next-generation experiments like the Deep Underground Neutrino Experiment (DUNE) \cite{Acciarri:2015uup}.

\subsection{Observability in X-ray Astronomy}

A sterile neutrino with nonzero mixing with active neutrinos has been known to have a radiative decay mode \cite{Shrock:1974nd,Pal:1981rm}. This led to constraints from diffuse X-ray backgrounds being considered \cite{Drees:2000qr,dolgov:2000ew}. Detectability of the signals by collapsed dark matter halos was first estimated in Ref.~\cite{abazajian:2001nj}, and a detailed analysis of detectability signals from X-ray observations of clusters of galaxies, field galaxies, and the diffuse X-ray background was done in Ref.~\cite{abazajian:2001vt}. It was shown in Ref.~\cite{abazajian:2001vt} that clusters of galaxies were promising targets, followed by field galaxies. Ref.~\cite{abazajian:2006jc} showed that the diffuse emission on the sky would likely be dominated by the intervening dark matter halo of the Milky Way. Ref.~\cite{abazajian:2001vt} specifically proposed that if the proposed high effective area and high energy resolution X-ray microcalorimetry missions in 2001 were not available that stacked observations of clusters from the at-the-time newly flown missions of  \textit{Chandra} and \textit{XMM-Newton} could achieve the exposures needed for significant sensitivities. 

In 2014, 8 Ms of stacked observations of clusters by \textit{XMM-Newton} saw an unidentified line at $(3.55\!-\!3.57)\pm0.03\,\rm keV$ at $4\sigma\!-\!5\sigma$, as well as toward the Perseus cluster of galaxies with \textit{Chandra} at lower significance \cite{bulbul:2014sua}. Simultaneously, an independent team saw  an unidentified line toward toward the center of M31 at $3.53\pm 0.025\rm\, keV$ at 3$\sigma$, and an unidentified line toward the Perseus cluster at $3.50^{+0.044}_{-0.036}\rm\, keV$ at 2.3$\sigma$ with the MOS spectrometer aboard  \textit{XMM-Newton} \cite{Boyarsky:2014jta}. There were several significant followup observations, with positive indications of the line and  significant detections using other space telescopes’ data, including a significant line at the same energy in blank sky data with zero-bounce photons in the \textit{NuSTAR} observatory \cite{Neronov:2016wdd}. The flux in that observed line was consistent with the dark matter in the field of view, though an instrumental line with unknown strength is at the same position. 

The flux expected from dark matter decay goes as
\begin{equation}
   \label{eq:flux_DM}
    \frac{d \Phi_{\chi}}{d \Omega d E} \propto \frac{\mathcal{L}_{\chi}}{4\pi} \frac{dN_{\chi}}{dE}  \, ,
\end{equation}
where $\mathcal{L}_{\chi}$ is the dark matter “luminosity” (often quoted in literature also as $D-$ and $J-$factor~\cite{Bergstrom:1997fj} for dark matter decay and annihilation, respectively) while $dN_{\chi}/dE$ is the photon energy spectrum from the dark matters’ final states. The dark matter decay luminosity is the line-of-sight integral of its density along any given Galactic coordinate $(\ell,b)$,
\begin{equation} \label{eq:decay_L}
    \mathcal{L}^{\text{dec.}}(\ell,b) = \int_{\text{los}} d\vec{s} \rho_{\chi}[r(\vec{s}\,)] \, ,
\end{equation}
where $\rho_{\chi}$ is the dark matter density at distance $r(\vec{s}\,) = \sqrt{R^{2}_{\odot}+s^2+ 2 \, \vec{s} \cdot \vec{R}_{\odot}}$ from the Galactic Center, and $\vec{s}$ identifies the direction of the observer line-of-sight -- specified by the Galactic $(\ell,b)$. 

\begin{figure}[htbp]
\begin{center}
\includegraphics[width=6in]{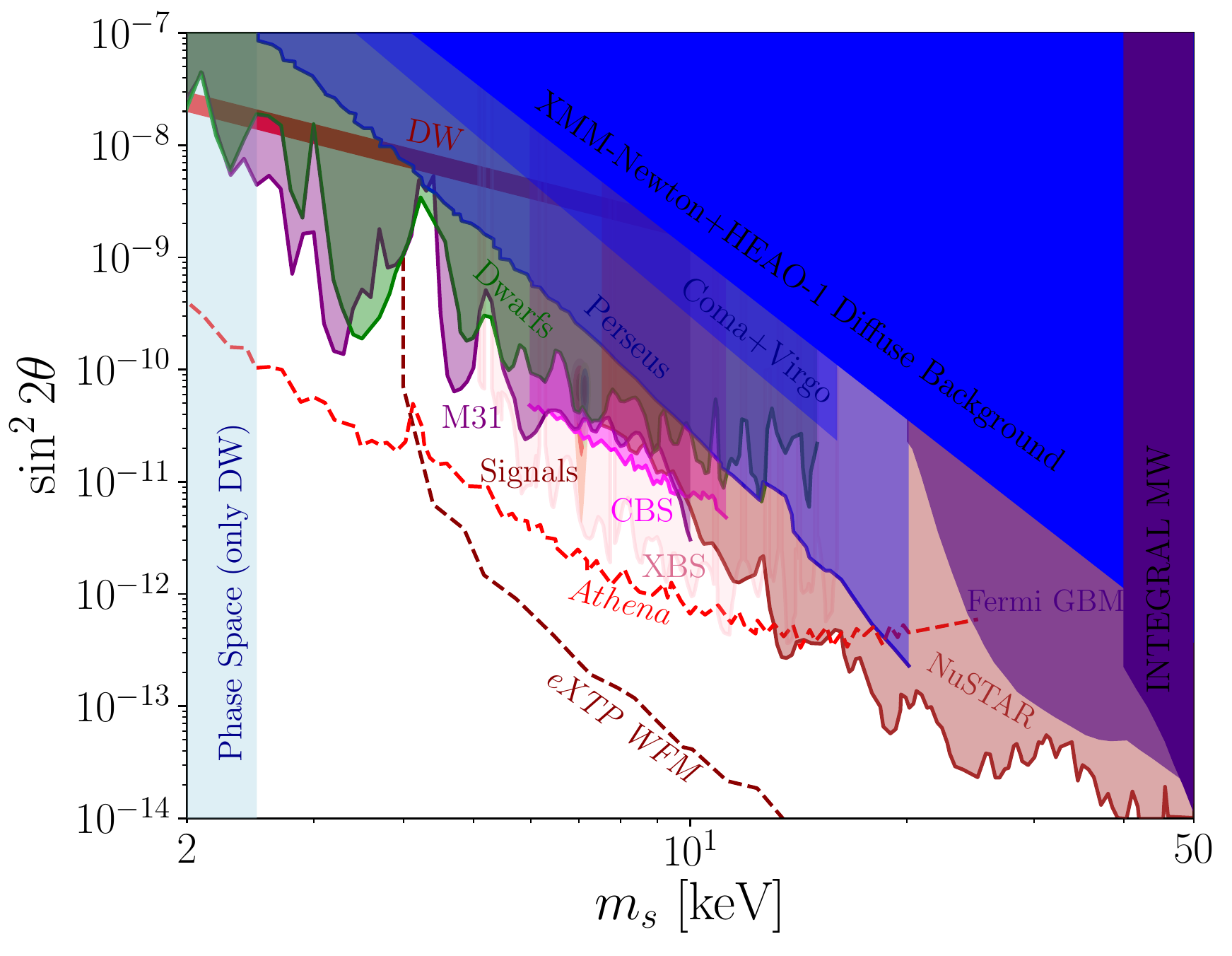}
\caption{The full parameter space for sterile neutrino dark matter,
  when it comprises all of the dark matter, is shown. Among the most
  stringent constraints at low energies and masses are constraints
  from X-ray observations M31 Horiuchi et al.~\cite{horiuchi:2013noa},
   stacked dwarfs \cite{malyshev:2014xqa}, 99\% upper limit from 51 Ms of blank sky data from \textit{Chandra} \cite{Sicilian:2020glg}, and the claimed blank sky 95\% limit from 547 Ms of \textit{XMM-Newton} data \cite{Foster:2021ngm}. Also shown are
  constraints from the diffuse X-ray background
  \cite{Boyarsky:2005us}, and individual clusters ``Coma+Virgo''
  \cite{Boyarsky:2006zi}. At higher masses and energies, I show the
  limits from NuSTAR \cite{Roach:2019ctw}, Fermi GBM \cite{Ng:2015gfa} and INTEGRAL
  \cite{Boyarsky:2007ge}. The signals near 3.55 keV from M31 and
  stacked clusters are also shown
  \cite{bulbul:2014sua,Boyarsky:2014jta}. The vertical mass constraint
  only directly applies to the Dodelson-Widrow model being all of the
  dark matter, labeled ``DW,'' which is now excluded as all of the
  dark matter. The Dodelson-Widrow model could still produce sterile
  neutrinos as a fraction of the dark matter. I also show forecast
  sensitivity of the planned {\it Athena X-ray Telescope}
  \cite{Neronov:2015kca}, and the potential optimistic-case reach of the WFM instrument aboard the {\it eXTP X-ray Telescope} \cite{Zhong:2020wre,Malyshev:2020hcc}.}
\label{fullparameters}
\end{center}
\end{figure}

\begin{figure}[htbp]
\begin{center}
\includegraphics[width=5.5in]{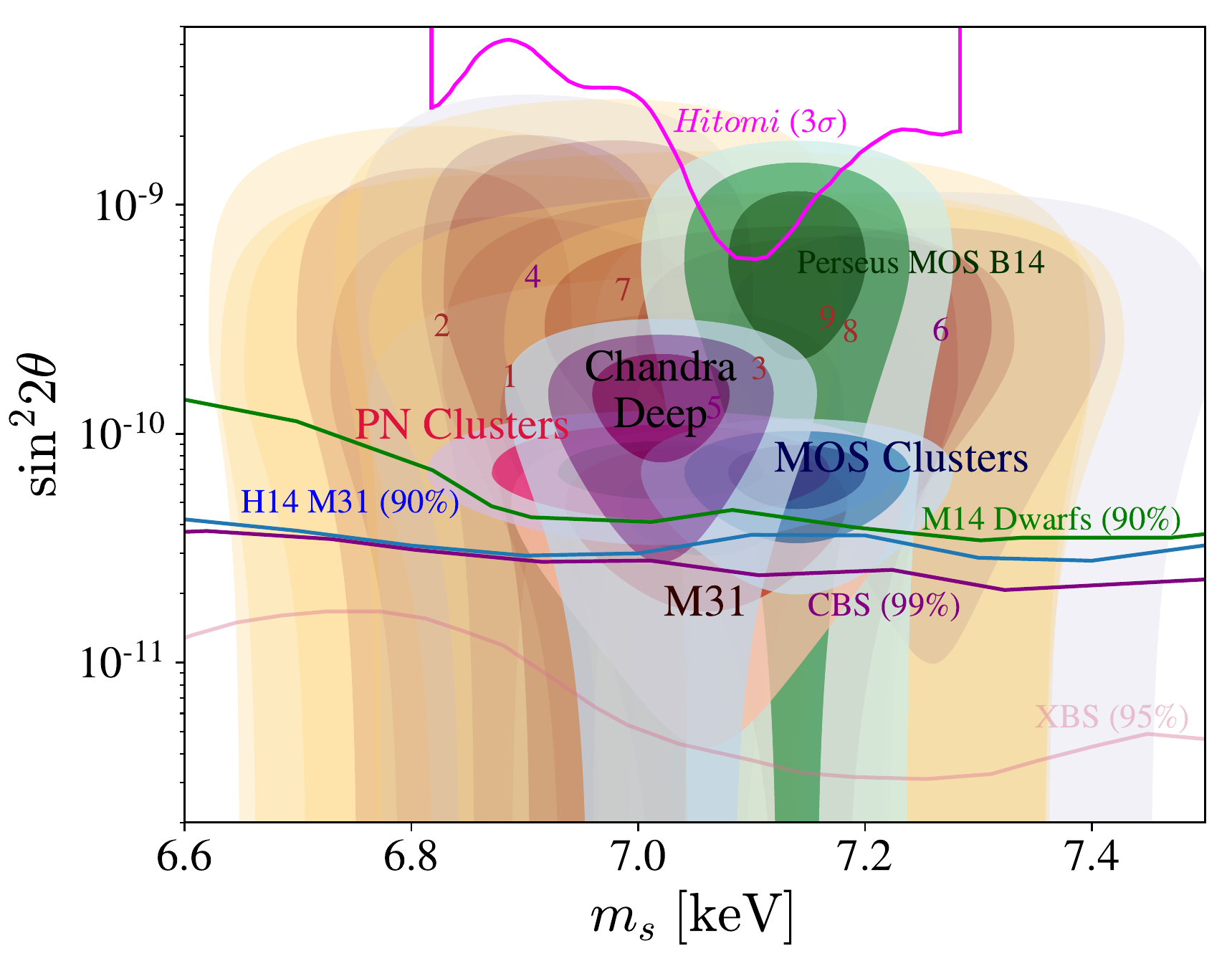}
\caption{X-ray line detections consistent with sterile neutrino dark
  matter are shown here. The dark colored regions are $1,2$ and $3$
  $\sigma$ from the MOS (blue) and PN (red) stacked clusters by Bulbul
  et al.~\cite{bulbul:2014sua}, the Bulbul et al. core-removed Perseus
  cluster (green), and M31 (orange) from Boyarsky et
  al.~\cite{Boyarsky:2014jta}. Also shown are the 1 and 2 $\sigma$
  regions of the detection in the Galactic Center (GC)
  \cite{Boyarsky:2014ska} as well as the $>\! 2\sigma$ line detections
  in 1.\ Abell 85; 2.\ Abell 2199; 3.\ Abell 496 (MOS); 4.\ Abell 496
  (PN); 5.\ Abell 3266; 6.\ Abell S805; 7.\ Coma; 8.\ Abell 2319;
  9.\ Perseus by Iakubovskyi et
  al.~\cite{Iakubovskyi:2015dna}. Numbers in the plot mark the
  centroid of the regions, with MOS detections in orange and PN in
  purple.  I also show, in purple, the region consistent with the
  signal in {\em Chandra} Deep Field observations, with errors given
  by the flux uncertainty, {\it i.e.}, not including dark matter
  profile uncertainties \cite{Cappelluti:2017ywp}. The lines show
  constraints at the $90\%$ level from {\it Chandra} observations of
  M31 (14) \cite{horiuchi:2013noa},  stacked dwarf galaxies (M14)
  \cite{malyshev:2014xqa}, the 99\% limit from 51 Ms of blank sky data from \textit{Chandra} \cite{Sicilian:2020glg}, and claimed blank sky 95\% limit from 547 Ms of \textit{XMM-Newton} data \cite{Foster:2021ngm}. }
\label{7kevparameters}
\end{center}
\end{figure}

A summary of X-ray detections and constraints through 2017 is available in Ref.~\cite{Abazajian:2017tcc}. Here, I highlight some new results. There were new constraints and a very weak signal in $\sim$1 Ms of  \textit{Chandra} X-ray observations of the Limiting Window   towards the Galactic bulge reported in Ref.~\cite{Hofmann:2019ihc}. A blank sky analysis using 30 Ms of \textit{XMM-Newton} observations in Ref.~\cite{Dessert:2018qih} claimed high sensitivity to the line and a lack of detection. However, Ref.~\cite{Dessert:2018qih} used a very narrow energy window, 0.5 keV, failed to include known instrumental and gas lines in the analysis, and adopted an excessively high local dark matter density. All of that led to their claimed stringent conclusions. Including the possible lines, proper dark matter density profile, and a broader-spectrum analysis, the claimed constraints of Ref.~\cite{Dessert:2018qih} are a factor of $\sim$20 weaker \cite{Abazajian:2020unr,Boyarsky:2020hqb}. The recently released analysis Ref.~\cite{Foster:2021ngm} starts with $\sim$18 times as much data (547 Ms), and methodology “built heavily off of” Ref.~\cite{Dessert:2018qih}, but uses a wider energy window, with incorporation of some astrophysical and instrumental line uncertainties, and a lower, dynamically motivated lower dark matter density. Despite having 18 times as much data, only $\sim$31 Ms qualifies for their signal region, and Ref.~\cite{Foster:2021ngm} finds weaker constraints than Ref.~\cite{Dessert:2018qih}. Importantly, Ref.~\cite{Foster:2021ngm} fails to include the 3.3 keV and 3.7 keV lines which have residuals just below their arbitrary threshold of 2$\sigma$ for taking into account astrophysical and instrumental lines. In addition, due to Ref.~\cite{Foster:2021ngm}’s choice of using Gaussian process modeling for the background, they cannot test the goodness-of-fit of their background model, which appears to have large unaccounted residuals. The work does perform a spurious-signal hyperparameter background mismodeling analysis. Given these issues,  the robustness of the limits from Ref.~\cite{Foster:2021ngm}---along with Ref.~\cite{Dessert:2018qih}---are questionable. Therefore, I leave the Ref.~\cite{Foster:2021ngm} claimed limit in a lighter color in Figs.~\ref{fullparameters} \& ~\ref{7kevparameters}.

A blank sky analysis of $\sim$51 Ms \textit{Chandra} blank-sky data, Ref.~\cite{Sicilian:2020glg} found no lines, and also provides one of the most stringent constraints. The \textit{NuSTAR X-ray Space Telescope} has sensitivity to dark matter lines from 3-25 keV, and searches for the line with that telescope have included observations of the Bullet Cluster \cite{Riemer-Sorensen:2015kqa}, blank-sky fields \cite{Neronov:2016wdd}, the Galactic center \cite{Perez:2016tcq}, the M31 galaxy \cite{Ng:2019gch}, and the Galactic Bulge \cite{Roach:2019ctw}. INTEGRAL
  \cite{Boyarsky:2007ge} and the \textit{Fermi Gamma-Ray Space Telescope’s} Gamma-Ray Burst monitor also provide constraints at high energy \cite{Ng:2015gfa}. All these signals and constraints are summarized in Fig.~\ref{fullparameters} and with a focus on the 7 keV region in Fig.~\ref{7kevparameters}.

Even in the initial work by Bulbul et al.~\cite{bulbul:2014sua}, it was shown that the lines could be due to charge exchange processes or metal lines, particularly K and Ar. However, for K, the abundance of K required would be more than an order of magnitude higher than that independently measured in X-ray clusters. For Ar, the lines that partner with one near 3.5 keV are independently constrained, and therefore Ar cannot contribute significantly near 3.5 keV. Related discussions on metal lines are in Refs~\cite{Jeltema:2014qfa,Bulbul:2014ala,Boyarsky:2014paa}.  In Ref.~\cite{Speckhard:2015eva} it was shown that dynamical differentiation between dark matter and metal lines from gas can be done by high energy resolution spectroscopy given the rotational motion of gas with the Milky Way’s disk and the expected quasi-isothermal velocity distribution of dark matter.  

In Ref.~\cite{Zhong:2020wre}, it was shown that this would likely require $\gtrsim 4\,\rm Ms$ of observations with the soon-to-be launched \textit{XRISM X-ray Space Telescope}, but with much more reasonable exposures of $\sim$10 ks to $\sim$100 ks with the \textit{ATHENA X-ray Space Telescope} \cite{Nandra:2013shg} or the \textit{Lynx Space Telescope} \cite{LynxTeam:2018usc}. In addition, Ref.~\cite{Zhong:2020wre} showed that the optics-free Wide Field Monitor (WFM) aboard the \textit{eXTP X-ray Space Telescope}  \cite{Zhang:2016ach}  could have a high sensitivity to dark matter decay due to its immense field-of-view, since the signal scales as $\sim$FOV and background flux uncertainty scales as $\sim\!\sqrt{\rm FOV}$. Depending on the nature of modeling the systematics of the WFM, which was not designed per-se as a dark matter detector, the WFM aboard \textit{eXTP} may be one of the most sensitive upcoming instruments in search for dark matter decay lines at the keV scale \cite{Malyshev:2020hcc}. See Fig.~\ref{fullparameters}. Clearly, with tens to hundreds of Ms of data from \textit{Chandra} and \textit{XMM Newton}, sensitivities to lines are not dramatically increasing and new technology like high-resolution and wide fields-of-view are needed. The sensitivity of the upcoming {\em Athena} X-ray Space Telescope is also promising (see Fig.~\ref{fullparameters}) \cite{Neronov:2015kca}. The {\em eROSITA}/{\em SRG} mission can also have sensitivity to the broader parameter space \cite{Zandanel:2015xca,Barinov:2020hiq}.

It is important to note that in low reheating-temperature Universe models \cite{Gelmini:2004ah,Gelmini:2019clw},  where the low reheating temperature and interaction rate allows for larger mixing angles and production of sterile neutrinos as a fraction of the total dark matter, the parameter space is very different from Fig.~\ref{fullparameters}. For example, see Fig.~5 in Ref.~\cite{Abazajian:2017tcc}. In such models, the fraction of dark matter can be very small but still produce X-ray signals but not have appreciable affects on structure formation, so that such a dark matter sterile neutrino  would be seen in X-ray observations or in the laboratory before they are constrained by structure formation.   Such production mechanisms occur at large mixing angles that can be probed in laboratory \cite{Smith:2016vku,Benso:2019jog}.

\section{Conclusions}

Overall, the topics touched upon by neutrinos in astrophysics and cosmology are very broad, including topics in the physics of the CMB, LSS, solar neutrinos, supernovae, dark matter production, small-scale structure formation, and even X-ray astronomy. With further specification of the properties of neutrino mass and mixing by current and upcoming experiments, discovery of any new nonstandard physics including new sterile states or nonstandard interactions, the bevy of places that neutrino properties affect astrophysics and cosmology will continue. I hope this set of lectures and this lecture write-up provide a resource for future enlightening investigations of neutrinos in astrophysics and cosmology. 

\acknowledgments
I would like to thank the discussions with students at the lectures, virtual coffee breaks, and via electronic messaging afterwards. Despite the unfortunate circumstances of a first-of-its-kind virtual TASI summer school, it was, I hope, a worthwhile program. I would also like to thank Kyle Cranmer, Kazunori Kohri, and Alex Kusenko for comments on an earlier draft. I also gratefully acknowledge support in part by NSF Theoretical Physics Program, Grant No. PHY-1915005.

\bibliographystyle{JHEP}
\bibliography{/Users/aba/Dropbox/master.bib}

\providecommand{\href}[2]{#2}\begingroup\raggedright\begin{thebibliography}{100}

\bibitem{Kajita1998}
{\scshape Super-Kamiokande} collaboration, T.~Kajita, \emph{Presentation at
  {XVIII}th {I}nternational {C}onference on {N}eutrino {P}hysics and
  {A}strophysics, {T}akayama, {J}apan,
  http://www-sk.icrr.u-tokyo.ac.jp/nu98/welcome\_nu98.html},  1998.

\bibitem{Venumadhav:2015pla}
T.~Venumadhav, F.-Y.~Cyr-Racine, K.N.~Abazajian and C.M.~Hirata, \emph{{Sterile
  neutrino dark matter: Weak interactions in the strong coupling epoch}},
  \href{https://doi.org/10.1103/PhysRevD.94.043515}{\emph{Phys. Rev.}
  {\bfseries D94} (2016) 043515}
  [\href{https://arxiv.org/abs/1507.06655}{{\ttfamily 1507.06655}}].

\bibitem{Wolfenstein:1977ue}
L.~Wolfenstein, \emph{{Neutrino Oscillations in Matter}},
  \href{https://doi.org/10.1103/PhysRevD.17.2369}{\emph{Phys. Rev.} {\bfseries
  D17} (1978) 2369}.

\bibitem{Mikheev:1986gs}
S.P.~Mikheev and A.Y.~Smirnov, \emph{{Resonance Amplification of Oscillations
  in Matter and Spectroscopy of Solar Neutrinos}}, {\emph{Sov. J. Nucl. Phys.}
  {\bfseries 42} (1985) 913}.

\bibitem{notzold:1987ik}
D.~Notzold and G.~Raffelt, \emph{Neutrino dispersion at finite temperature and
  density}, {\emph{Nucl. Phys.} {\bfseries B307} (1988) 924}.

\bibitem{abazajian:2002qx}
K.N.~Abazajian, J.F.~Beacom and N.F.~Bell, \emph{Stringent constraints on
  cosmological neutrino antineutrino asymmetries from synchronized flavor
  transformation}, {\emph{Phys. Rev.} {\bfseries D66} (2002) 013008}
  [\href{https://arxiv.org/abs/astro-ph/0203442}{{\ttfamily
  astro-ph/0203442}}].

\bibitem{wong:2002fa}
Y.Y.Y.~Wong, \emph{Analytical treatment of neutrino asymmetry equilibration
  from flavour oscillations in the early universe}, {\emph{Phys. Rev.}
  {\bfseries D66} (2002) 025015}
  [\href{https://arxiv.org/abs/hep-ph/0203180}{{\ttfamily hep-ph/0203180}}].

\bibitem{Robertson:2012ib}
W.C.~Haxton, R.G.~Hamish~Robertson and A.M.~Serenelli, \emph{{Solar Neutrinos:
  Status and Prospects}},
  \href{https://doi.org/10.1146/annurev-astro-081811-125539}{\emph{Ann. Rev.
  Astron. Astrophys.} {\bfseries 51} (2013) 21}
  [\href{https://arxiv.org/abs/1208.5723}{{\ttfamily 1208.5723}}].

\bibitem{dolgov:2002ab}
A.D.~Dolgov et~al., \emph{Cosmological bounds on neutrino degeneracy improved
  by flavor oscillations}, {\emph{Nucl. Phys.} {\bfseries B632} (2002) 363}
  [\href{https://arxiv.org/abs/hep-ph/0201287}{{\ttfamily hep-ph/0201287}}].

\bibitem{Steigman:1977kc}
G.~Steigman, D.N.~Schramm and J.E.~Gunn, \emph{{Cosmological Limits to the
  Number of Massive Leptons}},
  \href{https://doi.org/10.1016/0370-2693(77)90176-9}{\emph{Phys. Lett.}
  {\bfseries B66} (1977) 202}.

\bibitem{Langacker:1989sv}
P.~Langacker, \emph{{On the Cosmological Production of Light
  Sterile-Neutrinos}}, .

\bibitem{Enqvist:1991qj}
K.~Enqvist, K.~Kainulainen and M.J.~Thomson, \emph{{Stringent cosmological
  bounds on inert neutrino mixing}},
  \href{https://doi.org/10.1016/0550-3213(92)90442-E}{\emph{Nucl. Phys. B}
  {\bfseries 373} (1992) 498}.

\bibitem{abazajian:2002bj}
K.N.~Abazajian, \emph{Telling three from four neutrinos with cosmology},
  {\emph{Astropart. Phys.} {\bfseries 19} (2003) 303}
  [\href{https://arxiv.org/abs/astro-ph/0205238}{{\ttfamily
  astro-ph/0205238}}].

\bibitem{dibari:2001ua}
P.~Di~Bari, \emph{Update on neutrino mixing in the early universe},
  {\emph{Phys. Rev.} {\bfseries D65} (2002) 043509}
  [\href{https://arxiv.org/abs/hep-ph/0108182}{{\ttfamily hep-ph/0108182}}].

\bibitem{Cyburt:2015mya}
R.H.~Cyburt, B.D.~Fields, K.A.~Olive and T.-H.~Yeh, \emph{{Big Bang
  Nucleosynthesis: 2015}},
  \href{https://doi.org/10.1103/RevModPhys.88.015004}{\emph{Rev. Mod. Phys.}
  {\bfseries 88} (2016) 015004}
  [\href{https://arxiv.org/abs/1505.01076}{{\ttfamily 1505.01076}}].

\bibitem{foot:1995bm}
R.~Foot and R.R.~Volkas, \emph{Reconciling sterile neutrinos with big bang
  nucleosynthesis}, {\emph{Phys. Rev. Lett.} {\bfseries 75} (1995) 4350}
  [\href{https://arxiv.org/abs/hep-ph/9508275}{{\ttfamily hep-ph/9508275}}].

\bibitem{Bento:2001xi}
L.~Bento and Z.~Berezhiani, \emph{{Blocking active sterile neutrino
  oscillations in the early universe with a Majoron field}},
  \href{https://doi.org/10.1103/PhysRevD.64.115015}{\emph{Phys. Rev. D}
  {\bfseries 64} (2001) 115015}
  [\href{https://arxiv.org/abs/hep-ph/0108064}{{\ttfamily hep-ph/0108064}}].

\bibitem{Giudice:2000ex}
G.F.~Giudice, E.W.~Kolb and A.~Riotto, \emph{{Largest temperature of the
  radiation era and its cosmological implications}},
  \href{https://doi.org/10.1103/PhysRevD.64.023508}{\emph{Phys. Rev. D}
  {\bfseries 64} (2001) 023508}
  [\href{https://arxiv.org/abs/hep-ph/0005123}{{\ttfamily hep-ph/0005123}}].

\bibitem{Kawasaki:2000en}
M.~Kawasaki, K.~Kohri and N.~Sugiyama, \emph{{MeV scale reheating temperature
  and thermalization of neutrino background}},
  \href{https://doi.org/10.1103/PhysRevD.62.023506}{\emph{Phys. Rev. D}
  {\bfseries 62} (2000) 023506}
  [\href{https://arxiv.org/abs/astro-ph/0002127}{{\ttfamily
  astro-ph/0002127}}].

\bibitem{Hasegawa:2020ctq}
T.~Hasegawa, N.~Hiroshima, K.~Kohri, R.S.L.~Hansen, T.~Tram and S.~Hannestad,
  \emph{{MeV-scale reheating temperature and cosmological production of light
  sterile neutrinos}},
  \href{https://doi.org/10.1088/1475-7516/2020/08/015}{\emph{JCAP} {\bfseries
  08} (2020) 015} [\href{https://arxiv.org/abs/2003.13302}{{\ttfamily
  2003.13302}}].

\bibitem{Giovannini:2002qw}
M.~Giovannini, H.~Kurki-Suonio and E.~Sihvola, \emph{{Big bang nucleosynthesis,
  matter antimatter regions, extra relativistic species, and relic
  gravitational waves}},
  \href{https://doi.org/10.1103/PhysRevD.66.043504}{\emph{Phys. Rev. D}
  {\bfseries 66} (2002) 043504}
  [\href{https://arxiv.org/abs/astro-ph/0203430}{{\ttfamily
  astro-ph/0203430}}].

\bibitem{Chen:2000xxa}
X.-l.~Chen, R.J.~Scherrer and G.~Steigman, \emph{{Extended quintessence and the
  primordial helium abundance}},
  \href{https://doi.org/10.1103/PhysRevD.63.123504}{\emph{Phys. Rev. D}
  {\bfseries 63} (2001) 123504}
  [\href{https://arxiv.org/abs/astro-ph/0011531}{{\ttfamily
  astro-ph/0011531}}].

\bibitem{Murayama:2000hm}
H.~Murayama and T.~Yanagida, \emph{{LSND, SN1987A, and CPT violation}},
  \href{https://doi.org/10.1016/S0370-2693(01)01136-4}{\emph{Phys. Lett. B}
  {\bfseries 520} (2001) 263}
  [\href{https://arxiv.org/abs/hep-ph/0010178}{{\ttfamily hep-ph/0010178}}].

\bibitem{Barenboim:2001ac}
G.~Barenboim, L.~Borissov, J.D.~Lykken and A.Y.~Smirnov, \emph{{Neutrinos as
  the Messengers of CPT Violation}},
  \href{https://doi.org/10.1088/1126-6708/2002/10/001}{\emph{JHEP} {\bfseries
  10} (2002) 001} [\href{https://arxiv.org/abs/hep-ph/0108199}{{\ttfamily
  hep-ph/0108199}}].

\bibitem{Jacques:2013xr}
T.D.~Jacques, L.M.~Krauss and C.~Lunardini, \emph{{Additional Light Sterile
  Neutrinos and Cosmology}}, \href{https://doi.org/10.1103/PhysRevD.87.083515,
  10.1103/PhysRevD.88.109901}{\emph{Phys. Rev.} {\bfseries D87} (2013) 083515}
  [\href{https://arxiv.org/abs/1301.3119}{{\ttfamily 1301.3119}}].

\bibitem{Qian:1994wh}
Y.Z.~Qian and G.M.~Fuller, \emph{{Neutrino-neutrino scattering and matter
  enhanced neutrino flavor transformation in Supernovae}},
  \href{https://doi.org/10.1103/PhysRevD.51.1479}{\emph{Phys. Rev. D}
  {\bfseries 51} (1995) 1479}
  [\href{https://arxiv.org/abs/astro-ph/9406073}{{\ttfamily
  astro-ph/9406073}}].

\bibitem{Fuller:1993ry}
G.M.~Fuller, \emph{{Supernova dynamics and nucleosynthesis as a probe of
  closure mass neutrinos}},
  \href{https://doi.org/10.1016/0370-1573(93)90063-J}{\emph{Phys. Rept.}
  {\bfseries 227} (1993) 149}.

\bibitem{Pastor:2002we}
S.~Pastor and G.~Raffelt, \emph{{Flavor oscillations in the supernova hot
  bubble region: Nonlinear effects of neutrino background}},
  \href{https://doi.org/10.1103/PhysRevLett.89.191101}{\emph{Phys. Rev. Lett.}
  {\bfseries 89} (2002) 191101}
  [\href{https://arxiv.org/abs/astro-ph/0207281}{{\ttfamily
  astro-ph/0207281}}].

\bibitem{Hannestad:2006nj}
S.~Hannestad, G.G.~Raffelt, G.~Sigl and Y.Y.Y.~Wong, \emph{{Self-induced
  conversion in dense neutrino gases: Pendulum in flavour space}},
  \href{https://doi.org/10.1103/PhysRevD.74.105010}{\emph{Phys. Rev. D}
  {\bfseries 74} (2006) 105010}
  [\href{https://arxiv.org/abs/astro-ph/0608695}{{\ttfamily
  astro-ph/0608695}}].

\bibitem{Dasgupta:2007ws}
B.~Dasgupta and A.~Dighe, \emph{{Collective three-flavor oscillations of
  supernova neutrinos}},
  \href{https://doi.org/10.1103/PhysRevD.77.113002}{\emph{Phys. Rev. D}
  {\bfseries 77} (2008) 113002}
  [\href{https://arxiv.org/abs/0712.3798}{{\ttfamily 0712.3798}}].

\bibitem{Cherry:2011fn}
J.F.~Cherry, M.-R.~Wu, J.~Carlson, H.~Duan, G.M.~Fuller and Y.-Z.~Qian,
  \emph{{Neutrino Luminosity and Matter-Induced Modification of Collective
  Neutrino Flavor Oscillations in Supernovae}},
  \href{https://doi.org/10.1103/PhysRevD.85.125010}{\emph{Phys. Rev. D}
  {\bfseries 85} (2012) 125010}
  [\href{https://arxiv.org/abs/1109.5195}{{\ttfamily 1109.5195}}].

\bibitem{Johns:2020qsk}
L.~Johns, H.~Nagakura, G.M.~Fuller and A.~Burrows, \emph{{Fast oscillations,
  collisionless relaxation, and spurious evolution of supernova neutrino
  flavor}}, \href{https://doi.org/10.1103/PhysRevD.102.103017}{\emph{Phys. Rev.
  D} {\bfseries 102} (2020) 103017}
  [\href{https://arxiv.org/abs/2009.09024}{{\ttfamily 2009.09024}}].

\bibitem{Duan:2010bg}
H.~Duan, G.M.~Fuller and Y.-Z.~Qian, \emph{{Collective Neutrino Oscillations}},
  \href{https://doi.org/10.1146/annurev.nucl.012809.104524}{\emph{Ann. Rev.
  Nucl. Part. Sci.} {\bfseries 60} (2010) 569}
  [\href{https://arxiv.org/abs/1001.2799}{{\ttfamily 1001.2799}}].

\bibitem{Kohri:1996ke}
K.~Kohri, M.~Kawasaki and K.~Sato, \emph{{Big bang nucleosynthesis and lepton
  number asymmetry in the universe}},
  \href{https://doi.org/10.1086/512793}{\emph{Astrophys. J.} {\bfseries 490}
  (1997) 72} [\href{https://arxiv.org/abs/astro-ph/9612237}{{\ttfamily
  astro-ph/9612237}}].

\bibitem{Orito:2000zb}
M.~Orito, T.~Kajino, G.J.~Mathews and R.N.~Boyd, \emph{{Neutrino degeneracy and
  decoupling: New limits from primordial nucleosynthesis and the cosmic
  microwave background}},
  \href{https://arxiv.org/abs/astro-ph/0005446}{{\ttfamily astro-ph/0005446}}.

\bibitem{Orito:2002hf}
M.~Orito, T.~Kajino, G.J.~Mathews and Y.~Wang, \emph{{Constraints on neutrino
  degeneracy from the cosmic microwave background and primordial
  nucleosynthesis}},
  \href{https://doi.org/10.1103/PhysRevD.65.123504}{\emph{Phys. Rev. D}
  {\bfseries 65} (2002) 123504}
  [\href{https://arxiv.org/abs/astro-ph/0203352}{{\ttfamily
  astro-ph/0203352}}].

\bibitem{Bell:1998ds}
N.F.~Bell, R.R.~Volkas and Y.Y.Y.~Wong, \emph{Relic neutrino asymmetry
  evolution from first principles}, {\emph{Phys. Rev.} {\bfseries D59} (1999)
  113001} [\href{https://arxiv.org/abs/hep-ph/9809363}{{\ttfamily
  hep-ph/9809363}}].

\bibitem{Foot:1995qk}
R.~Foot, M.J.~Thomson and R.R.~Volkas, \emph{{Large neutrino asymmetries from
  neutrino oscillations}},
  \href{https://doi.org/10.1103/PhysRevD.53.R5349}{\emph{Phys. Rev. D}
  {\bfseries 53} (1996) 5349}
  [\href{https://arxiv.org/abs/hep-ph/9509327}{{\ttfamily hep-ph/9509327}}].

\bibitem{Shi:1996ic}
X.-D.~Shi, \emph{{Chaotic amplification of neutrino chemical potentials by
  neutrino oscillations in big bang nucleosynthesis}},
  \href{https://doi.org/10.1103/PhysRevD.54.2753}{\emph{Phys. Rev.} {\bfseries
  D54} (1996) 2753} [\href{https://arxiv.org/abs/astro-ph/9602135}{{\ttfamily
  astro-ph/9602135}}].

\bibitem{Foot:1996qc}
R.~Foot and R.R.~Volkas, \emph{{Studies of neutrino asymmetries generated by
  ordinary sterile neutrino oscillations in the early universe and implications
  for big bang nucleosynthesis bounds}},
  \href{https://doi.org/10.1103/PhysRevD.55.5147}{\emph{Phys. Rev. D}
  {\bfseries 55} (1997) 5147}
  [\href{https://arxiv.org/abs/hep-ph/9610229}{{\ttfamily hep-ph/9610229}}].

\bibitem{Abazajian:2008dz}
K.N.~Abazajian and P.~Agrawal, \emph{{Chaos, Determinacy and Fractals in
  Active-Sterile Neutrino Oscillations in the Early Universe}},
  \href{https://doi.org/10.1088/1475-7516/2008/10/006}{\emph{JCAP} {\bfseries
  0810} (2008) 006} [\href{https://arxiv.org/abs/0807.0456}{{\ttfamily
  0807.0456}}].

\bibitem{Bell:1998sr}
N.F.~Bell, R.~Foot and R.R.~Volkas, \emph{{Relic neutrino asymmetries and big
  bang nucleosynthesis in a four neutrino model}},
  \href{https://doi.org/10.1103/PhysRevD.58.105010}{\emph{Phys. Rev.}
  {\bfseries D58} (1998) 105010}
  [\href{https://arxiv.org/abs/hep-ph/9805259}{{\ttfamily hep-ph/9805259}}].

\bibitem{Shi:1999kg}
X.-D.~Shi, G.M.~Fuller and K.~Abazajian, \emph{{Neutrino mixing generated
  lepton asymmetry and the primordial He-4 abundance}},
  \href{https://doi.org/10.1103/PhysRevD.60.063002}{\emph{Phys. Rev. D}
  {\bfseries 60} (1999) 063002}
  [\href{https://arxiv.org/abs/astro-ph/9905259}{{\ttfamily
  astro-ph/9905259}}].

\bibitem{shi:1998km}
X.-d.~Shi and G.M.~Fuller, \emph{A new dark matter candidate: Non-thermal
  sterile neutrinos}, {\emph{Phys. Rev. Lett.} {\bfseries 82} (1999) 2832}
  [\href{https://arxiv.org/abs/astro-ph/9810076}{{\ttfamily
  astro-ph/9810076}}].

\bibitem{10.1093/ptep/ptaa104}
P.D.~Group, \emph{{Review of Particle Physics}},
  \href{https://doi.org/10.1093/ptep/ptaa104}{\emph{Progress of Theoretical and
  Experimental Physics} {\bfseries 2020} (2020) }
  [\href{https://arxiv.org/abs/https://academic.oup.com/ptep/article-pdf/2020/8/083C01/34673722/ptaa104.pdf}{{\ttfamily
  https://academic.oup.com/ptep/article-pdf/2020/8/083C01/34673722/ptaa104.pdf}}].

\bibitem{lesgourgues:2006nd}
J.~Lesgourgues and S.~Pastor, \emph{{Massive neutrinos and cosmology}},
  \href{https://doi.org/10.1016/j.physrep.2006.04.001}{\emph{Phys. Rept.}
  {\bfseries 429} (2006) 307}
  [\href{https://arxiv.org/abs/astro-ph/0603494}{{\ttfamily
  astro-ph/0603494}}].

\bibitem{Abazajian:2016hbv}
K.N.~Abazajian and M.~Kaplinghat, \emph{{Neutrino Physics from the Cosmic
  Microwave Background and Large-Scale Structure}},
  \href{https://doi.org/10.1146/annurev-nucl-102014-021908}{\emph{Ann. Rev.
  Nucl. Part. Sci.} {\bfseries 66} (2016) 401}.

\bibitem{lopez:1998vk}
R.E.~Lopez and M.S.~Turner, \emph{An accurate calculation of the big-bang
  prediction for the abundance of primordial helium}, {\emph{Phys. Rev.}
  {\bfseries D59} (1999) 103502}
  [\href{https://arxiv.org/abs/astro-ph/9807279}{{\ttfamily
  astro-ph/9807279}}].

\bibitem{mangano:2005cc}
G.~Mangano, G.~Miele, S.~Pastor, T.~Pinto, O.~Pisanti and P.D.~Serpico,
  \emph{{Relic neutrino decoupling including flavor oscillations}},
  \href{https://doi.org/10.1016/j.nuclphysb.2005.09.041}{\emph{Nucl. Phys.}
  {\bfseries B729} (2005) 221}
  [\href{https://arxiv.org/abs/hep-ph/0506164}{{\ttfamily hep-ph/0506164}}].

\bibitem{Grohs:2015tfy}
E.~Grohs, G.M.~Fuller, C.T.~Kishimoto, M.W.~Paris and A.~Vlasenko,
  \emph{{Neutrino energy transport in weak decoupling and big bang
  nucleosynthesis}},
  \href{https://doi.org/10.1103/PhysRevD.93.083522}{\emph{Phys. Rev. D}
  {\bfseries 93} (2016) 083522}
  [\href{https://arxiv.org/abs/1512.02205}{{\ttfamily 1512.02205}}].

\bibitem{tegmark:2003uf}
{\scshape SDSS} collaboration, \emph{The 3d power spectrum of galaxies from the
  sdss}, {\emph{Astrophys. J.} {\bfseries 606} (2004) 702}
  [\href{https://arxiv.org/abs/astro-ph/0310725}{{\ttfamily
  astro-ph/0310725}}].

\bibitem{Croft:2000hs}
R.A.C.~Croft, D.H.~Weinberg, M.~Bolte, S.~Burles, L.~Hernquist, N.~Katz et~al.,
  \emph{{Towards a precise measurement of matter clustering: Lyman alpha forest
  data at redshifts 2-4}},
  \href{https://doi.org/10.1086/344099}{\emph{Astrophys. J.} {\bfseries 581}
  (2002) 20} [\href{https://arxiv.org/abs/astro-ph/0012324}{{\ttfamily
  astro-ph/0012324}}].

\bibitem{Eisenstein:1997jh}
D.J.~Eisenstein and W.~Hu, \emph{{Power spectra for cold dark matter and its
  variants}}, \href{https://doi.org/10.1086/306640}{\emph{Astrophys. J.}
  {\bfseries 511} (1997) 5}
  [\href{https://arxiv.org/abs/astro-ph/9710252}{{\ttfamily
  astro-ph/9710252}}].

\bibitem{Hu:1997mj}
W.~Hu, D.J.~Eisenstein and M.~Tegmark, \emph{{Weighing neutrinos with galaxy
  surveys}}, \href{https://doi.org/10.1103/PhysRevLett.80.5255}{\emph{Phys.
  Rev. Lett.} {\bfseries 80} (1998) 5255}
  [\href{https://arxiv.org/abs/astro-ph/9712057}{{\ttfamily
  astro-ph/9712057}}].

\bibitem{Eisenstein:1998hr}
D.J.~Eisenstein, W.~Hu and M.~Tegmark, \emph{{Cosmic complementarity: Joint
  parameter estimation from CMB experiments and redshift surveys}},
  \href{https://doi.org/10.1086/307261}{\emph{Astrophys. J.} {\bfseries 518}
  (1999) 2} [\href{https://arxiv.org/abs/astro-ph/9807130}{{\ttfamily
  astro-ph/9807130}}].

\bibitem{croft:1999mm}
R.A.C.~Croft, W.~Hu and R.~Dave, \emph{{Cosmological Limits on the Neutrino
  Mass from the Lya Forest}},
  \href{https://doi.org/10.1103/PhysRevLett.83.1092}{\emph{Phys. Rev. Lett.}
  {\bfseries 83} (1999) 1092}
  [\href{https://arxiv.org/abs/astro-ph/9903335}{{\ttfamily
  astro-ph/9903335}}].

\bibitem{Tegmark:1999ke}
M.~Tegmark, D.J.~Eisenstein, W.~Hu and A.~de~Oliveira-Costa, \emph{{Foregrounds
  and forecasts for the cosmic microwave background}},
  \href{https://doi.org/10.1086/308348}{\emph{Astrophys. J.} {\bfseries 530}
  (2000) 133} [\href{https://arxiv.org/abs/astro-ph/9905257}{{\ttfamily
  astro-ph/9905257}}].

\bibitem{Cuesta:2015iho}
A.J.~Cuesta, V.~Niro and L.~Verde, \emph{{Neutrino mass limits: robust
  information from the power spectrum of galaxy surveys}},
  \href{https://doi.org/10.1016/j.dark.2016.04.005}{\emph{Phys. Dark Univ.}
  {\bfseries 13} (2016) 77} [\href{https://arxiv.org/abs/1511.05983}{{\ttfamily
  1511.05983}}].

\bibitem{Canac:2016smv}
N.~Canac, G.~Aslanyan, K.N.~Abazajian, R.~Easther and L.C.~Price,
  \emph{{Testing for New Physics: Neutrinos and the Primordial Power
  Spectrum}}, \href{https://doi.org/10.1088/1475-7516/2016/09/022}{\emph{JCAP}
  {\bfseries 1609} (2016) 022}
  [\href{https://arxiv.org/abs/1606.03057}{{\ttfamily 1606.03057}}].

\bibitem{Dodelson:1995es}
S.~Dodelson, E.~Gates and A.~Stebbins, \emph{{Cold + hot dark matter and the
  cosmic microwave background}},
  \href{https://doi.org/10.1086/177581}{\emph{Astrophys. J.} {\bfseries 467}
  (1996) 10} [\href{https://arxiv.org/abs/astro-ph/9509147}{{\ttfamily
  astro-ph/9509147}}].

\bibitem{Aghanim:2018eyx}
{\scshape Planck} collaboration, \emph{{Planck 2018 results. VI. Cosmological
  parameters}},
  \href{https://doi.org/10.1051/0004-6361/201833910}{\emph{Astron. Astrophys.}
  {\bfseries 641} (2020) A6}
  [\href{https://arxiv.org/abs/1807.06209}{{\ttfamily 1807.06209}}].

\bibitem{Kaplinghat:2003bh}
M.~Kaplinghat, L.~Knox and Y.-S.~Song, \emph{{Determining neutrino mass from
  the CMB alone}},
  \href{https://doi.org/10.1103/PhysRevLett.91.241301}{\emph{Phys. Rev. Lett.}
  {\bfseries 91} (2003) 241301}
  [\href{https://arxiv.org/abs/astro-ph/0303344}{{\ttfamily
  astro-ph/0303344}}].

\bibitem{Reid:2009xm}
B.A.~Reid et~al., \emph{{Cosmological Constraints from the Clustering of the
  Sloan Digital Sky Survey DR7 Luminous Red Galaxies}},
  \href{https://doi.org/10.1111/j.1365-2966.2010.16276.x}{\emph{Mon. Not. Roy.
  Astron. Soc.} {\bfseries 404} (2010) 60}
  [\href{https://arxiv.org/abs/0907.1659}{{\ttfamily 0907.1659}}].

\bibitem{Ade:2015xua}
{\scshape Planck} collaboration, \emph{{Planck 2015 results. XIII. Cosmological
  parameters}},  \href{https://arxiv.org/abs/1502.01589}{{\ttfamily
  1502.01589}}.

\bibitem{Addison:2015wyg}
G.E.~Addison, Y.~Huang, D.J.~Watts, C.L.~Bennett, M.~Halpern, G.~Hinshaw
  et~al., \emph{{Quantifying discordance in the 2015 Planck CMB spectrum}},
  \href{https://doi.org/10.3847/0004-637X/818/2/132}{\emph{Astrophys. J.}
  {\bfseries 818} (2016) 132}
  [\href{https://arxiv.org/abs/1511.00055}{{\ttfamily 1511.00055}}].

\bibitem{Riess:2016jrr}
A.G.~Riess et~al., \emph{{A 2.4\% Determination of the Local Value of the
  Hubble Constant}},
  \href{https://doi.org/10.3847/0004-637X/826/1/56}{\emph{Astrophys. J.}
  {\bfseries 826} (2016) 56}
  [\href{https://arxiv.org/abs/1604.01424}{{\ttfamily 1604.01424}}].

\bibitem{Riess:2018uxu}
A.G.~Riess et~al., \emph{{New Parallaxes of Galactic Cepheids from Spatially
  Scanning the Hubble Space Telescope: Implications for the Hubble Constant}},
  \href{https://doi.org/10.3847/1538-4357/aaadb7}{\emph{Astrophys. J.}
  {\bfseries 855} (2018) 136}
  [\href{https://arxiv.org/abs/1801.01120}{{\ttfamily 1801.01120}}].

\bibitem{Riess:2020fzl}
A.G.~Riess, S.~Casertano, W.~Yuan, J.B.~Bowers, L.~Macri, J.C.~Zinn et~al.,
  \emph{{Cosmic Distances Calibrated to 1\% Precision with Gaia EDR3 Parallaxes
  and Hubble Space Telescope Photometry of 75 Milky Way Cepheids Confirm
  Tension with LambdaCDM}},  \href{https://arxiv.org/abs/2012.08534}{{\ttfamily
  2012.08534}}.

\bibitem{DiValentino:2016hlg}
E.~Di~Valentino, A.~Melchiorri and J.~Silk, \emph{{Reconciling Planck with the
  local value of $H_0$ in extended parameter space}},
  \href{https://doi.org/10.1016/j.physletb.2016.08.043}{\emph{Phys. Lett.}
  {\bfseries B761} (2016) 242}
  [\href{https://arxiv.org/abs/1606.00634}{{\ttfamily 1606.00634}}].

\bibitem{Poulin:2018cxd}
V.~Poulin, T.L.~Smith, T.~Karwal and M.~Kamionkowski, \emph{{Early Dark Energy
  Can Resolve The Hubble Tension}},
  \href{https://doi.org/10.1103/PhysRevLett.122.221301}{\emph{Phys. Rev. Lett.}
  {\bfseries 122} (2019) 221301}
  [\href{https://arxiv.org/abs/1811.04083}{{\ttfamily 1811.04083}}].

\bibitem{Keeley:2019esp}
R.E.~Keeley, S.~Joudaki, M.~Kaplinghat and D.~Kirkby, \emph{{Implications of a
  transition in the dark energy equation of state for the $H_0$ and $\sigma_8$
  tensions}}, \href{https://doi.org/10.1088/1475-7516/2019/12/035}{\emph{JCAP}
  {\bfseries 12} (2019) 035}
  [\href{https://arxiv.org/abs/1905.10198}{{\ttfamily 1905.10198}}].

\bibitem{Battye:2013xqa}
R.A.~Battye and A.~Moss, \emph{{Evidence for Massive Neutrinos from Cosmic
  Microwave Background and Lensing Observations}},
  \href{https://doi.org/10.1103/PhysRevLett.112.051303}{\emph{Phys. Rev. Lett.}
  {\bfseries 112} (2014) 051303}
  [\href{https://arxiv.org/abs/1308.5870}{{\ttfamily 1308.5870}}].

\bibitem{Wyman:2013lza}
M.~Wyman, D.H.~Rudd, R.A.~Vanderveld and W.~Hu, \emph{{Neutrinos Help Reconcile
  Planck Measurements with the Local Universe}},
  \href{https://doi.org/10.1103/PhysRevLett.112.051302}{\emph{Phys. Rev. Lett.}
  {\bfseries 112} (2014) 051302}
  [\href{https://arxiv.org/abs/1307.7715}{{\ttfamily 1307.7715}}].

\bibitem{Dvorkin:2014lea}
C.~Dvorkin, M.~Wyman, D.H.~Rudd and W.~Hu, \emph{{Neutrinos help reconcile
  Planck measurements with both the early and local Universe}},
  \href{https://doi.org/10.1103/PhysRevD.90.083503}{\emph{Phys. Rev.}
  {\bfseries D90} (2014) 083503}
  [\href{https://arxiv.org/abs/1403.8049}{{\ttfamily 1403.8049}}].

\bibitem{Beutler:2014yhv}
{\scshape BOSS} collaboration, \emph{{The clustering of galaxies in the
  SDSS-III Baryon Oscillation Spectroscopic Survey: Signs of neutrino mass in
  current cosmological datasets}},
  \href{https://doi.org/10.1093/mnras/stu1702}{\emph{Mon. Not. Roy. Astron.
  Soc.} {\bfseries 444} (2014) 3501}
  [\href{https://arxiv.org/abs/1403.4599}{{\ttfamily 1403.4599}}].

\bibitem{Abazajian:2019ejt}
K.N.~Abazajian and A.~Kusenko, \emph{{Hidden treasures: Sterile neutrinos as
  dark matter with miraculous abundance, structure formation for different
  production mechanisms, and a solution to the $\sigma_8$ problem}},
  \href{https://doi.org/10.1103/PhysRevD.100.103513}{\emph{Phys. Rev. D}
  {\bfseries 100} (2019) 103513}
  [\href{https://arxiv.org/abs/1907.11696}{{\ttfamily 1907.11696}}].

\bibitem{dodelson:1993je}
S.~Dodelson and L.M.~Widrow, \emph{Sterile-neutrinos as dark matter},
  {\emph{Phys. Rev. Lett.} {\bfseries 72} (1994) 17}
  [\href{https://arxiv.org/abs/hep-ph/9303287}{{\ttfamily hep-ph/9303287}}].

\bibitem{Barbieri:1989ti}
R.~Barbieri and A.~Dolgov, \emph{{Bounds on Sterile-neutrinos from
  Nucleosynthesis}},
  \href{https://doi.org/10.1016/0370-2693(90)91203-N}{\emph{Phys. Lett.}
  {\bfseries B237} (1990) 440}.

\bibitem{abazajian:2002yz}
K.N.~Abazajian and G.M.~Fuller, \emph{Bulk qcd thermodynamics and sterile
  neutrino dark matter}, {\emph{Phys. Rev.} {\bfseries D66} (2002) 023526}
  [\href{https://arxiv.org/abs/astro-ph/0204293}{{\ttfamily
  astro-ph/0204293}}].

\bibitem{kusenko:1998bk}
A.~Kusenko and G.~Segre, \emph{Pulsar kicks from neutrino oscillations},
  {\emph{Phys. Rev.} {\bfseries D59} (1999) 061302}
  [\href{https://arxiv.org/abs/astro-ph/9811144}{{\ttfamily
  astro-ph/9811144}}].

\bibitem{fuller:2003gy}
G.M.~Fuller, A.~Kusenko, I.~Mocioiu and S.~Pascoli, \emph{Pulsar kicks from a
  dark-matter sterile neutrino}, {\emph{Phys. Rev.} {\bfseries D68} (2003)
  103002} [\href{https://arxiv.org/abs/astro-ph/0307267}{{\ttfamily
  astro-ph/0307267}}].

\bibitem{Gelmini:2004ah}
G.~Gelmini, S.~Palomares-Ruiz and S.~Pascoli, \emph{{Low reheating temperature
  and the visible sterile neutrino}},
  \href{https://doi.org/10.1103/PhysRevLett.93.081302}{\emph{Phys. Rev. Lett.}
  {\bfseries 93} (2004) 081302}
  [\href{https://arxiv.org/abs/astro-ph/0403323}{{\ttfamily
  astro-ph/0403323}}].

\bibitem{Gelmini:2019clw}
G.B.~Gelmini, P.~Lu and V.~Takhistov, \emph{{Cosmological Dependence of
  Resonantly Produced Sterile Neutrinos}},
  \href{https://doi.org/10.1088/1475-7516/2020/06/008}{\emph{JCAP} {\bfseries
  06} (2020) 008} [\href{https://arxiv.org/abs/1911.03398}{{\ttfamily
  1911.03398}}].

\bibitem{Smith:2016vku}
P.F.~Smith, \emph{{Proposed experiments to detect keV range sterile neutrinos
  using energy-momentum reconstruction of beta decay or K-capture events}},
  \href{https://doi.org/10.1088/1367-2630/ab1502}{\emph{New J. Phys.}
  {\bfseries 21} (2019) 053022}
  [\href{https://arxiv.org/abs/1607.06876}{{\ttfamily 1607.06876}}].

\bibitem{Benso:2019jog}
C.~Benso, V.~Brdar, M.~Lindner and W.~Rodejohann, \emph{{Prospects for Finding
  Sterile Neutrino Dark Matter at KATRIN}},
  \href{https://doi.org/10.1103/PhysRevD.100.115035}{\emph{Phys. Rev. D}
  {\bfseries 100} (2019) 115035}
  [\href{https://arxiv.org/abs/1911.00328}{{\ttfamily 1911.00328}}].

\bibitem{Adhikari:2016bei}
R.~Adhikari et~al., \emph{{A White Paper on keV Sterile Neutrino Dark Matter}},
  \href{https://doi.org/10.1088/1475-7516/2017/01/025}{\emph{JCAP} {\bfseries
  1701} (2017) 025} [\href{https://arxiv.org/abs/1602.04816}{{\ttfamily
  1602.04816}}].

\bibitem{shaposhnikov:2006xi}
M.~Shaposhnikov and I.~Tkachev, \emph{The numsm, inflation, and dark matter},
  {\emph{Phys. Lett.} {\bfseries B639} (2006) 414}
  [\href{https://arxiv.org/abs/hep-ph/0604236}{{\ttfamily hep-ph/0604236}}].

\bibitem{Kusenko:2006rh}
A.~Kusenko, \emph{{Sterile neutrinos, dark matter, and the pulsar velocities in
  models with a Higgs singlet}},
  \href{https://doi.org/10.1103/PhysRevLett.97.241301}{\emph{Phys. Rev. Lett.}
  {\bfseries 97} (2006) 241301}
  [\href{https://arxiv.org/abs/hep-ph/0609081}{{\ttfamily hep-ph/0609081}}].

\bibitem{Petraki:2007gq}
K.~Petraki and A.~Kusenko, \emph{{Dark-matter sterile neutrinos in models with
  a gauge singlet in the Higgs sector}},
  \href{https://doi.org/10.1103/PhysRevD.77.065014}{\emph{Phys. Rev.}
  {\bfseries D77} (2008) 065014}
  [\href{https://arxiv.org/abs/0711.4646}{{\ttfamily 0711.4646}}].

\bibitem{Bezrukov:2009yw}
F.~Bezrukov and D.~Gorbunov, \emph{{Light inflaton Hunter's Guide}},
  \href{https://doi.org/10.1007/JHEP05(2010)010}{\emph{JHEP} {\bfseries 05}
  (2010) 010} [\href{https://arxiv.org/abs/0912.0390}{{\ttfamily 0912.0390}}].

\bibitem{Kusenko:2012ch}
A.~Kusenko, M.~Loewenstein and T.T.~Yanagida, \emph{{Moduli dark matter and the
  search for its decay line using Suzaku X-ray telescope}},
  \href{https://doi.org/10.1103/PhysRevD.87.043508}{\emph{Phys. Rev.}
  {\bfseries D87} (2013) 043508}
  [\href{https://arxiv.org/abs/1209.6403}{{\ttfamily 1209.6403}}].

\bibitem{Merle:2015oja}
A.~Merle and M.~Totzauer, \emph{{keV Sterile Neutrino Dark Matter from Singlet
  Scalar Decays: Basic Concepts and Subtle Features}},
  \href{https://doi.org/10.1088/1475-7516/2015/06/011}{\emph{JCAP} {\bfseries
  1506} (2015) 011} [\href{https://arxiv.org/abs/1502.01011}{{\ttfamily
  1502.01011}}].

\bibitem{Petraki:2008ef}
K.~Petraki, \emph{{Small-scale structure formation properties of chilled
  sterile neutrinos as dark matter}},
  \href{https://doi.org/10.1103/PhysRevD.77.105004}{\emph{Phys. Rev.}
  {\bfseries D77} (2008) 105004}
  [\href{https://arxiv.org/abs/0801.3470}{{\ttfamily 0801.3470}}].

\bibitem{Kusenko:2010ik}
A.~Kusenko, F.~Takahashi and T.T.~Yanagida, \emph{{Dark Matter from Split
  Seesaw}}, \href{https://doi.org/10.1016/j.physletb.2010.08.031}{\emph{Phys.
  Lett.} {\bfseries B693} (2010) 144}
  [\href{https://arxiv.org/abs/1006.1731}{{\ttfamily 1006.1731}}].

\bibitem{Shuve:2014doa}
B.~Shuve and I.~Yavin, \emph{{Dark matter progenitor: Light vector boson decay
  into sterile neutrinos}},
  \href{https://doi.org/10.1103/PhysRevD.89.113004}{\emph{Phys. Rev.}
  {\bfseries D89} (2014) 113004}
  [\href{https://arxiv.org/abs/1403.2727}{{\ttfamily 1403.2727}}].

\bibitem{Abada:2014zra}
A.~Abada, G.~Arcadi and M.~Lucente, \emph{{Dark Matter in the minimal Inverse
  Seesaw mechanism}},
  \href{https://doi.org/10.1088/1475-7516/2014/10/001}{\emph{JCAP} {\bfseries
  1410} (2014) 001} [\href{https://arxiv.org/abs/1406.6556}{{\ttfamily
  1406.6556}}].

\bibitem{Asaka:2006ek}
T.~Asaka, M.~Shaposhnikov and A.~Kusenko, \emph{{Opening a new window for warm
  dark matter}},
  \href{https://doi.org/10.1016/j.physletb.2006.05.067}{\emph{Phys. Lett.}
  {\bfseries B638} (2006) 401}
  [\href{https://arxiv.org/abs/hep-ph/0602150}{{\ttfamily hep-ph/0602150}}].

\bibitem{Boyanovsky:2008nc}
D.~Boyanovsky, \emph{{Clustering properties of a sterile neutrino dark matter
  candidate}}, \href{https://doi.org/10.1103/PhysRevD.78.103505}{\emph{Phys.
  Rev.} {\bfseries D78} (2008) 103505}
  [\href{https://arxiv.org/abs/0807.0646}{{\ttfamily 0807.0646}}].

\bibitem{Patwardhan:2015kga}
A.V.~Patwardhan, G.M.~Fuller, C.T.~Kishimoto and A.~Kusenko, \emph{{Diluted
  equilibrium sterile neutrino dark matter}},
  \href{https://doi.org/10.1103/PhysRevD.92.103509}{\emph{Phys. Rev.}
  {\bfseries D92} (2015) 103509}
  [\href{https://arxiv.org/abs/1507.01977}{{\ttfamily 1507.01977}}].

\bibitem{abazajian:2006yn}
K.~Abazajian and S.M.~Koushiappas, \emph{{Constraints on sterile neutrino dark
  matter}}, \href{https://doi.org/10.1103/PhysRevD.74.023527}{\emph{Phys. Rev.}
  {\bfseries D74} (2006) 023527}
  [\href{https://arxiv.org/abs/astro-ph/0605271}{{\ttfamily
  astro-ph/0605271}}].

\bibitem{Menci:2017nsr}
N.~Menci, A.~Merle, M.~Totzauer, A.~Schneider, A.~Grazian, M.~Castellano
  et~al., \emph{{Fundamental physics with the Hubble Frontier Fields:
  constraining Dark Matter models with the abundance of extremely faint and
  distant galaxies}},
  \href{https://doi.org/10.3847/1538-4357/836/1/61}{\emph{Astrophys. J.}
  {\bfseries 836} (2017) 61}
  [\href{https://arxiv.org/abs/1701.01339}{{\ttfamily 1701.01339}}].

\bibitem{horiuchi:2013noa}
S.~Horiuchi, P.J.~Humphrey, J.~Onorbe, K.N.~Abazajian, M.~Kaplinghat et~al.,
  \emph{{Sterile neutrino dark matter bounds from galaxies of the Local
  Group}}, \href{https://doi.org/10.1103/PhysRevD.89.025017}{\emph{Phys.Rev.}
  {\bfseries D89} (2014) 025017}
  [\href{https://arxiv.org/abs/1311.0282}{{\ttfamily 1311.0282}}].

\bibitem{Abazajian:2017tcc}
K.N.~Abazajian, \emph{{Sterile neutrinos in cosmology}},
  \href{https://doi.org/10.1016/j.physrep.2017.10.003}{\emph{Phys. Rept.}
  {\bfseries 711-712} (2017) 1}
  [\href{https://arxiv.org/abs/1705.01837}{{\ttfamily 1705.01837}}].

\bibitem{colombi:1995ze}
S.~Colombi, S.~Dodelson and L.M.~Widrow, \emph{Large scale structure tests of
  warm dark matter}, {\emph{Astrophys. J.} {\bfseries 458} (1996) 1}
  [\href{https://arxiv.org/abs/astro-ph/9505029}{{\ttfamily
  astro-ph/9505029}}].

\bibitem{abazajian:2005xn}
K.~Abazajian, \emph{Linear cosmological structure limits on warm dark matter},
  {\emph{Phys. Rev.} {\bfseries D73} (2006) 063513}
  [\href{https://arxiv.org/abs/astro-ph/0512631}{{\ttfamily
  astro-ph/0512631}}].

\bibitem{abazajian:2014gza}
K.N.~Abazajian, \emph{{Resonantly-Produced 7 keV Sterile Neutrino Dark Matter
  Models and the Properties of Milky Way Satellites}},
  \href{https://doi.org/10.1103/PhysRevLett.112.161303}{\emph{Phys.Rev.Lett.}
  {\bfseries 112} (2014) 161303}
  [\href{https://arxiv.org/abs/1403.0954}{{\ttfamily 1403.0954}}].

\bibitem{boylankolchin:2011de}
M.~Boylan-Kolchin, J.S.~Bullock and M.~Kaplinghat, \emph{{Too big to fail? The
  puzzling darkness of massive Milky Way subhaloes}},
  {\emph{Mon.Not.Roy.Astron.Soc.} {\bfseries 415} (2011) L40}
  [\href{https://arxiv.org/abs/1103.0007}{{\ttfamily 1103.0007}}].

\bibitem{boylankolchin:2011dk}
M.~Boylan-Kolchin, J.S.~Bullock and M.~Kaplinghat, \emph{{The Milky Way's
  bright satellites as an apparent failure of LCDM}},
  \href{https://doi.org/10.1111/j.1365-2966.2012.20695.x}{\emph{Mon.Not.Roy.Astron.Soc.}
  {\bfseries 422} (2012) 1203}
  [\href{https://arxiv.org/abs/1111.2048}{{\ttfamily 1111.2048}}].

\bibitem{Bullock:2017xww}
J.S.~Bullock and M.~Boylan-Kolchin, \emph{{Small-Scale Challenges to the
  $\Lambda$CDM Paradigm}},
  \href{https://doi.org/10.1146/annurev-astro-091916-055313}{\emph{Ann. Rev.
  Astron. Astrophys.} {\bfseries 55} (2017) 343}
  [\href{https://arxiv.org/abs/1707.04256}{{\ttfamily 1707.04256}}].

\bibitem{lovell:2011rd}
M.R.~Lovell, V.~Eke, C.S.~Frenk, L.~Gao, A.~Jenkins et~al., \emph{{The Haloes
  of Bright Satellite Galaxies in a Warm Dark Matter Universe}},
  \href{https://doi.org/10.1111/j.1365-2966.2011.20200.x}{\emph{Mon.Not.Roy.Astron.Soc.}
  {\bfseries 420} (2012) 2318}
  [\href{https://arxiv.org/abs/1104.2929}{{\ttfamily 1104.2929}}].

\bibitem{anderhalden:2012jc}
D.~Anderhalden, A.~Schneider, A.V.~Maccio, J.~Diemand and G.~Bertone,
  \emph{{Hints on the Nature of Dark Matter from the Properties of Milky Way
  Satellites}},
  \href{https://doi.org/10.1088/1475-7516/2013/03/014}{\emph{JCAP} {\bfseries
  1303} (2013) 014} [\href{https://arxiv.org/abs/1212.2967}{{\ttfamily
  1212.2967}}].

\bibitem{Horiuchi:2015qri}
S.~Horiuchi, B.~Bozek, K.N.~Abazajian, M.~Boylan-Kolchin, J.S.~Bullock,
  S.~Garrison-Kimmel et~al., \emph{{Properties of resonantly produced sterile
  neutrino dark matter subhaloes}},
  \href{https://doi.org/10.1093/mnras/stv2922}{\emph{Mon. Not. Roy. Astron.
  Soc.} {\bfseries 456} (2016) 4346}
  [\href{https://arxiv.org/abs/1512.04548}{{\ttfamily 1512.04548}}].

\bibitem{Bozek:2015bdo}
B.~Bozek, M.~Boylan-Kolchin, S.~Horiuchi, S.~Garrison-Kimmel, K.~Abazajian and
  J.S.~Bullock, \emph{{Resonant Sterile Neutrino Dark Matter in the Local and
  High-z Universe}}, \href{https://doi.org/10.1093/mnras/stw688}{\emph{Mon.
  Not. Roy. Astron. Soc.} {\bfseries 459} (2016) 1489}
  [\href{https://arxiv.org/abs/1512.04544}{{\ttfamily 1512.04544}}].

\bibitem{polisensky:2010rw}
E.~Polisensky and M.~Ricotti, \emph{{Constraints on the Dark Matter Particle
  Mass from the Number of Milky Way Satellites}},
  \href{https://doi.org/10.1103/PhysRevD.83.043506}{\emph{Phys.Rev.} {\bfseries
  D83} (2011) 043506} [\href{https://arxiv.org/abs/1004.1459}{{\ttfamily
  1004.1459}}].

\bibitem{seljak:2006qw}
U.~Seljak, A.~Makarov, P.~McDonald and H.~Trac, \emph{{Can sterile neutrinos be
  the dark matter?}},
  \href{https://doi.org/10.1103/PhysRevLett.97.191303}{\emph{Phys. Rev. Lett.}
  {\bfseries 97} (2006) 191303}
  [\href{https://arxiv.org/abs/astro-ph/0602430}{{\ttfamily
  astro-ph/0602430}}].

\bibitem{Cherry:2017dwu}
J.F.~Cherry and S.~Horiuchi, \emph{{Closing in on Resonantly Produced Sterile
  Neutrino Dark Matter}},  \href{https://arxiv.org/abs/1701.07874}{{\ttfamily
  1701.07874}}.

\bibitem{Nadler:2020prv}
{\scshape DES} collaboration, \emph{{Milky Way Satellite Census. III.
  Constraints on Dark Matter Properties from Observations of Milky Way
  Satellite Galaxies}},  \href{https://arxiv.org/abs/2008.00022}{{\ttfamily
  2008.00022}}.

\bibitem{Gilman:2019nap}
D.~Gilman, S.~Birrer, A.~Nierenberg, T.~Treu, X.~Du and A.~Benson, \emph{{Warm
  dark matter chills out: constraints on the halo mass function and the
  free-streaming length of dark matter with eight quadruple-image strong
  gravitational lenses}},
  \href{https://doi.org/10.1093/mnras/stz3480}{\emph{Mon. Not. Roy. Astron.
  Soc.} {\bfseries 491} (2020) 6077}
  [\href{https://arxiv.org/abs/1908.06983}{{\ttfamily 1908.06983}}].

\bibitem{Nadler:2021dft}
E.O.~Nadler, S.~Birrer, D.~Gilman, R.H.~Wechsler, X.~Du, A.~Benson et~al.,
  \emph{{Dark Matter Constraints from a Unified Analysis of Strong
  Gravitational Lenses and Milky Way Satellite Galaxies}},
  \href{https://arxiv.org/abs/2101.07810}{{\ttfamily 2101.07810}}.

\bibitem{D_Aloisio_2010}
A.~D’Aloisio and P.~Natarajan, \emph{Cosmography with cluster strong lenses:
  the influence of substructure and line-of-sight haloes},
  \href{https://doi.org/10.1111/j.1365-2966.2010.17795.x}{\emph{Monthly Notices
  of the Royal Astronomical Society} {\bfseries 411} (2010) 1628–1640}.

\bibitem{Richardson:2021onm}
T.R.G.~Richardson, J.~St\"ucker, R.E.~Angulo and O.~Hahn, \emph{{Non-Halo
  Structures and their Effects on Gravitational Lensing}},
  \href{https://arxiv.org/abs/2101.07806}{{\ttfamily 2101.07806}}.

\bibitem{Hermans:2020skz}
J.~Hermans, N.~Banik, C.~Weniger, G.~Bertone and G.~Louppe, \emph{{Towards
  constraining warm dark matter with stellar streams through neural
  simulation-based inference}},
  \href{https://arxiv.org/abs/2011.14923}{{\ttfamily 2011.14923}}.

\bibitem{deGouvea:2019phk}
A.~De~Gouv\^ea, M.~Sen, W.~Tangarife and Y.~Zhang, \emph{{Dodelson-Widrow
  Mechanism in the Presence of Self-Interacting Neutrinos}},
  \href{https://doi.org/10.1103/PhysRevLett.124.081802}{\emph{Phys. Rev. Lett.}
  {\bfseries 124} (2020) 081802}
  [\href{https://arxiv.org/abs/1910.04901}{{\ttfamily 1910.04901}}].

\bibitem{Acciarri:2015uup}
{\scshape DUNE} collaboration, \emph{{Long-Baseline Neutrino Facility (LBNF)
  and Deep Underground Neutrino Experiment (DUNE)}: {Conceptual Design Report,
  Volume 2: The Physics Program for DUNE at LBNF}},
  \href{https://arxiv.org/abs/1512.06148}{{\ttfamily 1512.06148}}.

\bibitem{Shrock:1974nd}
R.~Shrock, \emph{{Decay l0 ---> nu(lepton) gamma in gauge theories of weak and
  electromagnetic interactions}},
  \href{https://doi.org/10.1103/PhysRevD.9.743}{\emph{Phys. Rev.} {\bfseries
  D9} (1974) 743}.

\bibitem{Pal:1981rm}
P.B.~Pal and L.~Wolfenstein, \emph{Radiative decays of massive neutrinos},
  {\emph{Phys. Rev.} {\bfseries D25} (1982) 766}.

\bibitem{Drees:2000qr}
M.~Drees, \emph{{Comment on `A New dark matter candidate: Nonthermal sterile
  neutrinos'}},  \href{https://arxiv.org/abs/hep-ph/0003127}{{\ttfamily
  hep-ph/0003127}}.

\bibitem{dolgov:2000ew}
A.D.~Dolgov and S.H.~Hansen, \emph{Massive sterile neutrinos as warm dark
  matter}, {\emph{Astropart. Phys.} {\bfseries 16} (2002) 339}
  [\href{https://arxiv.org/abs/hep-ph/0009083}{{\ttfamily hep-ph/0009083}}].

\bibitem{abazajian:2001nj}
K.~Abazajian, G.M.~Fuller and M.~Patel, \emph{Sterile neutrino hot, warm, and
  cold dark matter}, {\emph{Phys. Rev.} {\bfseries D64} (2001) 023501}
  [\href{https://arxiv.org/abs/astro-ph/0101524}{{\ttfamily
  astro-ph/0101524}}].

\bibitem{abazajian:2001vt}
K.~Abazajian, G.M.~Fuller and W.H.~Tucker, \emph{Direct detection of warm dark
  matter in the x-ray}, {\emph{Astrophys. J.} {\bfseries 562} (2001) 593}
  [\href{https://arxiv.org/abs/astro-ph/0106002}{{\ttfamily
  astro-ph/0106002}}].

\bibitem{abazajian:2006jc}
K.N.~Abazajian, M.~Markevitch, S.M.~Koushiappas and R.C.~Hickox, \emph{{Limits
  on the radiative decay of sterile neutrino dark matter from the unresolved
  cosmic and soft X-ray backgrounds}},
  \href{https://doi.org/10.1103/PhysRevD.75.063511}{\emph{Phys. Rev.}
  {\bfseries D75} (2007) 063511}
  [\href{https://arxiv.org/abs/astro-ph/0611144}{{\ttfamily
  astro-ph/0611144}}].

\bibitem{bulbul:2014sua}
E.~Bulbul, M.~Markevitch, A.~Foster, R.K.~Smith, M.~Loewenstein et~al.,
  \emph{{Detection of An Unidentified Emission Line in the Stacked X-ray
  spectrum of Galaxy Clusters}},
  \href{https://doi.org/10.1088/0004-637X/789/1/13}{\emph{Astrophys.J.}
  {\bfseries 789} (2014) 13} [\href{https://arxiv.org/abs/1402.2301}{{\ttfamily
  1402.2301}}].

\bibitem{Boyarsky:2014jta}
A.~Boyarsky, O.~Ruchayskiy, D.~Iakubovskyi and J.~Franse, \emph{{Unidentified
  Line in X-Ray Spectra of the Andromeda Galaxy and Perseus Galaxy Cluster}},
  \href{https://doi.org/10.1103/PhysRevLett.113.251301}{\emph{Phys.Rev.Lett.}
  {\bfseries 113} (2014) 251301}
  [\href{https://arxiv.org/abs/1402.4119}{{\ttfamily 1402.4119}}].

\bibitem{Neronov:2016wdd}
A.~Neronov, D.~Malyshev and D.~Eckert, \emph{{Decaying dark matter search with
  NuSTAR deep sky observations}},
  \href{https://doi.org/10.1103/PhysRevD.94.123504}{\emph{Phys. Rev.}
  {\bfseries D94} (2016) 123504}
  [\href{https://arxiv.org/abs/1607.07328}{{\ttfamily 1607.07328}}].

\bibitem{Bergstrom:1997fj}
L.~Bergstrom, P.~Ullio and J.H.~Buckley, \emph{{Observability of gamma-rays
  from dark matter neutralino annihilations in the Milky Way halo}},
  \href{https://doi.org/10.1016/S0927-6505(98)00015-2}{\emph{Astropart.Phys.}
  {\bfseries 9} (1998) 137}
  [\href{https://arxiv.org/abs/astro-ph/9712318}{{\ttfamily
  astro-ph/9712318}}].

\bibitem{Hofmann:2019ihc}
F.~Hofmann and C.~Wegg, \emph{{7.1 keV sterile neutrino dark matter constraints
  from a deep Chandra X-ray observation of the Galactic bulge Limiting
  Window}}, \href{https://doi.org/10.1051/0004-6361/201935561}{\emph{Astron.
  Astrophys.} {\bfseries 625} (2019) L7}
  [\href{https://arxiv.org/abs/1905.00916}{{\ttfamily 1905.00916}}].

\bibitem{Dessert:2018qih}
C.~Dessert, N.L.~Rodd and B.R.~Safdi, \emph{{The dark matter interpretation of
  the 3.5-keV line is inconsistent with blank-sky observations}},
  \href{https://doi.org/10.1126/science.aaw3772}{\emph{Science} {\bfseries 367}
  (2020) 1465} [\href{https://arxiv.org/abs/1812.06976}{{\ttfamily
  1812.06976}}].

\bibitem{Abazajian:2020unr}
K.N.~Abazajian, \emph{{Technical Comment on ``The dark matter interpretation of
  the 3.5-keV line is inconsistent with blank-sky observations''}},
  \href{https://arxiv.org/abs/2004.06170}{{\ttfamily 2004.06170}}.

\bibitem{Boyarsky:2020hqb}
A.~Boyarsky, D.~Malyshev, O.~Ruchayskiy and D.~Savchenko, \emph{{Technical
  comment on the paper of Dessert et al. ''The dark matter interpretation of
  the 3.5 keV line is inconsistent with blank-sky observations''}},
  \href{https://arxiv.org/abs/2004.06601}{{\ttfamily 2004.06601}}.

\bibitem{Foster:2021ngm}
J.W.~Foster, M.~Kongsore, C.~Dessert, Y.~Park, N.L.~Rodd, K.~Cranmer et~al.,
  \emph{{A deep search for decaying dark matter with XMM-Newton blank-sky
  observations}},  \href{https://arxiv.org/abs/2102.02207}{{\ttfamily
  2102.02207}}.

\bibitem{Sicilian:2020glg}
D.~Sicilian, N.~Cappelluti, E.~Bulbul, F.~Civano, M.~Moscetti and
  C.S.~Reynolds, \emph{{Probing the Milky Way\textquoteright{}s Dark Matter
  Halo for the 3.5 keV Line}},
  \href{https://doi.org/10.3847/1538-4357/abbee9}{\emph{Astrophys. J.}
  {\bfseries 905} (2020) 146}
  [\href{https://arxiv.org/abs/2008.02283}{{\ttfamily 2008.02283}}].

\bibitem{Riemer-Sorensen:2015kqa}
S.~Riemer-S\o{}rensen et~al., \emph{{Dark matter line emission constraints from
  NuSTAR observations of the Bullet Cluster}},
  \href{https://doi.org/10.1088/0004-637X/810/1/48}{\emph{Astrophys. J.}
  {\bfseries 810} (2015) 48}
  [\href{https://arxiv.org/abs/1507.01378}{{\ttfamily 1507.01378}}].

\bibitem{Perez:2016tcq}
K.~Perez, K.C.Y.~Ng, J.F.~Beacom, C.~Hersh, S.~Horiuchi and R.~Krivonos,
  \emph{{Almost closing the \ensuremath{\nu}MSM sterile neutrino dark matter
  window with NuSTAR}},
  \href{https://doi.org/10.1103/PhysRevD.95.123002}{\emph{Phys. Rev. D}
  {\bfseries 95} (2017) 123002}
  [\href{https://arxiv.org/abs/1609.00667}{{\ttfamily 1609.00667}}].

\bibitem{Ng:2019gch}
K.C.Y.~Ng, B.M.~Roach, K.~Perez, J.F.~Beacom, S.~Horiuchi, R.~Krivonos et~al.,
  \emph{{New Constraints on Sterile Neutrino Dark Matter from $NuSTAR$ M31
  Observations}}, \href{https://doi.org/10.1103/PhysRevD.99.083005}{\emph{Phys.
  Rev. D} {\bfseries 99} (2019) 083005}
  [\href{https://arxiv.org/abs/1901.01262}{{\ttfamily 1901.01262}}].

\bibitem{Roach:2019ctw}
B.M.~Roach, K.C.Y.~Ng, K.~Perez, J.F.~Beacom, S.~Horiuchi, R.~Krivonos et~al.,
  \emph{{NuSTAR Tests of Sterile-Neutrino Dark Matter: New Galactic Bulge
  Observations and Combined Impact}},
  \href{https://doi.org/10.1103/PhysRevD.101.103011}{\emph{Phys. Rev. D}
  {\bfseries 101} (2020) 103011}
  [\href{https://arxiv.org/abs/1908.09037}{{\ttfamily 1908.09037}}].

\bibitem{Boyarsky:2007ge}
A.~Boyarsky, D.~Malyshev, A.~Neronov and O.~Ruchayskiy, \emph{{Constraining DM
  properties with SPI}},
  \href{https://doi.org/10.1111/j.1365-2966.2008.13003.x}{\emph{Mon. Not. Roy.
  Astron. Soc.} {\bfseries 387} (2008) 1345}
  [\href{https://arxiv.org/abs/0710.4922}{{\ttfamily 0710.4922}}].

\bibitem{Ng:2015gfa}
K.C.Y.~Ng, S.~Horiuchi, J.M.~Gaskins, M.~Smith and R.~Preece, \emph{{Improved
  Limits on Sterile Neutrino Dark Matter using Full-Sky Fermi Gamma-Ray Burst
  Monitor Data}}, \href{https://doi.org/10.1103/PhysRevD.92.043503}{\emph{Phys.
  Rev.} {\bfseries D92} (2015) 043503}
  [\href{https://arxiv.org/abs/1504.04027}{{\ttfamily 1504.04027}}].

\bibitem{malyshev:2014xqa}
D.~Malyshev, A.~Neronov and D.~Eckert, \emph{{Constraints on 3.55 keV line
  emission from stacked observations of dwarf spheroidal galaxies}},
  \href{https://doi.org/10.1103/PhysRevD.90.103506}{\emph{Phys.Rev.} {\bfseries
  D90} (2014) 103506} [\href{https://arxiv.org/abs/1408.3531}{{\ttfamily
  1408.3531}}].

\bibitem{Boyarsky:2005us}
A.~Boyarsky, A.~Neronov, O.~Ruchayskiy and M.~Shaposhnikov, \emph{{Constraints
  on sterile neutrino as a dark matter candidate from the diffuse x-ray
  background}},
  \href{https://doi.org/10.1111/j.1365-2966.2006.10458.x}{\emph{Mon.Not.Roy.Astron.Soc.}
  {\bfseries 370} (2006) 213}
  [\href{https://arxiv.org/abs/astro-ph/0512509}{{\ttfamily
  astro-ph/0512509}}].

\bibitem{Boyarsky:2006zi}
A.~Boyarsky, A.~Neronov, O.~Ruchayskiy and M.~Shaposhnikov, \emph{{Restrictions
  on parameters of sterile neutrino dark matter from observations of galaxy
  clusters}},
  \href{https://doi.org/10.1103/PhysRevD.74.103506}{\emph{Phys.Rev.} {\bfseries
  D74} (2006) 103506} [\href{https://arxiv.org/abs/astro-ph/0603368}{{\ttfamily
  astro-ph/0603368}}].

\bibitem{Neronov:2015kca}
A.~Neronov and D.~Malyshev, \emph{{Toward a full test of the $\nu$MSM sterile
  neutrino dark matter model with Athena}},
  \href{https://doi.org/10.1103/PhysRevD.93.063518}{\emph{Phys. Rev.}
  {\bfseries D93} (2016) 063518}
  [\href{https://arxiv.org/abs/1509.02758}{{\ttfamily 1509.02758}}].

\bibitem{Zhong:2020wre}
D.~Zhong, M.~Valli and K.N.~Abazajian, \emph{{Near to long-term forecasts in
  x-ray and gamma-ray bands: Are we entering the era of dark matter
  astronomy?}}, \href{https://doi.org/10.1103/PhysRevD.102.083008}{\emph{Phys.
  Rev. D} {\bfseries 102} (2020) 083008}
  [\href{https://arxiv.org/abs/2003.00148}{{\ttfamily 2003.00148}}].

\bibitem{Malyshev:2020hcc}
D.~Malyshev, C.~Thorpe-Morgan, A.~Santangelo, J.~Jochum and S.-N.~Zhang,
  \emph{{$eXTP$ perspectives for the $\nu$MSM sterile neutrino dark matter
  model}}, \href{https://doi.org/10.1103/PhysRevD.101.123009}{\emph{Phys. Rev.
  D} {\bfseries 101} (2020) 123009}
  [\href{https://arxiv.org/abs/2001.07014}{{\ttfamily 2001.07014}}].

\bibitem{Boyarsky:2014ska}
A.~Boyarsky, J.~Franse, D.~Iakubovskyi and O.~Ruchayskiy, \emph{{Checking the
  Dark Matter Origin of a 3.53 keV Line with the Milky Way Center}},
  \href{https://doi.org/10.1103/PhysRevLett.115.161301}{\emph{Phys. Rev. Lett.}
  {\bfseries 115} (2015) 161301}
  [\href{https://arxiv.org/abs/1408.2503}{{\ttfamily 1408.2503}}].

\bibitem{Iakubovskyi:2015dna}
D.~Iakubovskyi, E.~Bulbul, A.R.~Foster, D.~Savchenko and V.~Sadova,
  \emph{{Testing the origin of ~3.55 keV line in individual galaxy clusters
  observed with XMM-Newton}},
  \href{https://arxiv.org/abs/1508.05186}{{\ttfamily 1508.05186}}.

\bibitem{Cappelluti:2017ywp}
N.~Cappelluti, E.~Bulbul, A.~Foster, P.~Natarajan, M.C.~Urry, M.W.~Bautz
  et~al., \emph{{Searching for the 3.5 keV Line in the Deep Fields with
  Chandra: the 10 Ms observations}},
  \href{https://arxiv.org/abs/1701.07932}{{\ttfamily 1701.07932}}.

\bibitem{Jeltema:2014qfa}
T.E.~Jeltema and S.~Profumo, \emph{{Discovery of a 3.5 keV line in the Galactic
  Centre and a critical look at the origin of the line across astronomical
  targets}}, \href{https://doi.org/10.1093/mnras/stv768}{\emph{Mon. Not. Roy.
  Astron. Soc.} {\bfseries 450} (2015) 2143}
  [\href{https://arxiv.org/abs/1408.1699}{{\ttfamily 1408.1699}}].

\bibitem{Bulbul:2014ala}
E.~Bulbul, M.~Markevitch, A.R.~Foster, R.K.~Smith, M.~Loewenstein and
  S.W.~Randall, \emph{{Comment on ``Dark matter searches going bananas: the
  contribution of Potassium (and Chlorine) to the 3.5 keV line''}},
  \href{https://arxiv.org/abs/1409.4143}{{\ttfamily 1409.4143}}.

\bibitem{Boyarsky:2014paa}
A.~Boyarsky, J.~Franse, D.~Iakubovskyi and O.~Ruchayskiy, \emph{{Comment on the
  paper ``Dark matter searches going bananas: the contribution of Potassium
  (and Chlorine) to the 3.5 keV line'' by T. Jeltema and S. Profumo}},
  \href{https://arxiv.org/abs/1408.4388}{{\ttfamily 1408.4388}}.

\bibitem{Speckhard:2015eva}
E.G.~Speckhard, K.C.Y.~Ng, J.F.~Beacom and R.~Laha, \emph{{Dark Matter Velocity
  Spectroscopy}},
  \href{https://doi.org/10.1103/PhysRevLett.116.031301}{\emph{Phys. Rev. Lett.}
  {\bfseries 116} (2016) 031301}
  [\href{https://arxiv.org/abs/1507.04744}{{\ttfamily 1507.04744}}].

\bibitem{Nandra:2013shg}
K.~Nandra et~al., \emph{{The Hot and Energetic Universe: A White Paper
  presenting the science theme motivating the Athena+ mission}},
  \href{https://arxiv.org/abs/1306.2307}{{\ttfamily 1306.2307}}.

\bibitem{LynxTeam:2018usc}
{\scshape Lynx Team} collaboration, \emph{{The Lynx Mission Concept Study
  Interim Report}},  \href{https://arxiv.org/abs/1809.09642}{{\ttfamily
  1809.09642}}.

\bibitem{Zhang:2016ach}
{\scshape eXTP} collaboration, \emph{{eXTP -- enhanced X-ray Timing and
  Polarimetry Mission}}, \href{https://doi.org/10.1117/12.2232034}{\emph{Proc.
  SPIE Int. Soc. Opt. Eng.} {\bfseries 9905} (2016) 99051Q}
  [\href{https://arxiv.org/abs/1607.08823}{{\ttfamily 1607.08823}}].

\bibitem{Zandanel:2015xca}
F.~Zandanel, C.~Weniger and S.~Ando, \emph{{The role of the eROSITA all-sky
  survey in searches for sterile neutrino dark matter}},
  \href{https://doi.org/10.1088/1475-7516/2015/09/060}{\emph{JCAP} {\bfseries
  1509} (2015) 060} [\href{https://arxiv.org/abs/1505.07829}{{\ttfamily
  1505.07829}}].

\bibitem{Barinov:2020hiq}
V.V.~Barinov, D.S.~Gorbunov, R.A.~Burenin and R.A.~Krivonos, \emph{{Towards
  Testing Sterile Neutrino Dark Matter with SRG Mission}},
  \href{https://arxiv.org/abs/2007.07969}{{\ttfamily 2007.07969}}.

\end{thebibliography}\endgroup

\end{document}